\documentclass[times]{nmeauth}

\usepackage{moreverb}

\usepackage[colorlinks,bookmarksopen,bookmarksnumbered,citecolor=red,urlcolor=red]{hyperref}

\newcommand\BibTeX{{\rmfamily B\kern-.05em \textsc{i\kern-.025em b}\kern-.08em
T\kern-.1667em\lower.7ex\hbox{E}\kern-.125emX}}

\usepackage[tight]{subfigure}		% enable pictures side-by-side
\usepackage{stmaryrd} %jump operator
\usepackage{xcolor,graphicx}                                    % basic graphic packages

\usepackage{algorithm}     % LOAD ALGORITHM PACKAGES
\usepackage{algorithmic}

\usepackage{amssymb,amsmath,amstext, amsfonts, mathrsfs, mathtools} %   basic math packages
\usepackage{arydshln} % dashlines in array (for matrix)
\newtheorem*{remark}{Remark}

\usepackage{cite}

\usepackage{anyfontsize}

\graphicspath{{fig/}}
\DeclareGraphicsExtensions{.pdf}

\begin{document}

\runningheads{C.~Ager et al.}{A consistent approach for FSCI}

\title{A consistent approach for fluid-structure-contact interaction based on a porous flow model for rough surface contact}

\author{C.~Ager\textsuperscript{1}\corrauth,B.~Schott\textsuperscript{1},A.-T.~Vuong\textsuperscript{1},A.~Popp\textsuperscript{2},W.A.~Wall\textsuperscript{1}}
\address{\textsuperscript{1} Institute for Computational Mechanics ,
				Technische Universit\"a{}t M\"u{}nchen,\linebreak
				Boltzmannstr. 15, 85747 Garching b. M\"u{}nchen\linebreak
				\textsuperscript{2} Institute for Mathematics and Computer-Based Simulation,
				University of the Bundeswehr Munich,\linebreak
				Werner-Heisenberg-Weg 39, 85577 Neubiberg, Germany}

\corraddr{C.~Ager, Institute for Computational Mechanics, Technical University of Munich, Boltzmannstra{\ss}e 15, D-85747 Garching, Germany. E-mail:~ager@lnm.mw.tum.de}

\begin{abstract}
Simulation approaches for fluid-structure-contact interaction, especially if requested to be consistent even down to the real contact scenarios, 
belong to the most challenging and still unsolved problems in computational mechanics. 
The main challenges are twofold -- 
one is to have a correct physical model for this scenario,
and the other one is to have a numerical method that is capable of working and being consistent down to a zero gap. 
And when analyzing such
challenging setups of fluid-structure interaction that include contact of submersed solid components it gets obvious 
that the influence of surface roughness effects is essential for a physical consistent modeling of such configurations.
To capture this system behavior, we present a continuum mechanical model which is able to include the effects of the surface microstructure in a fluid-structure-contact interaction framework.
An averaged representation for the mixture of fluid and solid on the rough surfaces, which is of major interest for the macroscopic response of such a system, is introduced therein.
The inherent coupling of the macroscopic fluid flow and the flow inside the rough surfaces, the stress exchange of all contacting solid bodies involved, and the interaction between fluid and solid 
is included in the construction of the model.
Although the physical model is not restricted to finite element based methods, a numerical approach with its core based on the Cut Finite Element Method (CutFEM), enabling topological changes of the fluid domain to solve the presented model numerically, is introduced.
Such a CutFEM based approach is able to deal with the numerical challenges mentioned above.
Different test cases give a perspective towards the potential capabilities of the presented physical model and numerical approach.

\vspace{1cm}
\end{abstract}


\maketitle

\definecolor{darkgreen}{rgb}{0,0.5,0}
\definecolor{green}{rgb}{0,1,0}
\definecolor{orange}{rgb}{1.0,0.5,0.0}
\definecolor{darkorange}{rgb}{0.6,0.3,0.0}

\newcommand{\tns}[1]{\underline{\boldsymbol{#1}}}
\newcommand{\hattns}[1]{\hat{\underline{\boldsymbol{#1}}}}
\newcommand{\disctns}[1]{\underline{\mathfrak{#1}}}
\newcommand{\mat}[1]{\underline{\boldsymbol{#1}}}
\newcommand{\partiald}[2]{\dfrac{ \partial #1}{\partial #2}}
\newcommand{\partialdt}[2]{\dfrac{ \partial^2 #1}{\partial {#2}^2}}
\newcommand{\pddt}[1]{\dfrac{\partial {#1}}{\partial t}}
\newcommand{\pddtX}[1]{\left.\dfrac{\partial {#1}}{\partial t}\right|_{\tns{X}}}
\newcommand{\pddtx}[1]{\left.\dfrac{\partial {#1}}{\partial t}\right|_{\tns{x}}}
\newcommand{\innerp}[3]{\left(#1, #2\right)_{#3}}
\newcommand{\innerpb}[3]{\left\langle #1, #2 \right\rangle_{#3}}
\newcommand{\jump}[1]{\ensuremath{\left[\!\left[#1\right]\!\right]} }
\newcommand{\grad}   { \boldsymbol{\nabla}    }
\newcommand{\gradRef}   { \boldsymbol{\nabla}_0    }
\newcommand{\laplace}   { \boldsymbol{\Delta}}
\renewcommand{\div} { \grad \! \cdot \!} 
\newcommand{\divRef} { \gradRef \! \cdot \!} 
\newcommand{\tr}   { \rm{tr}\,    }
\newcommand{\ccprod}   { \boldsymbol{:}   }
\newcommand{\zerovec}{\tns{0}}
\newcommand{\dzerovec}{\disctns{0}}
\newcommand{\unity}{\tns{I}}
\newcommand{\abs}[1]{\ensuremath{\left|#1\right|}}
\newcommand{\averg}[1]{\ensuremath{\left\lbrace#1\right\rbrace} }
\newcommand{\avergt}[1]{\ensuremath{<#1> }}
\newcommand{\norminf}[1]{\ensuremath{\left|\left|#1\right|\right|_{\infty}}}
\newcommand{\normL}[2]{\ensuremath{\left|\left|#1\ensuremath\right|\right|_{L^2\left(#2\right)}}}

\newcommand{\testfkt}{\delta \tns{\velocityf}}
\newcommand{\stress}{\tns{\sigma}}
\newcommand{\normal}{\tns{n}}
\newcommand{\tangent}{\tns{t}}
\newcommand{\geninterface}{\interface^*}
\newcommand{\residual}{\underline{\mathcal{R}}}
\newcommand{\btime}{t_0}
\newcommand{\etime}{t_E}
\newcommand{\timesfulltime}{\times [\btime, \etime]}
\newcommand{\Pnormallong}{\left(\normal \otimes \normal\right)}
\newcommand{\Ptangentlong}{\left(\unity - \normal \otimes \normal\right)}
\newcommand{\Pnormal}{\tns P_n}
\newcommand{\Ptangent}{\tns P_t}
\newcommand{\WeakB}[1][]{B_{#1}}
\newcommand{\WeakC}[1][]{C_{#1}}
\newcommand{\WeakD}[1][]{D_{#1}}
\newcommand{\Weakc}[1][]{c_{#1}}
\newcommand{\Weakd}[1][]{d_{#1}}
\newcommand{\Weakf}[1][]{f_{#1}}
\newcommand{\strainenergy}[1]{\psi^{#1}}

\newcommand{\domain}{\Omega}
\newcommand{\fullbound}{\partial \domain}
\newcommand{\refdomain}{\Omega_0}
\newcommand{\fullrefbound}{\partial \refdomain}

\newcommand{\domainf}{\Omega^F}
\newcommand{\velocityf}{v}
\newcommand{\displacementf}{u}
\newcommand{\pressuref}{p}
\newcommand{\densityf}{\rho^F}
\newcommand{\velf}[1][]{\tns{\velocityf}^{F #1}}
\newcommand{\velfn}{\velf\tns{\velocityf}^{F,n}}
\newcommand{\velfnp}{\tns{\velocityf}^{F,n+1}}
\newcommand{\velgridf}{\tns{\velocityf}^G}
\newcommand{\velfD}{\tns{\hat{\velocityf}}^F}
\newcommand{\velfB}{\tns{\mathring{\velocityf}}^F}
\newcommand{\pf}{\pressuref^F}
\newcommand{\pfn}{\pressuref^{F,n}}
\newcommand{\pfnp}{\pressuref^{F,n+1}}
\newcommand{\viscf}{\mu^F}
\newcommand{\epsf}{\tns{\epsilon}^F}
\newcommand{\stressf}{\tns{\sigma}^F}
\newcommand{\stressfn}{\tns{\sigma}^{F,n}}
\newcommand{\stressfnp}{\tns{\sigma}^{F,n+1}}
\newcommand{\bodyff}{\hat{\tns b}^F}
\newcommand{\bodyffn}{\hat{\tns b}^{F,n}}
\newcommand{\bodyffnp}{\hat{\tns b}^{F,n+1}}
\newcommand{\normalf}{\tns n^F}
\newcommand{\tangf}{\tns t^F}
\newcommand{\tractionf}{\tns{h}^{F}}
\newcommand{\tractionfN}{\tns{\hat{h}}^{F,N}}
\newcommand{\tractionfNn}{\tns{\hat{h}}^{F,N,n}}
\newcommand{\tractionfNnp}{\tns{\hat{h}}^{F,N,n+1}}
\newcommand{\timef}{t}
\newcommand{\testvelf}[1][]{\delta \tns{\velocityf}^{F #1}}
\newcommand{\testpf}{\delta \pressuref^F}
\newcommand{\thetaf}{\theta}
\newcommand{\timestepf}{\Delta t}
\newcommand{\nboundf}{\interface^{F,N}}
\newcommand{\dboundf}{\interface^{F,D}}
\newcommand{\fullboundf}{\partial \domainf}
\newcommand{\restboundf}{\interface^{F,*}}
\newcommand{\incvelf}{\Delta \disctns{\velocityf}^F}
\newcommand{\incdispf}{\Delta \disctns{\displacementf}^A}
\newcommand{\dispf}{\tns{\displacementf}^A}
\newcommand{\testfktspacef}{V^F}

\newcommand{\timep}{t}
\newcommand{\domainp}{\Omega^P}
\newcommand{\refdomainp}{\Omega^P_0}
\newcommand{\displacementp}{u}
\newcommand{\velocityp}{v}
\newcommand{\pressurep}{p}
\newcommand{\porosity}{\phi}
\newcommand{\porosityB}{\mathring{\porosity}}
\newcommand{\velp}{\tns{\velocityp}^{P}}
\newcommand{\dispp}{\tns{\displacementp}^{P}}
\newcommand{\disppB}{\tns{\mathring{\displacementp}}^P}
\newcommand{\velpsB}{\tns{\mathring{\velocityp}}^{P^S}}
\newcommand{\velpB}{\tns{\mathring{\velocityp}}^{P}}
\newcommand{\velps}{ \partiald{\dispp}{\timep}}
\newcommand{\velpss}{ \dot{\dispp}}
\newcommand{\accps}{ \partialdt{\dispp}{\timep}}
\newcommand{\pp}{\pressurep^{P}}
\newcommand{\densitypf}{\densityf}
\newcommand{\refdensitypf}{\densityf}%\rho^{P^F}_0}
\newcommand{\refdensityps}{\rho^{P^S}_0}
\newcommand{\refmassps}{m^{P^S}_0}%m^{P^S}_0}
\newcommand{\refavdensityps}{\tilde{\rho}^{P^S}_0}%m^{P^S}_0}
\newcommand{\stresspkp}{\tns{S}^P}
\newcommand{\stressp}{\tns{\sigma}^P}
\newcommand{\bodyfpf}{\hat{\tns b}^{P^F}}
\newcommand{\bodyfp}{\hat{\tns b}^{P}}
\newcommand{\refbodyfp}{\bodyfp_0}
\newcommand{\tractionpf}{\tns{h}^{P^F}}
\newcommand{\tractionps}{\tns{h}^{P^S}}
\newcommand{\reftractionps}{\tractionps_0}
\newcommand{\tractionp}{\tns{h}^{P}}
\newcommand{\reftractionp}{\tractionp_0}
\newcommand{\tractionpfN}{\hat{h}^{P^F,N}}
\newcommand{\tractionpN}{\tns{\hat{h}}^{P,N}}
\newcommand{\reftractionpN}{\tractionpN_0}
\newcommand{\viscp}{\viscf}
\newcommand{\permeabpscalar}{k}
\newcommand{\permeabp}{\tns{\permeabpscalar}}
\newcommand{\matpermeabp}{\tns{\matpermeabpscalar}}
\newcommand{\matpermeabpscalar}{K}
\newcommand{\initmatpermeabpscalar}{\mathring{\matpermeabpscalar}}
\newcommand{\Jp}{J^P}
\newcommand{\strainenergyp}{\strainenergy{P}}
\newcommand{\strainenergyps}{\psi^{P^S}}
\newcommand{\strainglp}{\tns{E}^{P}}
\newcommand{\straincgp}{\tns{C}^{P}}
\newcommand{\defgradp}{\tns{F}^{P}}
\newcommand{\testvelp}{\delta \tns{\velocityp}^P}
\newcommand{\testdispp}{\delta \tns{\displacementp}^P}
\newcommand{\testvelpss}{\delta  \dot{\dispp}}
\newcommand{\testpp}{\delta \pressurep^P}
\newcommand{\dboundpf}{\interface^{P^F,D}}
\newcommand{\nboundpf}{\interface^{P^F,N}}
\newcommand{\dboundp}{\interface^{P,D}}
\newcommand{\nboundp}{\interface^{P,N}}
\newcommand{\refdboundp}{\dboundp_0}
\newcommand{\refnboundp}{\nboundp_0}
\newcommand{\restboundp}{\interface^{P,*}}
\newcommand{\refrestboundp}{\restboundp_0}
\newcommand{\restboundpf}{\interface^{P^F,*}}
\newcommand{\fullboundp}{\partial \domainp}
\newcommand{\reffullboundp}{\interface^P_0}
\newcommand{\normalp}{\tns n^P}
\newcommand{\tangp}{\tns t^P}
\newcommand{\snormalp}{\tilde{\tns n}}
\newcommand{\stangp}{\tilde{\tns t}}
\newcommand{\refnormalp}{\normalp_0}
\newcommand{\velpnD}{\hat{\velocityp}^P_n}
\newcommand{\disppD}{\tns{\hat{\displacementp}}^P}
\newcommand{\coordp}{\tns x^P}
\newcommand{\refcoordp}{\tns X^P}
\newcommand{\incvelp}{\Delta \disctns{\velocityp}^P}
\newcommand{\incdispp}{\Delta \disctns{\displacementp}^P}

\newcommand{\displacements}{u}
\newcommand{\velocitys}{v}
\newcommand{\domains}{\Omega^S}
\newcommand{\refdomains}{\domains_0}
\newcommand{\disps}{\tns{\displacements}^{S}}
\newcommand{\vels}{ \partiald{\disps}{\timep}}
\newcommand{\accs}{ \partialdt{\disps}{\timep}}
\newcommand{\defgrads}{\tns{F}^{S}}
\newcommand{\Js}{J^{S}}
\newcommand{\stresspks}{\tns{S}^S}
\newcommand{\stresss}{\tns{\sigma}^S}
\newcommand{\strainenergyNHs}{\strainenergys_{NH}}
\newcommand{\strainenergys}{\psi^{S}}
\newcommand{\straingls}{\tns{E}^{S}}
\newcommand{\densitys}{\rho^{S}_0}
\newcommand{\refdensitys}{\rho^{S}_0}
\newcommand{\refbodyfs}{\hat{\tns b}^{S}_0}
\newcommand{\dispsD}{\tns{\hat{\displacements}}^S}
\newcommand{\dispsB}{\tns{\mathring{\displacements}}^S}
\newcommand{\velsB}{\tns{\mathring{\velocitys}}^S}
\newcommand{\tractions}{\tns{h}^{S}}
\newcommand{\reftractions}{\tractions_0}
\newcommand{\tractionsN}{\tns{\hat{h}}^{S,N}}
\newcommand{\reftractionsN}{\tractionsN_0}
\newcommand{\refnbounds}{\interface^{S,N}_0}
\newcommand{\refdbounds}{\interface^{S,D}_0}
\newcommand{\restbounds}{\interface^{S,*}}
\newcommand{\refrestbounds}{\restbounds_0}
\newcommand{\fullbounds}{\partial \domains}
\newcommand{\fullrefbounds}{\partial \refdomains}
\newcommand{\refnormals}{\normals_0}
\newcommand{\testdisps}{\delta \tns{\displacements}^S}
\newcommand{\coords}{\tns x^S}
\newcommand{\refcoords}{\tns X^S}
\newcommand{\incdisps}{\Delta \disctns{\displacements}^S}
\newcommand{\normals}{\tns n^S}
\newcommand{\timeso}{t}

\newcommand{\BJfac}{{\color{red}\beta_{BJ} \cdot}}
\newcommand{\scaleFt}[1][]{{\color{darkgreen}c^{F #1}_t}}
\newcommand{\scalePt}[1][]{{\color{darkgreen}c^{P #1}_t}}
\newcommand{\scaleSt}[1][]{{\color{darkgreen}c^{S #1}_t}}
\newcommand{\scaleFn}[1][]{{\color{darkgreen}c^{F #1}_n}}
\newcommand{\scalePn}[1][]{{\color{darkgreen}c^{P #1}_n}}
\newcommand{\scaleSn}[1][]{{\color{darkgreen}c^{S #1}_n}}
\newcommand{\interface}{\Gamma}
\newcommand{\refinterface}{\interface_0}
\newcommand{\fsiinterface}{\interface^{FS}}
\newcommand{\reffsiinterface}{\fsiinterface_0}
\newcommand{\fpiinterface}{\interface^{FP}}
\newcommand{\reffpiinterface}{\fpiinterface_0}
\newcommand{\psiinterface}{\interface^{PS}}
\newcommand{\psciinterface}{\interface^{PS,c}}
\newcommand{\pscpiinterfaceh}{\interface^{P,c}_h}
\newcommand{\pscsiinterfaceh}{\interface^{S,c}_h}
\newcommand{\pscpiinterface}{\interface^{P,c}}
\newcommand{\pscsiinterface}{\interface^{S,c}}
\newcommand{\refpsiinterface}{\psiinterface_0}
\newcommand{\refpsciinterface}{\psciinterface_0}
\newcommand{\sliplengh}{\kappa}
\newcommand{\lagmultpss}{\tns \lambda}
\newcommand{\lagmultpssnormal}{\lambda_n}
\newcommand{\lagmultpsp}{\tns \lambda_2}
\newcommand{\testnlagmultpss}{\delta \lambda_{n}}
\newcommand{\testtlagmultpss}{\delta \lambda_{t_i}}
\newcommand{\testlagmultpss}{\delta \tns{\lambda}}
\newcommand{\testnlagmultpsp}{\delta \lambda_{2,n}}
\newcommand{\testtlagmultpsp}{\delta \lambda_{2,t_i}}
\newcommand{\dlagmultpss}{\tns{\lambda}}
\newcommand{\dlagmultpsp}{\disctns{\lambda}_2}
\newcommand{\testdualpss}{\Phi}
\newcommand{\dmatrix}{\mat{D}}
\newcommand{\mmatrix}{\mat{M}}
\newcommand{\pmat}{\mat{P}}
\newcommand{\timefac}{\theta}
\newcommand{\prescontact}{p^{PS,c}}

\newcommand{\identity}{\mat{I}}
\newcommand{\itraction}{\tns h^I}
\newcommand{\itractionf}{\tns h^{F,I}}
\newcommand{\itractions}{\tns h^{S,I}}
\newcommand{\refitractions}{\itractions_0}
\newcommand{\itractionpf}{h^{P^F,I}}
\newcommand{\itractionp}{\tns h^{P,I}}

\newcommand{\dummycite}[1]{{\color{red}[:-)#1] }\color{black}}
\newcommand{\todo}[1]{{\color{blue}(TODO: #1)}\color{black}}

\section{Introduction}
\label{sec:intro}
The interaction of two surfaces with fluid in between is of great interest in various engineering and biomechanical applications.
Any surface, no matter if manufactured or naturally occurring, is not completely smooth but has a microstructure with an average roughness height, which is magnitudes smaller than the object size itself.
As soon as the height of the fluid film between these surfaces is in the same order of magnitude as surface asperities, surface roughness has an essential impact on the physical response of such a system.
Occurring leakage, lubrication, friction, and wear are of great importance for the performance of valves, bearings, gears, and tires, for example.

The challenge to develop a consistent physical model for fluid-structure interaction including contact lies in the multiscale nature of the considered problem.
While at the macroscopic scale the fluid physics are governed by the well-known Navier-Stokes equations supplemented by a no-slip condition on the fluid-structure interface,
this does not necessarily hold true when solid bodies come into contact.
Solving simply this macroscopic physical model does not lead to contact of submersed smooth solid bodies due to the increasing viscous stress when the fluid gap gets smaller.
Therefore a finite fluid gap is retained, which is not in agreement with the observation of contacting bodies.
A deeper insight into the underlying physical process of contacting submersed bodies allows to identify two ways out of this dilemma.

First, the limit of absolutely smooth contacting solid bodies is considered.
Taking into account this assumption, the fluid gap between approaching solid bodies will fall below the validity limit of the macroscopic fluid dynamic models.
As a first consequence a transition of the no-slip interface condition to a slip interface condition 
(including in general a surface roughness dependent sliplength)
into the so called ``Slip-Flow Regime'' (see e.g. \cite{barber2006}),
while retaining the validity of the Navier-Stokes equations in the fluid domain, can be observed. 
Therefore this local relaxation of the tangential no-slip condition has to be incorporated in the physical model,
while retaining the no-slip condition for the remaining part of the interface.
This results in a consistent physical model to consider contact of smooth solid bodies with fluid between the contacting surfaces.
It should be pointed out, that the only exception is the case of two parallel plates, 
where due to the required acceleration of the fluid mass no contact occurs in finite time.

Second, taking into account that any real surface has a microstructure, this contacting process can change fundamentally.
If single asperities come into contact before the validity of the classical fluid equations is lost, the  contacting process described first cannot hold anymore.
From a macroscopic point of view, contact is enabled in this case via a fluid mass transfer from the fluid domain into the rough microstructure.
Therefore considering the rough microstructure of contacting surfaces can be essential for a consistent fluid-structure-contact interaction model.
This is the case when the characteristic roughness height is larger than the limiting size of classical fluid equations, which depends on 
the considered type of fluid and its molecular mean free path length.
In the following we will focus on the modeling of this second process as it is the relevant one for many problems of interest.

To analyze and predict such tribological systems for thin fluid films, the Reynolds equation \cite{reynolds1886},
which can be derived from the Navier-Stokes equation by utilizing assumptions valid for thin film flows, is widely used.
To incorporate the effect of surface roughness without resolving the surfaces, an averaged Reynolds equation is often used to solve for the averaged fluid pressure, see e.g.
\cite{christensen1971,patir1978,tripp1983,bayada1989,prat2002}. In \cite{jai2002,bou2004}, it is shown that significantly fewer degrees of freedom are required for solving the homogenized equations compared to 
the direct equations in order to obtain the pressure field between rough surfaces.
A framework to consider the effects of deformation of structural bodies interacting via a thin fluid film was presented in \cite{yang2009,budt2012}.

A comparison of numerical solutions for the full spatially discretized fluid momentum and continuity equations and the Reynolds approach, tested for a problem setup with
valid thin film approximation presented in \cite{almqvist2004}, shows that there is no significant deviation of the results between both approaches.
Nevertheless, with increasing film size this result does not hold any longer, as the underlying geometrical assumptions of the Reynolds equation become invalid.
In this case, the solution of the full fluid equations seems to be absolutely essential,
even though the computational cost is higher due to the increased number of degrees of freedom as compared to the Reynolds approach.

In fluid-structure interaction problems, structures of arbitrary shape are deformed by the fluid traction acting on the interfaces. 
Herein, the geometry of the fluid domain and all corresponding interface or boundary conditions, which are given by the structural boundaries and outer boundaries, are not known in advance.
Therefore, fluid equations, like the Navier-Stokes equations, are solved spatially resolved without including information concerning the fluid domain or boundary/interface conditions beforehand
and are coupled to the structural deformation.
Details on such methods can be found e.g. in \cite{farhat2004,fernandez2011, Gee2011}, and in \cite{loon2006,santos2008,astorino2009,mayer20103,wick2014,kamensky2015} 
with attempts to treat contact of submersed bodies.
However, none of these methods is able to meet the challenges mentioned above concerning physical modeling and practicability of the numerical approach.
While several of those approaches are numerically inconsistent down to a zero gap or cannot represent the fluid stress discontinuity 
(especially the discontinuity of the fluid pressure) across considered slender bodies,
also surface roughness is typically not considered and the ``no-slip'' condition
is applied on the overall fluid-structure interface. This is only a reasonable assumption as long as the distance between these surfaces is large compared to the microstructure of them.

As soon as contact between solid bodies cannot be ruled out, there is no lower limit for the size of the gap between the surfaces involved.
Therefore, effects arising from the roughness of these surfaces can even dominate the macroscopic overall physical response. 
Two examples are the analysis of leakage in seals or the opening pressure in valves.
Another example arising in physical modeling is the ``no-collision'' paradox that contact between smooth surfaces with ``no-slip'' cannot occur in an
incompressible, viscous fluid in finite time (see \cite{hillairet2009,gerard2015}); this is contrary to the macroscopic observation.
For ``slip'' interface conditions on the colliding surfaces (shown in e.g. \cite{Hocking1973}) or non-smooth surfaces (analyzed in e.g. \cite{cawthorn2010,gerard2010}), contact is possible. 
A physical explanation for this paradox is the lack of consideration of the surface microstructure.
As soon as the microscopic roughness is treated, solid-solid contact can occur \cite{davis2003}.

This aspect highlights the importance of considering the effects of surface roughness to extend the validity within a fluid-structure interaction framework for specific configurations.
However, a direct resolution of the microstructure with a computational discretization is not practicable for engineering applications,
as the focus of interest is mostly on averaged quantities such as the average velocity in the fluid gap of a valve or the average pressure in the fluid gap of a bearing. With such computed averaged quantities, predictions for global quantities such as the leakage flow of a valve or the load capacity of a bearing can be deduced.
Resolving the potentially complex fluid flow between single roughness asperities is generally not necessary for these types of applications 
besides the fact that the exact microstructure is not known at all in most cases.

To reduce the high computational demands arising from a fully-resolving computation, 
as well as the necessity to know the exact microstructure, 
we propose a homogenized model to include the average physical behavior of the roughness layer into the fluid-structure-contact interaction framework.
While this has not been done for real FSI so far, similar ideas have 
been successfully applied to consider roughness within Reynolds equation based formulations.
Homogenizing a domain which consists of a fluid-filled deformable microstructure results in a poroelastic, fluid-saturated averaged domain.
The basic idea of modeling surface roughness as a porous layer can already be found in \cite{Tichy1995, li1999}.
Our novel approach is based on a similar idea but also works for general fluid-structure interaction problems,
applicable for arbitrary shaped domains, including finite deformations of the solid, topological changes of the fluid domain and rearrangement of all interface conditions.

Averaging the surface roughness as a poroelastic layer still allows us to account for many physically essential effects on a macroscopic level, thus in an averaged sense.
Examples are the fluid pressure distribution between contacting bodies, stresses exchanged between contacting solids, 
the deformation of the roughness layer and the resulting fluid to solid fraction within the layer.
Having this physical information available allows us to also include additional models to treat specific physical phenomena of the general problem of colliding bodies in fluid such as, for instance, friction of mixed lubrication contact or wear, which, however, are not in the focus of this contribution.

In addition to the physical modeling, a potential numerical approximation of the proposed model will be presented. 
It is based on spatial finite element discretizations for the structural, the fluid, as well as the poroelastic domain.
It should be pointed out that this is just one possible discretization technique which allows for approximating the proposed physical model.
Alternative approaches, for example based on the finite volume framework, might be also possible.
In the present work, contact occurring between structural bodies is handled by the dual mortar contact approach \cite{hueber2008} for finite deformations \cite{popp2010}.
Topological changes in the fluid domain due to the occurring contact are enabled by the Cut-Finite Element Method (CutFEM) \cite{schott2014,burman2015cutfem,massing2016} applied to the Navier-Stokes equations.
Interface coupling conditions are introduced weakly by Nitsche-based methods \cite{nitsche1971,burman2014,schott2017}.
Finally, a so-called monolithic scheme \cite{heil2004,Kuttler2010,Gee2011,verdugo2016} is applied to solve the global system of equations, which is beneficial in the case of a strong interaction between all involved domains.

The paper is organized as follows:
In Section~\ref{sec:rough_contact_model}, we depict the porous flow based model for rough fluid-structure-contact interaction, followed by the presentation of all governing equations for the different physical domains in Section~\ref{sec:fields}.
In Section~\ref{sec:interfaces}, the coupling conditions on all interfaces between the occurring physical domains,
as well as the interactions of all involved interfaces, are discussed.
Section \ref{sec:nummethod} presents the numerical method applied to solve different exemplary problem configurations.
Three configurations, including a leakage test, a rough surface contacting stamp and a non-return valve, are presented and analyzed in Section~\ref{sec:num_ex}.
Finally, a short summary and conclusion is given in Section~\ref{sec:conclusion}.

\section{Rough surface contact model}
\label{sec:rough_contact_model}
\begin{figure}[t]
\centering
\def\svgwidth{1.0\textwidth}
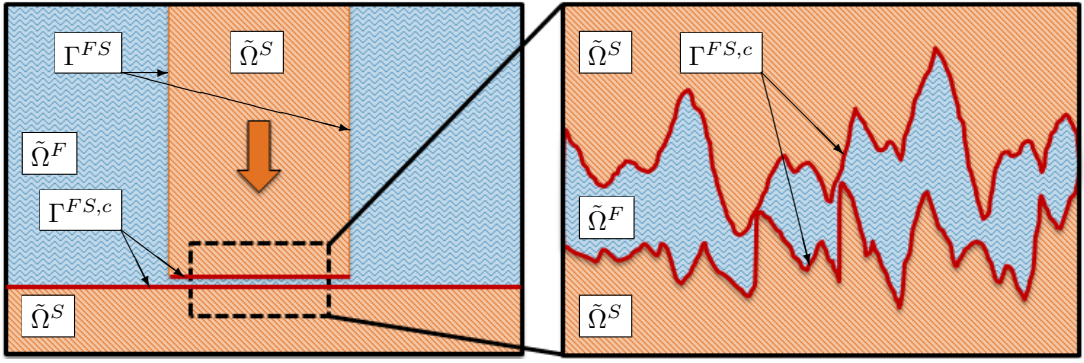
\caption{Typical configuration of an FSCI problem (left). Real surface geometry including the microstructure of rough bodies before contact  (right).}
\label{fig:model1}
\end{figure}

\begin{figure}[t]
\centering
\def\svgwidth{1.0\textwidth}
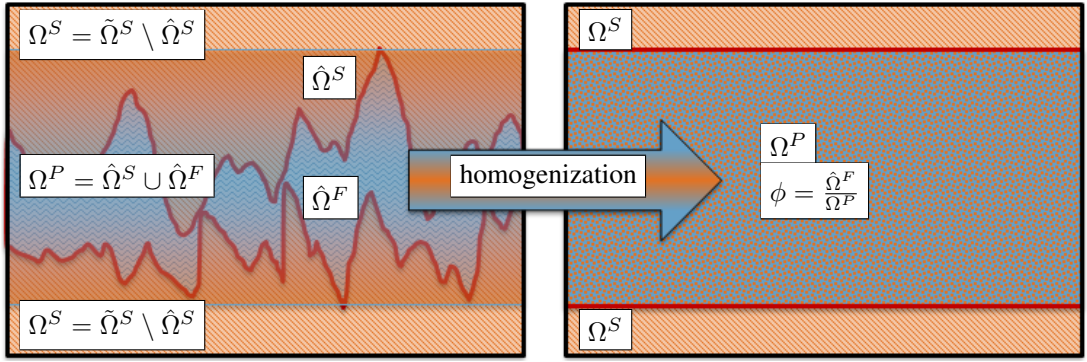
\caption{Detailed view of a potential microstructure between the contacting surfaces (left), represented by a homogenized poroelastic domain $\domainp$ with porosity $\porosity$ in the proposed model (right).}
\label{fig:model2}
\end{figure}

A typical configuration of fluid-structure-contact interaction (FSCI) problems, with fluid domain $\tilde{\Omega}^F$ and structural bodies occupying domain $\tilde{\Omega}^S$, is shown in Figure \ref{fig:model1} (left).
Herein, solid surfaces do not include any information about their present microstructure. 
This simplification is a good approximation regarding interfaces between structural bodies and fluid like $\fsiinterface$, 
where the distance to the next interface or boundary is large compared to the average height of surface asperities. 
In this case, the influence of surface roughness on the physical response of the fluid-structure interaction system is completely negligible.
However, for interfaces $\Gamma^{FS,c}$, where the surfaces of solid bodies approach each other and the size of the fluid gap in between can get very small or even vanish, this argumentation does not hold anymore.
In the following, the term ``gap'' or ``fluid gap'' is exclusively used to specify the normal distance between approaching or contacting surfaces.
As soon as the gap is in the same order of magnitude as the roughness height (see Figure \ref{fig:model1} (right)), effects caused by the microstructure of the surfaces will start to influence the macroscopic physical behavior of the system.
Finally, when first surface asperities start to contact, assuming smooth surfaces is definitely far off from the physical response of such an FSCI system.
As the prediction of effects dominated by rough surfaces such as, e.g. leakage or lubrication is of great importance, 
a model to include surface roughness of these contacting surfaces into an FSCI framework in an efficient computational way will be presented in the following.
The domain of interest $\Omega^P$ 
consists of a part of the fluid domain $\hat{\Omega}^F$ between the contacting surfaces, as well as the structural domain $\hat{\Omega}^S$ in the neighborhood of the rough surface (see Figure \ref{fig:model2}).
The overall domain $\Omega = \tilde{\Omega}^S \cup \tilde{\Omega}^F = \domains \cup \domainf \cup \domainp$ is composed of the remaining solid domain $\domains= \tilde{\Omega}^S \setminus \hat{\Omega}^S$, fluid domain $\domainf = \tilde{\Omega}^F \setminus \hat{\Omega}^F$ , and 
the fluid filled poroelastic domain $\Omega^P=\hat{\Omega}^S\cup\hat{\Omega}^F$.

We propose to model this region in a homogenized manner, an approach well known from porous media mechanics, see e.g. \cite{schrefler1998,schrefler2001,Coussy:04,Chapelle2010b,Vuong2015}.
We obtain a poroelastic layer, which describes the fluid flow and the structural elastodynamics in an averaged sense in every point of the poroelastic domain $\Omega^P$.
Herein, the porosity $\porosity = \left.\text{vol}\left(\hat{\Omega}^F\right)\right/\text{vol}\left(\Omega^P\right)$ specifies the ratio of fluid volume inside the poroelastic layer.
The model describing this poroelastic layer should be able to represent the influence of structural deformation on the fluid flow and vice versa.
Furthermore, the deformation of the roughness layer, arising from applied external stress or external deformation on the outer boundaries of the poroelastic layer, has to be considered.
A coupling of the poroelastic layer to the outer fluid flow has to be guaranteed as well.
We propose to add a poroelastic layer (at least on one of the contacting interfaces) to all interfaces $\Gamma^{FS,c}$, which potentially come into contact.
As the influence of this porous layer is negligibly small if bodies are not close to contact, a porous layer for all fluid structure interfaces $\fsiinterface$ does not seem to be necessary.
\begin{figure}[t]
\centering
\def\svgwidth{0.49\textwidth}
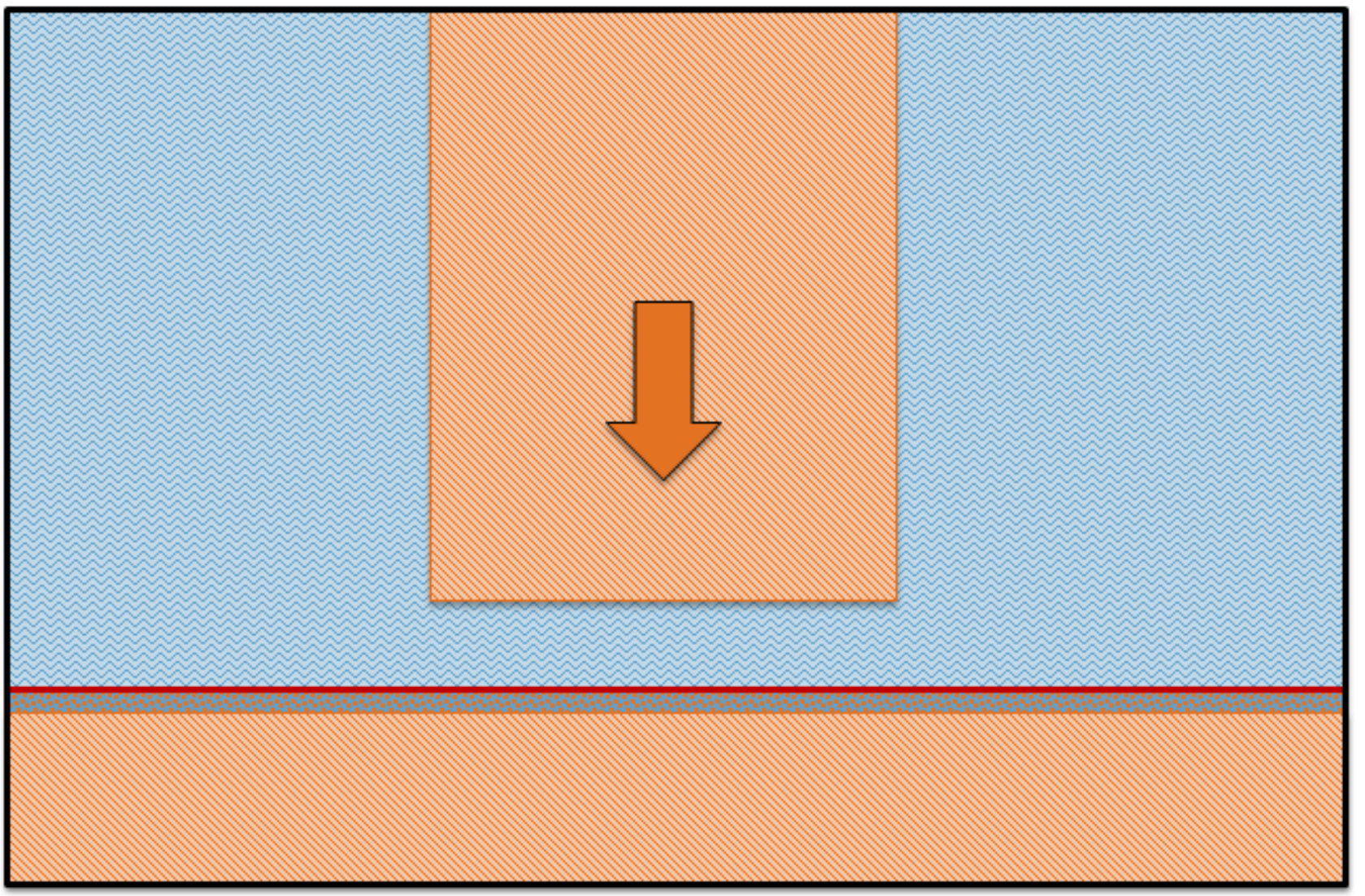
\def\svgwidth{0.49\textwidth}
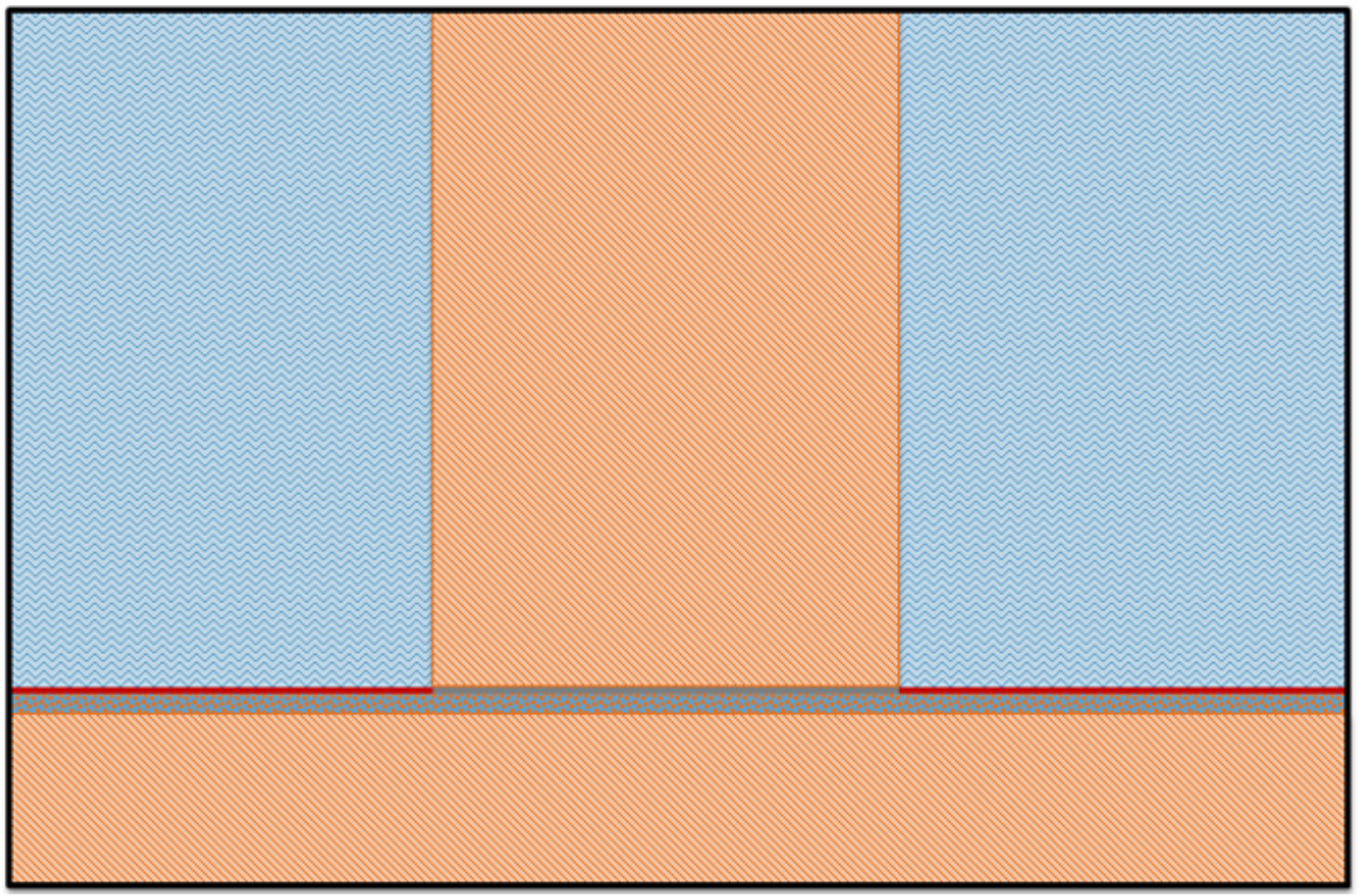
\caption{Proposed rough surface contact model for fluid-structure interaction, including all computationally essential domains $\domains, \domainf, \domainp$ and interfaces between these domains 
$\fsiinterface, \fpiinterface, \psiinterface, \psciinterface$; not in contact (left), in contact (right).}
\label{fig:model3}
\end{figure}
In Figure \ref{fig:model3}, a schematic of the final model for the rough surface FSCI is presented, where all involved physical domains with different physical principles and the interfaces between these domains are shown.
In the following sections, the governing equations for the whole domain $\domain$, which is split into three resulting domains, namely the structural domain $\Omega^S$, the fluid domain $\Omega^F$ and the poroelastic domain $\Omega^P$ with appropriate boundary conditions on the outer boundary $\fullbound$, are presented. 
It should be pointed out that the fluid as well as the structural domain no longer include the volume occupied by the rough surface, as this is represented by the poroelastic model.
All interfaces occurring between these domains and the appropriate coupling conditions will be presented in Section \ref{sec:interfaces}. These are the interfaces $\Gamma^{FS}$ between the fluid domain and structural domain,
the interfaces $\Gamma^{FP}$ between the fluid domain and poroelastic domain, and the interfaces $\Gamma^{PS}$, $\Gamma^{PS,c}$ between the poroelastic domain and the structural domain.

In contrast to the widely used Reynolds equation in elastohydrodynamic lubrication (EHL), where the no-slip condition is incorporated directly, the averaging procedure does not introduce any boundary conditions onto the fluid flow.
This allows the incorporation of spatial and temporal changing boundary/coupling conditions during the computation.
As a result, a continuous in time representation of the different states of colliding solid bodies in surrounding fluid is possible.

To illustrate this feature of the presented model, the impact of two submerged bodies serves as an example.
As long as the distance between both structural bodies is large, the influence of the poroelastic layer is small and a ``free-flowing'' fluid similar to classical fluid-structure interaction problems is described.
Once the gap between both contacting surfaces is in the same order of magnitude as the roughness height amplitude, a continuously increasing part of the fluid volume flow passes through the poroelastic layer.
Further reduction of the fluid gap until first surface asperities start to contact directly is represented by contact between the poroelastic layer and the solid body. 
As soon as all of the fluid in the gap is inside the microstructure of the surfaces, the entire fluid flow passes through the poroelastic layer.
Further increase of the contact pressure will result in progressing deformation of the poroelastic layer, which results in a reduction of the fluid fraction $\porosity$ inside the layer.
This goes along with an increase of fluid resistance and therefore reduced flow or ``leakage rate'' for a specific fluid pressure gradient in the flow direction.
Finally, for high loads in relation to the structural stiffness, the fluid fraction will approach the vanishing limit, whereby only the structural part is left between the contacting surfaces.
It should be mentioned that this exemplary sequence of steps which was described in detail, occurs just in the case that small variations in the fluid pressure and velocity far-fields arise during the approaching process of both structural bodies.
If this is not the case, different pathways will be passed, for instance when an increasing fluid pressure deforms the poroelastic layer such that the contained fluid fraction is increased, which finally can lead to a ``lift-off'' of the contacting surfaces.

\section{Governing equations for all involved physical domains}
\label{sec:fields}
The following section presents the governing equations as well as appropriate boundary conditions for the different physical domains, i.e. structure, fluid and poroelastic domain.\\
The time interval of interest should be bounded by the initial point in time $\btime$ and the end time $\etime$.
Below, all quantities $* \, , \tns *$ with additional index $*_0\, , \tns *_0$  are described in the undeformed material configuration.
The ``hat'' symbol $\hat{*}\, , \tns{\hat{*}}$ indicates prescribed time-dependent quantities at boundaries and in the domains.

Finally, $\mathring{*}\, , \tns{\mathring{*}}$ specifies prescribed initial quantities at the initial point in time.
\subsection{Structural domain $\domains$}
\label{sec:struct}
The well-known initial boundary value problem (IBVP) for non-linear elastodynamics (see e.g. \cite{Holzapfel2000}) is described by 
\begin{align}
 \refdensitys \accs - \divRef \left(\defgrads \cdot \stresspks \right) - \refdensitys \refbodyfs = \zerovec  \quad \text{in} \;  \refdomains \times [\btime,\etime],
 \label{eq:structure_eq}
\end{align}
\begin{align}
\quad \stresspks = \partiald{\strainenergys}{\straingls}, \quad \straingls = \frac{1}{2}\left[\left(\defgrads\right)^T\cdot\defgrads - \identity\right], \quad \defgrads = \identity + \partiald{\disps}{\refcoords}.
 \label{eq:structure_constitutive}
\end{align}
Therein, $\disps = \coords - \refcoords$ denotes the displacement vector, which describes the motion of a material point (with position $\refcoords$ at initial time $\timeso = \btime$) 
due to deformation of the elastic body to the current position $\coords$.
Next, $\refdensitys$ is the structural density in material configuration, $\divRef \tns *$ is the material divergence operator,  $\defgrads$ the deformation gradient, $\stresspks$ the second Piola-Kirchhoff stress tensor, 
and $\refbodyfs$ the body force per unit mass. The nonlinear, compressible material behavior is given by the hyperelastic strain energy function $\strainenergys$,
which expresses the stress-strain relation. Therein, $\straingls$ is the Green-Lagrange strain tensor.
For the interaction with other physical fields described in the current configuration, the Cauchy stress is given by $\stresss = \frac{1}{\Js} \defgrads \cdot \stresspks \cdot \left( \defgrads \right)^T$, 
with $\Js$ being the determinate of the deformation gradient.
Further, adequate initial conditions with a prescribed displacement field $\dispsB$ and velocity field $\velsB$ are defined
\begin{align}
\disps= \dispsB \quad \text{in} \;  \refdomains \times \left\lbrace t_0 \right\rbrace, \qquad
\vels = \velsB \quad \text{in} \;  \refdomains \times \left\lbrace t_0 \right\rbrace.
\label{eq:structure_init}
\end{align}
Finally, to complete the description of the structural problem, suitable boundary conditions on the outer boundary $\fullrefbound\cap\fullrefbounds$ must be specified with the predefined displacement $\dispsD$ on Dirichlet boundaries $\refdbounds$ and the
given traction $\reftractionsN$ on Neumann boundaries $\refnbounds$:
\begin{align}
\disps = \dispsD \quad\text{on} \;  \refdbounds \times [\btime,\etime], \qquad
\left( \defgrads \cdot \stresspks \right) \cdot \refnormals = \reftractionsN \quad \text{on} \;  \refnbounds \times [\btime,\etime].
\label{eq:structure_bcs}
\end{align}
Therein, $\refnormals$  is the outward-pointing unit normal vector on the boundary.
Still missing are the conditions on the subset of the structural boundary $\refrestbounds = \fullrefbounds \setminus \left( \refdbounds \cup \refnbounds \right)$ 
where the structural domain is coupled to the other fields. They are not part of the outer boundary of the FSCI problem $\fullrefbound\cap\refrestbounds=\emptyset$ and will be discussed in Sections \ref{sec:fsinterface} and \ref{sec:psinterface} .

\subsection{Fluid domain $\domainf$}
Transient, incompressible, viscous flow should be considered in the fluid domain. Considering that the balance of mass and momentum is given by the Navier-Stokes equations (see e.g. \cite{donea2003})
\begin {align}
\densityf \partiald{\velf}{\timef} + \densityf \velf \cdot \grad \velf + \grad \pf - \div( 2 \viscf \epsf (\velf)) - \densityf\bodyff  &= \zerovec   \quad \text{in} \;  \domainf \times [\btime,\etime],\label{eq:fluidm_eq}\\
\div \velf &= 0   \quad \text{in} \;  \domainf \times [\btime,\etime].\label{eq:fluidc_eq}
\end{align}
Herein, $\velf$ and $\pf$ are the fluid velocity and pressure, $\densityf$ is the constant fluid density, $\viscf$ the dynamic viscosity,
$\epsf (\velf) = \frac{1}{2}\left[\grad \velf + \left(\grad \velf\right)^T\right]$ the strain-rate tensor, and $\bodyff$ the body force per unit mass.
Again, adequate initial conditions with the given initial velocity field $\velfB$ are prescribed:
\begin{align}
\velf = \velfB \quad \text{in} \;  \domainf \times \left\lbrace t_0 \right\rbrace.
\label{eq:fluid_init}
\end{align}
To finalize the description of the fluid problem, boundary conditions on the outer boundary  $\fullbound\cap\fullboundf$  are considered. Subsequently, the fluid velocity $\velfD$ on Dirichlet boundaries $\dboundf$ 
or the fluid traction $\tractionfN$ on Neumann boundaries $\nboundf$ is prescribed. Here, $\stressf = -\pf \identity + 2 \viscf \epsf (\velf)$ is the Cauchy stress and 
$\normalf$ the outward-pointing unit normal vector to the boundary
\begin{align}
\velf = \velfD \quad \text{on} \;  \dboundf \times [\btime,\etime], \qquad
\stressf \cdot \normalf = \tractionfN \quad \text{on} \;  \nboundf\times [\btime,\etime].
\label{eq:fluid_bcs}
\end{align}
Again, conditions on the boundary subset $\restboundf = \fullboundf\setminus\left(\dboundf\cup\nboundf\right)$, which is not part of the outer boundary $\fullbound~\cap~\restboundf=\emptyset$, 
will be discussed in Sections \ref{sec:fsinterface} and \ref{sec:fpinterface}.

\subsection{Poroelastic domain $\domainp$}
\label{sec:poro}
As presented in Section \ref{sec:rough_contact_model}, modeling the fluid-saturated rough surface domain as homogenized poroelastic media (see e.g. \cite{Coussy:04}), leads to specific requirements on the applied model.
The following governing equations were developed and successfully applied in \cite{Chapelle2010b, Vuong2015,vuong2016, vuong2017} and are capable to represent all essential physical effects, such as incompressible flow on the micro-scale, finite deformations of the poroelastic matrix, deformation dependent and variable porosity, as well as arbitrary strain energy functions for the skeleton.
Due to the high flexibility of this formulation for the homogenized roughness layer, a broad range of different microstructures as well as material behavior of fluid and structures are applicable.
Nevertheless, the focus of this contribution should be on including general modeling of rough surface contact into a fluid-structure interaction framework.
A brief outline of parameter estimation for specific rough surfaces is provided in a following remark.

In the presented formulation, macroscopic flow through the deformable porous media is modeled by a Darcy flow based equation \eqref{eq:porom_eq}.
The balance of mass and momentum of the fluid phase in the current configuration, 
the balance of mass (included implicitly, see \cite{Chapelle2010b,Vuong2015}) and momentum for the whole poroelastic mixture (consisting of fluid and solid) in material configuration on macroscopic scale can be expressed as:
\begin{align}
\left.
\partiald{\porosity}{\timep}
\right|_{\refcoordp}
+ \porosity \div \velps + \div \left[ \porosity \left(\velp - \velps \right) \right] = 0    \quad \text{in} \;  \domainp \times [\btime,\etime]
\label{eq:poroc_eq},\\
\left.
\densitypf \partiald{\velp}{\timep}
\right|_{\refcoordp}
- \densityf\velps\cdot\grad\velp +\grad \pp - \densitypf \bodyfpf \nonumber\\+\viscp  \porosity \permeabp^{-1} \cdot \left(\velp-\velps\right) = \zerovec    \quad \text{in} \;  \domainp \times [\btime,\etime]
\label{eq:porom_eq},\\
\refavdensityps \accps - \divRef \, \left( \defgradp \cdot \stresspkp \right) - \refavdensityps \refbodyfp
- \Jp \porosity \left(\defgradp \right)^{-T} \cdot \gradRef \pp 
\nonumber\\ - \viscp \Jp \porosity^2 \permeabp^{-1} \cdot \left( \velp - \velps \right) = \zerovec   \quad \text{in} \;  \refdomainp \times [\btime,\etime]
\label{eq:mixturem_eq}.
\end{align}
These equations are valid on the macro-scale and represent an average microscopic state in the poroelastic media.
Fluctuations due to the microstructure are not represented.
Therefore, all quantities occurring in equations \eqref{eq:poroc_eq}-\eqref{eq:mixturem_eq} also describe the average state from a macroscopic view.
Herein, $\porosity$ is the porosity already introduced in Section \ref{sec:rough_contact_model}, $\velp$ the velocity and $\pp$ the pressure of the fluid phase, 
$\dispp = \coordp - \refcoordp$ the macroscopic displacement of the poroelastic domain analogous to Section \ref{sec:struct},
$\bodyfpf$ the body force acting on the embedded fluid per unit mass, $\permeabp=\left(\Jp\right)^{-1}\defgradp \cdot \matpermeabp \cdot \left(\defgradp\right)^T$ the spatial permeability of the poroelastic matrix 
with $\matpermeabp$ being the corresponding material permeability.
Further, $\refavdensityps = (1-\porosity) \refdensitys$ is the macroscopic averaged initial density of the solid phase with $\refdensitys$ the associated averaged initial density in the solid domain. $\refbodyfp$ is the body force acting on the poroelastic mixture per unit averaged solid mass,
$\defgradp$ the macroscopic deformation gradient, $\Jp$ the determinant of the macroscopic deformation gradient, and $\stresspkp$ the homogenized second Piola-Kirchhoff stress tensor. 
The material behavior of the poroelastic mixture is given by the macroscopic strain energy function
$\strainenergyp(\strainglp, \Jp (1-\porosity)) = \strainenergy{P,skel}(\strainglp)+\strainenergy{P,vol}(\Jp (1-\porosity))+\strainenergy{P,pen}(\strainglp, \Jp (1-\porosity))$,
whereas $\strainenergy{P,skel}$ accounts for the strain energy due to macroscopic deformation of the solid phase, 
$\strainenergy{P,vol}$ arises from the volume change of the solid phase due to changing fluid pressure, and finally 
$\strainenergy{P,pen}$ guarantees positive porosity of the poroelastic model (see \cite{Chapelle2010b,Vuong2015}).
Using the Green-Lagrange strain tensor $\strainglp$ as the strain measure gives two constitutive relations to complete the system of equations for poroelasticity:
\begin{align}
\stresspkp &= 
\partiald{\strainenergyp(\strainglp, \Jp (1-\porosity)=\text{const.})}{\strainglp}
- \pp\Jp\left(\defgradp\right)^{-1}\cdot\left(\defgradp\right)^{-T},
\label{eq:poro_constitutive1}\\
\pp &= 
-\partiald{\strainenergyp(\strainglp=\text{const.}, \Jp(1-\porosity))}{(\Jp(1-\porosity))},
\label{eq:poro_constitutive2}
\\
\quad \strainglp &= \frac{1}{2}\left[\left(\defgradp\right)^T\cdot\defgradp - \identity\right], 
\quad \defgradp = \identity + \partiald{\dispp}{\refcoordp}. \nonumber
\end{align}
The necessary initial conditions for the poroelastic problem are:
\begin{align}
&\dispp = \disppB \quad \text{in} \;  \refdomainp \times \left\lbrace t_0 \right\rbrace, \qquad
&&\velps = \velpsB \quad \text{in} \;  \refdomainp \times \left\lbrace t_0 \right\rbrace, \qquad \nonumber\\
&\porosity = \porosityB \quad \text{in} \;  \domainp \times \left\lbrace t_0 \right\rbrace, \qquad
&&\velp = \velpB \quad \text{in} \;  \domainp \times \left\lbrace t_0 \right\rbrace.
\label{eq:poro_init}
\end{align}
Here, $\disppB$, $\velpsB$, $\porosityB$ and $\velpB$ are the initial displacement, initial solid phase velocity, porosity, and fluid velocity field, respectively.
To complete the problem description of poroelasticity, adequate boundary conditions on the outer boundary  $\fullbound\cap\fullboundp$  have to be prescribed:
\begin{alignat}{2}
\velp \cdot \normalp &= \velpnD\quad \text{on} \;  \dboundpf\times [\btime,\etime],\qquad
-\pp \normalp = \tractionpfN \normalp \quad \text{on} \;  \nboundpf\times [\btime,\etime],\label{eq:porof_bcs}\\
\dispp &= \disppD \quad \text{on} \;  \refdboundp\times [\btime,\etime],\qquad
\left( \defgradp \cdot \stresspkp \right) \cdot \refnormalp = \reftractionpN \quad \text{on} \;  \refnboundp\times [\btime,\etime].\label{eq:poro_bcs}
\end{alignat}
Therein, $\velpnD$ is the scalar normal fluid velocity of the Darcy-like flow on Dirichlet boundaries $\dboundpf$, 
$\tractionpfN$ the traction in normal direction on Neumann boundaries $\nboundpf$,
$\disppD$ the displacement of the poroelastic domain on Dirichlet boundaries $\refdboundp$, and 
$\reftractionpN$ the traction acting onto the poroelastic mixture on Neumann boundaries $\refnboundp$, with $\normalp$ being the outward-pointing unit normal vector on the boundary.
Conditions on the still not considered part of the poroelastic boundary $\restboundp = \fullboundp \setminus \left( \dboundpf \cup \nboundpf \right) = \fullboundp \setminus \left( \dboundp \cup \nboundp \right)$
will be presented in Sections \ref{sec:fpinterface} and \ref{sec:psinterface}.
For further details on this poroelastic formulation, the reader is referred to \cite{Chapelle2010b,Vuong2015,vuong2016}.

\begin{remark}[A brief outline on the estimation of the poroelastic material parameters]
\label{rem:poro_matpar}
In the following, a computationally assisted way to determine a proper set of parameters of the poroelastic layer for a specific roughness layer is presented.
To estimate the material parameters of the poroelastic layer for specific surfaces, 
the macroscopic material behavior of both contacting solid bodies as well as a resolved, representative microstructure geometry of the rough surfaces is required.
First, performing direct contact simulations without any fluid pressure contribution for characteristic parts of the resolved rough surfaces, 
allows us to specify the material parameters of strain energy function $\strainenergy{P,skel}$ of the poroelastic layer.
In general, a comparison to the bulk material shows that the initial tangent stiffness of the poroelastic layer is smaller due to the deformation of single contacting asperities.
Increasing contact stress leads to a rapid increase in the tangent stiffness, as the fluid fraction decreases.
This behavior is reflected by an increased non-linearity of the strain energy function $\strainenergy{P,skel}$ compared to the bulk material.
Measuring the void space allows the determination of the porosity $\porosity$.
Furthermore, by including a predefined normal load, which represents the fluid pressure, on the contact interface of the direct contact simulations, the correlation of fluid pressure and the solid phase compression can be analyzed.
This allows us to identify the parameters of the strain energy function $\strainenergy{P,vol}$.
To specify the permeability $\permeabp$ numerically, a resolved computation of the fluid flow between the rough surfaces can be consulted.
Performing these fluid flow simulations for several deformation states, allows us to specify the relation of permeability and porosity $\permeabp = \permeabp (\porosity)$.
Besides this proposed computational approach, experimental determination for single parameters or the whole set of parameters is also possible.
\end{remark}

\section{Interfacial coupling constraints}
\label{sec:interfaces}
In this section, appropriate coupling conditions on the interfaces occurring between all physical domains are discussed independently of their interaction,
followed by an examination of the change of these different interface coupling conditions in context of the entire FSCI problem.
In the following, $\normal$ is a uniquely chosen unit normal vector on each considered interface.

\subsection{Interface between fluid domain and structural domain}
\label{sec:fsinterface}
The interface between viscous fluid domain and structural domain $\fsiinterface = \fullbounds \cap \fullboundf = \restbounds \cap \restboundf$ needs to fulfill the dynamic stress balance \eqref{eq:fsi_dyneq}
as well as the no-slip condition \eqref{eq:fsi_noslip}:
\begin{align}
\stressf\cdot \normal &= \stresss\cdot  \normal \quad &&\text{on} \; \fsiinterface \timesfulltime,
\label{eq:fsi_dyneq}\\
\velf &= \vels \quad &&\text{on} \; \fsiinterface \timesfulltime.
\label{eq:fsi_noslip}
\end{align}

\subsection{Interface between fluid domain and poroelastic domain}
\label{sec:fpinterface}
On the interface between viscous fluid and the poroelastic domain $\fpiinterface = \fullboundp \cap \fullboundf = \restboundp \cap \restboundf$, the following conditions need to be fulfilled 
(see \cite{discacciati2009,ager2018b} for details):
\begin{align}
\stressf \cdot \normal &= \stressp \cdot \normal \quad &&\text{on} \; \fpiinterface \timesfulltime,
\label{eq:fpi_dyneq}\\
\normal \cdot  \stressf \cdot \normal &= -\pp \quad &&\text{on} \; \fpiinterface \timesfulltime,
\label{eq:fpi_dyneqp}\\
\left(\velf - \velps \right)\cdot \normal &= \porosity \left( \velp - \velps \right)\cdot \normal \quad &&\text{on} \; \fpiinterface \timesfulltime,
\label{eq:fpi_massb}\\
- \sliplengh \normal\cdot \stressf \cdot \tangent_i &= \left[\velf - \velps - \porosity \left( \velp - \velps \right) \right]\cdot \tangent_i \quad i=1,2 \, &&\text{on} \; \fpiinterface \timesfulltime\label{eq:fpi_bj}.
\end{align}
Herein, \eqref{eq:fpi_dyneq} is the dynamic stress balance in current configuration between the Cauchy stresses of fluid and the entire poroelastic mixture, 
with $\stressp = \frac{1}{\Jp} \defgradp \cdot \stresspkp \cdot \left( \defgradp \right)^T$ being the Cauchy stress of the poroelastic mixture.
Additionally, the dynamic stress balance \eqref{eq:fpi_dyneqp} between the normal components of viscous fluid and the fluid inside of the poroelastic layer is required.
Mass balance on the interface leads to a kinematic constraint in normal direction \eqref{eq:fpi_massb}.
Finally, the kinematic constraint in tangential direction for the viscous fluid is still missing.
Equation \eqref{eq:fpi_bj} is the so-called Beavers-Joseph condition \cite{beavers1967}, which proposes a proportionality (with factor $\kappa$) of the viscous fluid shear stress and the relative velocity slip in tangential direction between the adjacent fluids on both sides of the interface.
Therein, the tangential plane is specified by two tangential vectors $\tangent_i$, orthogonal to vector $\normal$.
It should be mentioned that in many cases a simplified condition can be used, where the seepage velocity $\porosity \left(\velp-\velps\right)$ is neglected as proposed in \cite{saffman1971}.
Analyses on the Beavers-Joseph condition can be found in \cite{gartling1996,cao2010,cao2010b,ager2018b}.

\subsection{Interface between poroelastic domain and structural domain in contact case}
\label{sec:psinterface}
The interfaces between poroelastic layer and structural domain $\psiinterface \cup \psciinterface = \fullboundp \cap \fullbounds = \restboundp \cap \restbounds$
 consist of a matching part $\psiinterface$ and the contacting subset $\psciinterface$ (see Figure \ref{fig:model3} (right)), which is only non-zero in the contact case.
Here we want to focus onto the contacting part only, as the physical and numerical treatment of interfaces such as $\psiinterface$ 
can be found in the domain decomposition literature (see e.g. \cite{wohlmuth2001}) for the coupling of displacement degrees of freedom and \cite{vuong2016} for the impermeability constraint.
Considering frictionless contact, the following conditions need to be fulfilled:
\begin{align}
\left( \coords_{\Gamma} - \coordp_{\Gamma} \right) \cdot \normal &= 0 \quad &&\text{on} \; \psciinterface \timesfulltime,
\label{eq:psci_gap}\\
\porosity\left( \velp - \velps \right)\cdot\normal &= 0 \quad &&\text{on} \; \psciinterface \timesfulltime,
\label{eq:psci_nopen}\\
\normal \cdot \stressp \cdot \normal  &= \normal \cdot \stresss  \cdot \normal  \quad &&\text{on} \; \psciinterface \timesfulltime.
\label{eq:psci_dyneq}\\
\tangent_i \cdot \stressp \cdot \normal  &= \tangent_i \cdot \stresss  \cdot \normal = 0 \quad i=1,2  \quad &&\text{on} \; \psciinterface \timesfulltime.
\label{eq:psci_dyneq_t}
\end{align}
Herein, \eqref{eq:psci_gap} enforces the zero gap between the structural body and porous layer, with $\coordp_{\Gamma}$ being the position on the interface $\restboundp$ and $\coords_{\Gamma}$ 
the projection of $\coordp_{\Gamma}$ in direction $\normal$ onto the interface $\restbounds$. 
Equation \eqref{eq:psci_nopen} represents the mass flow balance on the interface, where the relative fluid flow has to vanish due to the impermeability of the solid.
Finally, \eqref{eq:psci_dyneq} and \eqref{eq:psci_dyneq_t} reflect the dynamic equilibrium between poroelastic and solid domain in normal and tangential direction, respectively.

\subsection{Change of interface conditions in the coupled problem}
\begin{figure}[htb]
\centering
\hspace*{-0.5cm}
\def\svgwidth{1\textwidth}
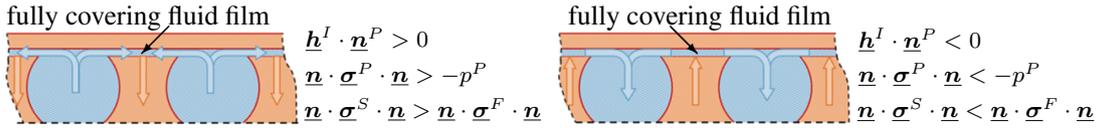
\caption{Detailed view on the condition transition point $\psciinterface\cap\fsiinterface\cap\fpiinterface$.
Arrows indicate the process of generation (left) or degeneration (right) of a fully covering fluid film, depending on the traction difference between total contact traction and the ambient fluid stress.
}
\label{fig:interface_ffilm}
\end{figure}

In the previous sections, interfaces and their corresponding conditions were considered to be independent of each other. 
It is obvious that occurring contact between poroelastic layer and structural domain modifies not only the ``active'' contact interface $\psciinterface$,
but also the interfaces between fluid and poroelastic/structural domain $\fpiinterface, \, \fsiinterface$.
The union of all three interfaces $\interface$  is given by the current configuration of the solid and poroelastic domain, in particular of the respective parts of the outer boundaries
$\interface = \psciinterface \cup \fsiinterface \cup \fpiinterface = \restboundf \cup \restboundp \cup \restbounds$.
The criteria specifying the different interface types on the interface $\interface$, are given by the following Karush-Kuhn-Tucker conditions, which need to be fulfilled:
\begin{align}
\left( \coords_{\Gamma} - \coordp_{\Gamma} \right) \cdot \normalp&\geq 0 \quad \text{on} \quad \interface \timesfulltime,
\label{eq:gapcond}\\
\itraction \cdot \normalp&\leq 0 \quad \text{on} \quad  \interface \timesfulltime,
\label{eq:contacttractioncond}\\
\left[\left( \coords_{\Gamma} - \coordp_{\Gamma} \right) \cdot \normalp\right]\left[\itraction \cdot \normalp\right] &= 0 \quad \text{on} \quad  \interface \timesfulltime.\label{eq:contactexclusioncond}
\end{align}
Condition \eqref{eq:gapcond} guarantees that there is always a positive gap or no gap between potentially contacting bodies.
Additionally, \eqref{eq:contacttractioncond} restricts the minimal stress by compression transferred directly between the solid bodies to the surrounding fluid normal traction.
Herein, $\itraction$ (e.g. $=\stressp \cdot \normalp + \pp \normalp$ or $=\stresss \cdot\normalp - \stressf \cdot \normalp$) is the traction difference between total contact traction and the ambient fluid stress.
This condition is a result of the assumption 
that the time interval for the formation of a fully covering fluid film on top of the rough microstructure is negligible and this process does not need to be modeled.
As soon as the local (e.g. at single asperities) structural stress is smaller than the fluid stress, this fluid film will develop.
In Figure \ref{fig:interface_ffilm}, a schematic visualization of this process is given.
Finally, \eqref{eq:contactexclusioncond} enforces exclusively either a zero gap between both interfaces (see equation \eqref{eq:gapcond}) or a vanishing relative traction (see equation \eqref{eq:contacttractioncond}).

The subset of the interface $\interface$, where the first condition \eqref{eq:gapcond} is exactly zero and therefore fulfills \eqref{eq:psci_gap}, is the contact interface between poroelastic domain and structural domain $\psciinterface$. 
The remaining parts, where the second condition \eqref{eq:contacttractioncond} is exactly zero, are the interfaces between fluid domain and structural domain $\fsiinterface$ or between fluid domain and poroelastic domain $\fpiinterface$, 
depending on if it is a subset of the structural or poroelastic boundary.

\begin{figure}[htb]
\centering
\def\svgwidth{0.8\textwidth}
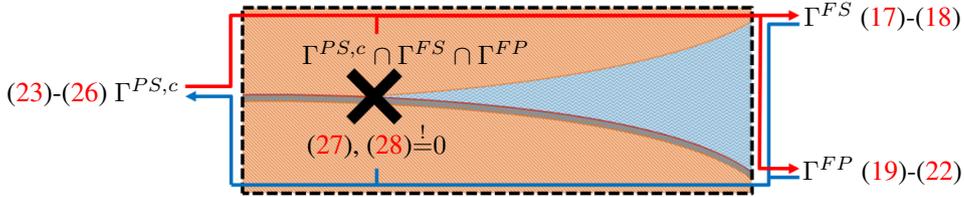
\caption{Visualization of the continuous transition of normal conditions at the point $\psciinterface\cap\fsiinterface\cap\fpiinterface$ which is marked by the black cross.
Following the upper red path indicates the equivalence of the contact conditions by making use of the transition conditions and the normal conditions 
on the interfaces $\fsiinterface$ and $\fpiinterface$.
Following the lower blue path indicates the equivalence of the conditions on the interfaces $\fsiinterface$ and $\fpiinterface$ in normal direction by making use of the transition conditions 
and the contact conditions on $\psciinterface$.}
\label{fig:interface_paths}
\end{figure}

In the following paragraph, the continuity of the formulation at the point or line of changing interface conditions (marked by the black cross in Figure \ref{fig:interface_paths}) is discussed. 
To ensure continuity of the problem at this specific position, the conditions of all adjacent interfaces have to be fulfilled simultaneously. 
To prove this, we assume one type of interface conditions (e.g. contact) plus the criteria for transition to hold and verify the fulfillment of the other conditions (e.g. fluid-structure and fluid-poroelastic conditions).
We analyze the fulfillment of the different interface conditions in normal direction at the transition points from a contact interface ($\psciinterface$) to non-contact interfaces ($\fsiinterface$ and $\fpiinterface$) and vice versa. 
Those points are formed by the intersection of the interfaces $\psciinterface$, $\fsiinterface$ and $\fpiinterface$ and they satisfy the transition conditions, stated as conditions \eqref{eq:gapcond} and \eqref{eq:contacttractioncond} satisfied equal to zero.
In order to enable a continuous transition of interface types, combining conditions on the contact interface $\psciinterface$ (conditions \eqref{eq:psci_gap}-\eqref{eq:psci_dyneq_t}) 
with the transition conditions has to fulfill conditions on the interfaces $\fsiinterface$ and $\fpiinterface$ (conditions \eqref{eq:fsi_dyneq}-\eqref{eq:fsi_noslip} and \eqref{eq:fpi_dyneq}-\eqref{eq:fpi_bj}) by default 
(see upper red path in Figure \ref{fig:interface_paths}).
Therefore, let's assume that the conditions of contact are satisfied and even for a vanishing fluid film we consider fluid state vectors ($\velf$, $\pf$) and implicitly the corresponding fluid stress ($\stressf$) to result in continuous fulfillment of all interface conditions.
Then, for the normal components of velocity and normal traction difference, the following relations hold:
\begin{align}
\text{kinematic constraints: } \quad
\velp \cdot \normal
\stackrel{\text{condition } \eqref{eq:psci_gap}}{=}
\vels \cdot \normal
\stackrel{\text{condition } \eqref{eq:psci_nopen}}{=}
\velps \cdot \normal
\stackrel{\text{(*1)} }{=}
\velf \cdot \normal
\label{eq:int_kineq}
\end{align}
\hspace*{\fill}(*1) = normal velocity of emerging fluid film equals contact interface normal velocity
\begin{align}
\text{stress equilibrium: } \quad
\itraction \cdot \normal
&\stackrel{\text{def.}}{=}
0
\stackrel{\text{transition point of }\eqref{eq:contacttractioncond}}{=}
\normal \cdot \stressp \cdot \normal + \pp
\stackrel{\text{condition } \eqref{eq:psci_dyneq}}{=}
\normal\cdot  \stresss \cdot\normal + \pp\nonumber\\
&\stackrel{\text{(*2)} }{=}
\normal\cdot \stresss \cdot\normal - \normal \cdot \stressf \cdot \normal
\stackrel{\text{condition } \eqref{eq:psci_dyneq}}{=}
\normal\cdot \stressp\cdot \normal - \normal \cdot \stressf \cdot \normal
\label{eq:int_dyneq} 
\end{align}
\hspace*{\fill}(*2) = normal fluid stress of emerging fluid film in balance with fluid pressure in poroelastic layer\\
\\
As a consequence of \eqref{eq:int_kineq} and \eqref{eq:int_dyneq} the conditions on the interfaces $\fsiinterface$ and $\fpiinterface$ (conditions \eqref{eq:fsi_dyneq}-\eqref{eq:fsi_noslip} and \eqref{eq:fpi_dyneq}-\eqref{eq:fpi_massb}) in normal direction are fulfilled naturally.
For the change from the non-contact interfaces $\fsiinterface$ and $\fpiinterface$  to the contact interface $\psciinterface$ this can be shown analogously
(see lower blue path in Figure \ref{fig:interface_paths}).

\begin{remark}[Continuity of the formulation for changing interface conditions in tangential direction]
\label{rem:continuity}
Considering frictionless contact at the contact interface $\psciinterface$ in combination with the no-slip condition at the interface $\fsiinterface$ and the Beavers-Joseph condition at the interface $\fpiinterface$ leads to a non-continuous change in the tangential component of the interface conditions.
A continuous change between contact and non-contact interfaces for frictionless contact would require zero tangential stress (``full-slip'' conditions) close to the contacting zone on the interfaces $\fsiinterface$ and $\fpiinterface$ as well.
Nevertheless, contrary to the continuity of the normal component of interface conditions, the continuity of tangential components turned out to be less essential in our numerical computations.
\end{remark}

It should be pointed out that for a pure structural contact case with vanishing fluid, conditions \eqref{eq:gapcond}, \eqref{eq:contacttractioncond}, and \eqref{eq:contactexclusioncond} reduce to the classical contact conditions, 
as the relative traction equals the absolute traction for vanishing fluid pressure.
\section{Discretization and solution approach}
\label{sec:nummethod}
In this section, the applied methods for solving the coupled rough FSCI problem are presented.
The discretization of the continuous problem described in the previous Sections \ref{sec:fields} and \ref{sec:interfaces} is based on the Finite Element Method. 
First, the spatially discretized semi-discrete formulations for the structural domain, the fluid domain as well as the poroelastic layer are given.
Topological changes of the fluid domain in the rough FSCI problem due to occurring contact of surfaces are enabled via a CutFEM applied onto the fluid domain.
The embedded interfaces between fluid domain and poroelastic domain together with the interfaces between fluid domain and structural domain are imposed by Nitsche-based methods.
To incorporate contact into the formulation, the dual mortar method is used. Finally, all contributions are considered in one global system and solved monolithically.

As the focus of this contribution should not be on the specific numerical methods, these are just presented briefly, since further details can be found in the referenced literature.
In the following sections, all quantities, including the primary unknowns, the test functions in the weak form as well as the domains and interfaces are discretized in space.
No additional index $h$ is added to these discrete quantities since this double meaning of notation is accepted for the sake of simplicity of presentation.
Below, the expressions $\innerp{*}{*}{\Omega}$ and $\innerpb{*}{*}{\partial \Omega}$ denote the inner $\mathcal{L}_2$ product integrated in the domain $\Omega$ and on the boundary/interface $\partial \Omega$, respectively.

\subsection{Weak forms of domains $\domains, \domainf, \domainp$}
The weak forms of the structural, fluid, and poroelastic domains can be derived from equations \eqref{eq:structure_eq}, \eqref{eq:fluidm_eq} - \eqref{eq:fluidc_eq} and \eqref{eq:poroc_eq} - \eqref{eq:mixturem_eq}, respectively:

\begin{align}
&\mathcal{W}^S\left[\testdisps,\disps\right] = \innerp{\testdisps}{\refdensitys \accs}{\refdomains} + \innerp{\gradRef \testdisps}{\defgrads \cdot \stresspks}{\refdomains} 
\nonumber\\&\qquad- \innerp{\testdisps}{\refdensitys \refbodyfs}{\refdomains}
-\innerpb{\testdisps}{\reftractionsN}{\refnbounds} 
,\label{eq:w_solid}\\
&\mathcal{W}^F\left[\left(\testvelf, \testpf\right),\left(\velf, \pf\right)\right] = \innerp{\testvelf}{\densityf \partiald{\velf}{t}}{\domainf} + \innerp{\testvelf}{\densityf \velf \cdot \grad \velf}{\domainf} 
\nonumber\\&\qquad-\innerp{\div \testvelf}{\pf}{\domainf}
+\innerp{\epsf (\testvelf)}{2 \viscf \epsf (\velf)}{\domainf}
-\innerp{\testvelf}{\densityf\bodyff}{\domainf} 
\nonumber\\&\qquad-\innerpb{\testvelf}{\tractionfN}{\nboundf}
+\innerp{\testpf}{\div \velf}{\domainf}  ,\label{eq:w_fluid}\\
&\mathcal{W}^P\left[\left(\testvelp, \testdispp, \testpp\right),\left(\velp,\dispp,\pp\right)\right] = \innerp{\testpp}{\partiald{\porosity}{\timep}}{\domainp} 
+ \innerp{\testpp}{\porosity \div \velps}{\domainp}
\nonumber\\&\qquad-\innerp{\grad \testpp}{\porosity \left( \velp - \velps \right) }{\domainp}
+ \innerpb{\testpp}{\porosity \normalp \cdot \left( \velp - \velps \right) }{\fullboundp}
\nonumber\\&\qquad+\innerp{\testvelp}{\densitypf \partiald{\velp}{\timep}}{\domainp} 
- \innerp{\div \testvelp}{\pp}{\domainp} 
- \innerp{\testvelp}{\densitypf \partiald{\dispp}{\timep}\grad \velp}{\domainp}
\nonumber\\&\qquad+\innerp{\testvelp}{\viscp  \porosity \permeabp^{-1} \cdot  \velp}{\domainp}
-\innerp{\testvelp}{\viscp \porosity \permeabp^{-1} \cdot \velps}{\domainp} 
- \innerp{\testvelp}{\densitypf \bodyfpf}{\domainp}
\nonumber\\&\qquad- \innerpb{\testvelp}{\tractionpfN \normalp}{\nboundpf}
+\innerp{\testdispp}{\refavdensityps \accps}{\refdomainp} 
+\innerp{\gradRef \testdispp}{\defgradp\cdot\stresspkp}{\refdomainp}
\nonumber\\&\qquad+\innerp{\testdispp}{\viscp \Jp \porosity^2 \permeabp^{-1} \cdot \velps}{\refdomainp}
-\innerp{\testdispp}{\viscp \Jp  \porosity^2 \permeabp^{-1} \cdot \velp}{\refdomainp}
\nonumber\\&\qquad-\innerp{\testdispp}{\Jp \porosity \left(\defgradp\right)^{-T} \gradRef \pp}{\refdomainp}
-\innerp{\testdispp}{\refavdensityps \refbodyfp}{\refdomainp} 
-\innerpb{\testdispp}{\reftractionpN}{\refnboundp}
.\label{eq:w_poro}
\end{align}
Herein, $\left(\testdisps,\testvelf, \testpf,\testvelp, \testdispp, \testpp\right)$ are the corresponding test functions of the primary unknowns $\left(\disps,\velf, \pf,\velp,\right.\left.\dispp,\pp\right)$.
It should be pointed out that the treatment of the coupling conditions for the different interfaces was omitted here and will be handled in 
Sections \ref{sec:num_fsinterface}, \ref{sec:num_fpinterface} and \ref{sec:num_psinterface} by incoporating the still missing interface terms on $\restbounds, \restboundf, \restboundpf, \restboundp$ arising in the derivation of the weak form.

In order to control convective instabilities, to ensure discrete mass conservation, and to guarantee inf-sup stability for equal order interpolation of velocity and pressure, stabilization operators are added to 
the discrete weak forms of both fluid equations \eqref{eq:w_fluid} and \eqref{eq:w_poro}.
\begin{align}
\mathcal{W}^F_{\mathcal{S}}\left[\left(\testvelf, \testpf\right),\left(\velf, \pf\right)\right] &= \mathcal{S}^F_v\left[\testvelf, \left(\velf,\pf\right)\right]+ \mathcal{S}^F_p\left[\testpf, \left(\velf,\pf\right)\right]\\
\mathcal{W}^P_{\mathcal{S}}\left[\left(\testvelp, \testpp\right),\left(\velp, \pp\right)\right] &= \mathcal{S}^P_v\left[\testvelp, \velp\right]+\mathcal{S}^P_p\left[\testpp, \left(\velp,\pp\right)\right]
\end{align}
This can be achieved by residual-based stabilization or face-oriented stabilization. A comparison of different stabilization techniques for incompressible flow problems is presented in \cite{braack2007}. 
Face-oriented stabilizations were applied for all numerical examples presented in Section \ref{sec:num_ex}.
\subsubsection{The CutFEM applied onto the fluid domain}
As the fluid domain $\domainf$ changes in time and performs even topological changes for the coupled rough FSCI problem, it is beneficial to use a method 
where the boundary of computational discretization does not necessarily match the boundary of the physical domain.
The CutFEM applied onto the Navier-Stokes equations makes this essential split-up possible.
The following presents a brief overview of the most important aspects for applying the CutFEM on the presented model.
For a general overview of the method, the reader is referred to \cite{burman2015cutfem} and the references therein.

\begin{figure}[t]
\centering
\def\svgwidth{0.9\textwidth}
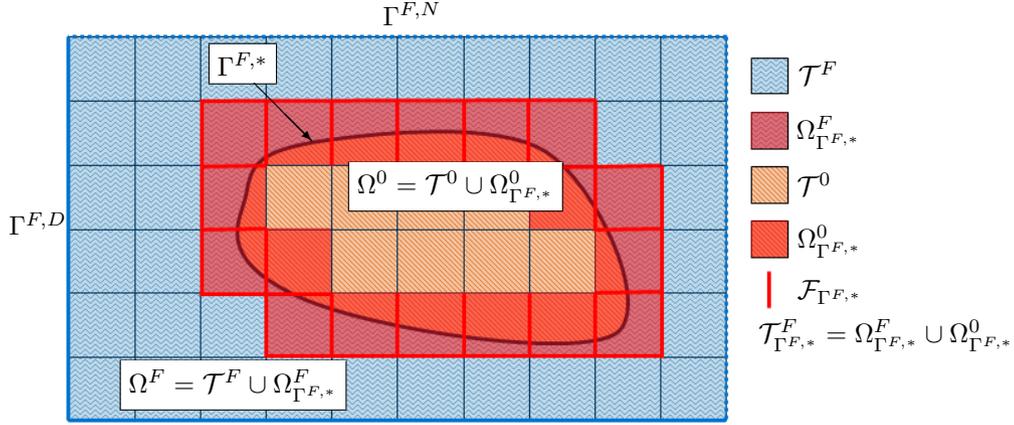
\caption{Basic concept for the applied CutFEM, where structural or poroelastic domain $\Omega^0 = \domains \cup \domainp$ are embedded in the fluid domain $\domainf$. 
The fluid domain consists of the uncut elements in $\mathcal{T}^F$ and just a subset $\Omega^F_{\Gamma^{F,*}}$ of the intersected elements in $\mathcal{T}^F_{\Gamma^{F,*}}$.
For all element faces $\mathcal{F}_{\Gamma^{F,*}}$ that are connected to at least one element in $\mathcal{T}^F_{\Gamma^{F,*}}$, a ``ghost penalty'' stabilization is applied.}
\label{fig:cutfem}
\end{figure}

Figure \ref{fig:cutfem} shows an exemplary configuration for a computational mesh, where all elements span the domain $\domainf \cup \Omega^0$, which is constant in time. 
The non-physical part $\Omega^0$ of the fluid computational mesh includes the structural domain $\domains$ as well as the poroelastic domain $\domainp$.
The governing equations for the latter domains are discretized with separate interface- and boundary-matching computational meshes.
The approach to consider all coupling conditions on the interfaces between these fields and to therefore gain equilibrium between all fields will be presented in Sections \ref{sec:num_fsinterface}, \ref{sec:num_fpinterface}, \ref{sec:num_psinterface} and \ref{sec:num_globalsystem} .
For the fluid field, the integration of the $\mathcal{L}_2$-inner products (see equation \eqref{eq:w_fluid}) is only performed on the physical fluid domain $\domainf$. 
This numerical integration is realized with standard Gaussian quadrature rules on all uncut elements in $\mathcal{T}^F$.
For the physical fluid domain $\domainf_{\Gamma^{F,*}}$ in all intersected elements $\mathcal{T}^F_{\Gamma^{F,*}}$, the method described in \cite{sudhakar2014} is applied, which utilizes the divergence theorem.
For elements which are completely covered by the domain $\domain^0$, the fluid weak form does not have to be integrated.

The intersection of the computational fluid mesh by the interface $\Gamma^{F,*}$ is given by the position of the structural or poroelastic domain.
In fact, the deformed current configuration of the matching computational meshes for these domains, which is part of the solution of the overall system, describes these interfaces.
Therefore, intersections which lead to a very small contribution of single discrete degrees of freedom to the fluid weak form are critical.
This can result in an ill-conditioned system matrix or a loss of discrete stability of the method caused by the imposition of boundary/coupling conditions (see Sections \ref{sec:num_fsinterface} or \ref{sec:num_fpinterface}).
To overcome these issues, a weakly consistent stabilization operator is added to the fluid weak form.
Therein, basically deviations of an extension of the physical solution into the non-physical domain $\Omega^0_{\Gamma^{F,*}}$ from a smooth solution are penalized.
This so-called ``ghost penalty'' stabilization was first presented in \cite{burman2010} for the Poisson's problem.
Details on the ``ghost penalty'' stabilization for the fluid equation, describing the applied method, can be found in \cite{massing2016}.
Herein, operators \eqref{eq:w_ghostpenalty} are added, which penalize jumps of normal derivatives of velocities and pressures integrated on a selected set of element faces $\mathcal{F}_{\Gamma^{F,*}}$
(see Figure~\ref{fig:cutfem}).
\begin{align}
\mathcal{W}^F_{\mathcal{G}}\left[\left(\testvelf, \testpf\right),\left(\velf, \pf\right)\right] =
\mathcal{G}_v\left(\testvelf,\velf\right)+\mathcal{G}_p\left(\testpf,\pf\right).
\label{eq:w_ghostpenalty}
\end{align}

\subsection{Nitsche-based method on the interface between fluid domain and structural domain}
\label{sec:num_fsinterface}
The geometry of the discrete interface between the fluid domain and structural domain is given by the deformed configuration of the computational mesh of the structural domain.
Therefore, $\fsiinterface$ does not match the element boundaries of the computational grid for the fluid domain.
Thus, a weak imposition of the coupling conditions \eqref{eq:fsi_dyneq} and \eqref{eq:fsi_noslip} is applied.
Using a Nitsche-based method leads to the following consistent contributions to the weak form (see \cite{burman2014,schott2017}):
\begin{align}
\mathcal{W}^{FSI}\left[\left(\testdisps,\testvelf, \testpf\right),\left(\disps,\velf, \pf\right)\right]=
\innerpb{\testdisps-\testvelf}{\stressf \normalf}{\fsiinterface} \nonumber\\
-\innerpb{\testpf \normalf - 2 \viscf \epsf(\testvelf) \normalf}{\velf - \vels}{\fsiinterface}
+\frac{\phi^F_{v}}{\gamma h_E}\innerpb{\testvelf-\testdisps}{\velf - \vels}{\fsiinterface}.
\label{eq:w_fsi}
\end{align}
The contributions in the first line arise from integration by parts of the strong forms \eqref{eq:structure_eq} and \eqref{eq:fluidm_eq},
by utilizing the dynamic equilibrium on the interface \eqref{eq:fsi_dyneq} and choosing the fluid stress to represent the interface traction (fluid-sided weighting).
The second line includes a skew-symmetric adjoint consistency term as well as a penalty term with a sufficiently small positive constant $\gamma$, 
an appropriate element length $h_E$, and the stabilization parameter $\phi^F_v$, which considers the viscous, convective and temporal components of the discrete fluid equations (see \cite{massing2016}).
These terms are consistent as they include the kinematic constraint \eqref{eq:fsi_noslip} and are essential for the stability of the numerical scheme as well as the enforcement of the coupling conditions.

\subsection{Nitsche-based method on the interface between fluid domain and poroelastic domain}
\label{sec:num_fpinterface}
Numerical treatment for interfaces between a viscous fluid domain and Darcy flow or poroelastic domain can be found in \cite{Badia2009} and in \cite{dangelo2011,bukavc2015,ager2018b} based on Nitsche's method.
Herein, similar to the discrete interface $\fsiinterface$, the discrete interface $\fpiinterface$ is given by the deformed configuration of the computational mesh of the poroelastic domain,
which in general is non-matching to the computational mesh of the fluid domain.

All constraints \eqref{eq:fpi_dyneq}--\eqref{eq:fpi_bj} are imposed weakly by a Nitsche-based method.
Fundamental mathematical analysis on Nitsche's method for general boundary conditions, such as the Beavers-Joseph condition, can be found in \cite{Juntunen2009} and \cite{winter2017}.
The present formulation is based on \cite{winter2017}, where a strict mathematical analysis for the Oseen problem with general Navier boundary conditions is performed.
All extensions to this formulation are performed in a way to keep the main argumentation of \cite{winter2017} valid, but without giving a strict mathematical proof for this formulation.
This Nitsche-based formulation is presented and analyzed for an unfitted fluid-poroelastic interaction in \cite{ager2018b}, where evidence of the suitability of this method for the 
problem setup considered herein is given.
\begin{align}
\mathcal{W}^{FPI}\left[\left(\testvelf, \testpf,\testvelp, \testdispp, \testpp\right),\left(\velf, \pf,\velp, \dispp, \pp\right)\right]=\nonumber\\
\innerpb{\testvelp+\testdispp-\testvelf}{\stressf  \cdot \normalf \cdot \Pnormal}{\fpiinterface}\nonumber\\
-\innerpb{\testpf \cdot \normalf - 2 \viscf \epsf (\testvelf) \cdot \normalf }{\left[{\velf} {- \velps} - \porosity \left( {\velp} {- \velps} \right)\right]\cdot \Pnormal}{\fpiinterface}\nonumber\\
 +\frac{\phi^F_v}{\gamma_n h_E}\innerpb{\testvelf - \testvelp - \testdispp}{\left[{\velf} - {\velps} - \porosity \left( {\velp} - {\velps} \right)\right]\cdot \Pnormal}{\fpiinterface}\nonumber\\
 +\innerpb{\testdispp-\testvelf}{ \stressf  \cdot \normalf \cdot \Ptangent}{\fpiinterface}\nonumber\\
+\frac{\gamma_t h_E}{\kappa\viscf+\gamma_t h_E}\innerpb{2 \viscf \epsf (\testvelf) \cdot \normalf }
 {\left[{\velf} {- \velps} - \porosity \left( {\velp} {- \velps} \right) + {\kappa \stressf \cdot \normalf} \right]\cdot \Ptangent}{\fpiinterface}\nonumber\\
 +\frac{\viscf}{\kappa\viscf+\gamma_t h_E}\innerpb{\testvelf - \testdispp}
 {\left[{\velf} - {\velps} - \porosity \left( {\velp} - {\velps} \right) + {\kappa \stressf \cdot \normalf} \right]\cdot \Ptangent}{\fpiinterface}.
 \label{eq:w_fpi}
\end{align}
As the kinematic constraints in normal and tangential direction are different, the interface terms are split by the normal $\Pnormal:=\Pnormallong$ and the tangential $\Ptangent:=\Ptangentlong$ projection operator.

The first three lines consider the coupling constraints in normal direction. The contribution in the first line originates from the normal components of integration by parts of equations \eqref{eq:fluidm_eq}, 
\eqref{eq:porom_eq} and \eqref{eq:mixturem_eq}, exploiting the dynamic equilibrium on the interface \eqref{eq:fpi_dyneq} and \eqref{eq:fpi_dyneqp}. 
Again, the interface traction is represented by the fluid stress.
In the following two lines, the so-called adjoint consistency and the penalty terms are given, which are consistent due to the included mass conservation on the interface \eqref{eq:fpi_massb}.
The constant $\gamma_n$, the element length $h_E$, and the stability parameter $\phi^F_v$ are specified analogously to Section \ref{sec:num_fsinterface}.
These terms are crucial to balance the destabilizing effect of the consistency terms in the first line and enforce mass balance on the interface.

All terms starting from line four enforce the coupling conditions in tangential direction. 
The fourth line includes the standard consistency boundary integrals in tangential direction, which arise in the derivation of the weak form equations \eqref{eq:fluidm_eq} and \eqref{eq:mixturem_eq}. 
Using the dynamic equilibrium \eqref{eq:fpi_dyneq} and choosing the fluid stress to represent also the tangential interface traction leads to the presented contributions.
In line five, the skew-symmetric adjoint consistency term, which balances not only the viscous consistency terms in line four, but also the ``consistency-like'' contribution of the penalty term in line six.
This is the reason for the prefactor of the term, which consists of the sufficiently small positive constant $\gamma_t$, the appropriate element length $h_E$, and the slip-coefficient $\kappa$ (see equation \eqref{eq:fpi_bj}).
Finally, the last term is the tangential penalty, just as the adjoint consistency, this term is consistent due to the included kinematic constraint \eqref{eq:fpi_bj}.
For further details on this formulation the reader is referred to \cite{ager2018b}.

\subsection{Dual Mortar method on the contact interface between poroelastic domain and structural domain}
\label{sec:num_psinterface}
To handle contact between poroelastic domain and structural domain on interface $\psciinterface$, 
a mortar segment-to-segment approach, where the no-penetration constraint \eqref{eq:psci_gap} is enforced in a weak sense by a Lagrange multiplier, is applied.
This approach is combined with a Nitsche-based method to guarantee fluid mass balance \eqref{eq:psci_nopen} between poroelastic domain and impermeable solid (see Section \ref{sec:num_psinterface_fluid}).

Details on mortar based contact approaches can be found in, e.g., \cite{puso2004} or for dual mortar based methods in \cite{hueber2008, popp2010}, whereby the letter reference gives details on the specific method applied here.
In the following, just a brief overview of the most important aspects of the method is presented.

In contrast to the continuous interfaces, the discrete contact interfaces arising from the poroelastic domain $\restboundp$ and the structural domain $\restbounds$ are geometrically not exact overlapping. 
Therefore, we introduce $\pscpiinterfaceh$ and $\pscsiinterfaceh$, which are the potential contact interfaces arising from the poroelastic domain $\domainp$ and the structural domain $\domains$, respectively.
In addition, we also introduce the interfaces $\pscpiinterface = \restboundp \setminus \fpiinterface$ and $\pscsiinterface = \restbounds \setminus \fsiinterface$, which are the contact interfaces restricted to the ``active'' contact zone.
A Lagrange multiplier $\lagmultpss$, discretized on the interface $\pscpiinterfaceh$, which represents the total contact traction between the interfaces, is introduced.

The Karush-Kuhn-Tucker conditions \eqref{eq:gapcond} - \eqref{eq:contactexclusioncond} can be expressed by a complementary function $C\left(\lagmultpssnormal^j,g^j \right)$ (with an algorithmic constant $c_n>0$):
\begin{align}
\label{eq:w_KKT}
\lagmultpssnormal^j = \lagmultpss^j \cdot \snormalp^j ,\quad g^j = \innerpb{\psi^j_\delta}{\left( \coords_{\Gamma} - \coordp_{\Gamma} \right) \cdot \snormalp}{\pscpiinterfaceh},\quad\nonumber\\
C\left(\lagmultpssnormal^j,g^j \right) = \left(\lagmultpssnormal^j-f^j_n\right) - \text{max}\left(0,\left(\lagmultpssnormal^j-f^j_n\right)-c_n g^j\right) = 0.
\end{align}
Herein, the index $j$ specifies quantities which correspond to a specific computational node $j$.
The smoothed normal vector field $\snormalp$ is evaluated on the interface $\pscpiinterfaceh$ based on the unit outward-pointing normal $\normalp$ of the poroelastic domain $\domainp$.
This normal vector field is also utilized for all projections between the interfaces $\pscpiinterfaceh$ and $\pscsiinterfaceh$.
Subsequently, $\snormalp^j$ is the nodal smoothed normal vector, $\lagmultpss^j$ the nodal discrete component of the Lagrange multiplier $\lagmultpss$, $g^j$ the nodal weighted gap,
and $\psi^j_{\delta}$ the nodal shape function of the test function $\testlagmultpss$ on the interface $\pscpiinterfaceh$.

In order to take into account the change in the contact state depending on the relative traction (see $\itraction$ in \eqref{eq:contacttractioncond}) 
and to allow for a continuous transition between ``active'' dual mortar contact nodes and the Nitsche coupling method on $\fsiinterface$ or $\fpiinterface$, the nodal fluid pressure force and its normal component is evaluated:
\begin{align}
\tns f^j = \frac{\innerpb{N^j}{\pp \normalp}{\pscpiinterface}}{\innerpb{N^j}{\psi^j}{\pscpiinterfaceh}}, \quad f^j_n = \tns f^j \cdot \snormalp^j.
\end{align}
Herein, $N^j$ are the nodal shape functions applied for the discretization of the test function $\testdispp$ on the discrete boundary of the porous domain $\domainp$.
It should be pointed out that $\tns f^j$ and $f^j_n$ is non-zero just for nodes, which are adjacent to contacting boundary elements, due to the integration on the ``active'' part of the contact interface $\pscpiinterface$.

By incorporating the complementary function and restricting the discrete Lagrange multiplier with nodal shape function $\psi^j$, as well as its corresponding test function to the ``active set'' and in normal direction:
\begin{align}
\label{eq:w_alambda}
\lagmultpss^{\mathcal{A}} = \sum_{j} \psi^j\lagmultpss^j \, \,
\text{with}\,\, \lagmultpss^j \cdot \left(\unity - \snormalp^j \otimes \snormalp^j\right) = \zerovec  \,\, \text{and} \,\, \lagmultpss^j = \tns f^j \,\, \text{if}\,\, (\lagmultpssnormal^j -f^j_n - c_n g^j) < 0,\nonumber\\
\testlagmultpss^{\mathcal{A}} = \sum_{j} \psi^j_\delta\testlagmultpss^j \, \,
\text{with}\,\, \testlagmultpss^j \cdot \left(\unity - \snormalp^j \otimes \snormalp^j\right) = \zerovec  \,\, \text{and} \,\, \testlagmultpss^j = \zerovec\,\, \text{if}\,\, (\lagmultpssnormal^j -f^j_n - c_n g^j) < 0,
\end{align}
the contribution of the contact constraints can be written as:
\begin{align}
\label{eq:w_contact}
\mathcal{W}^{PS,c}\left[\left(\testdisps, \testdispp, \testlagmultpss^\mathcal{A} \right),\left(\disps, \dispp,\dlagmultpss^\mathcal{A}\right)\right]=
\nonumber\\\qquad-\innerpb{\testdispp-\testdisps}{-\lagmultpss^{\mathcal{A}}}{\pscpiinterfaceh} 
+\innerpb{\testlagmultpss^{\mathcal{A}}}{\left( \coords_{\Gamma} - \coordp_{\Gamma} \right)}{\pscpiinterfaceh}.
\end{align}
Herein, the first term originates from the standard Galerkin interface terms of \eqref{eq:structure_eq} and \eqref{eq:mixturem_eq} by inserting the dynamic equilibrium \eqref{eq:psci_dyneq} and representing the interface traction as $\lagmultpss^\mathcal{A}$.
It should be pointed out that the second term includes the constraint \eqref{eq:psci_gap} due to the restriction into normal direction \eqref{eq:w_alambda} of the test function.

For the Lagrange multiplier, dual shape functions are applied.
This allows for an efficient condensation of the Lagrange multiplier from the final system and, therefore, to a reduction of the final system size being solved, as well as removal of the saddlepoint structure of the final system.
Details on these aspects can be found in \cite{popp2010}. For the discretization of the test function $\testlagmultpss$, standard shape
functions were used, leading to a Petrov-Galerkin type of Lagrange multiplier interpolation, details can be found in \cite{popp2013}.

\subsubsection{Nitsche-based method for fluid mass balance on contact interface}
\label{sec:num_psinterface_fluid}
In addition to the classical contact constraints, the fluid mass balance \eqref{eq:psci_nopen} on the contact interface has to be fulfilled.
For this purpose, a Nitsche-based method with the following contributions to the weak form is applied once again:
\begin{align}
\label{eq:w_contact_f}
\mathcal{W}^{PS,f}\left[\left(\testvelp, \testdispp,\testpp\right),\left(\velp, \dispp,\pp\right)\right] = 
\innerpb{\testvelp}{\pp \normalp}{\pscpiinterface},
\nonumber\\
-\innerpb{\testpp}{\porosity \left( \velp - \velps \right)\cdot \normalp  }{\pscpiinterface}
+\frac{1}{\gamma_P}\innerpb{\left(\testvelp-\testdispp\right)\cdot\normalp}{\left(\velp - \velps\right)\cdot\normalp}{\pscpiinterface}.
\end{align}
Herein, the first term is the consistency term arising from the derivation of the weak form from \eqref{eq:porom_eq}.
The second line includes a skew-symmetric adjoint consistency term and finally a penalty term 
with a sufficiently small positive penalty parameter $\gamma_P$.
The motivation for adding these different contributions is already depicted in Sections \ref{sec:num_fsinterface} and \ref{sec:num_fpinterface}.

\subsection{Final coupled system for the rough FSCI problem}
\label{sec:num_globalsystem}
Finally, summing up all contributions arising from the different domains and interfaces described in the last sections yields the following discrete weak form of the rough FSCI problem that is to be solved.\\
Find $\left(\disps,\velf, \pf ,\velp,\dispp,\pp,\lagmultpss^{\mathcal{A}}\right)$ such that $\forall \left(\testdisps,\testvelf, \testpf,\testvelp, \testdispp, \testpp,\testlagmultpss^{\mathcal{A}}\right)$:
\begin{align}
\mathcal{W}^S\left[\testdisps,\disps\right] \nonumber\\
+ \mathcal{W}^F\left[\left(\testvelf, \testpf\right),\left(\velf, \pf\right)\right] + \mathcal{W}^F_{\mathcal{S}}\left[\left(\testvelf, \testpf\right),\left(\velf, \pf\right)\right]
+ \mathcal{W}^F_{\mathcal{G}}\left[\left(\testvelf, \testpf\right),\left(\velf, \pf\right)\right]\nonumber\\
+ \mathcal{W}^P\left[\left(\testvelp, \testdispp, \testpp\right),\left(\velp,\dispp,\pp\right)\right] + \mathcal{W}^P_{\mathcal{S}}\left[\left(\testvelp, \testpp\right),\left(\velp, \pp\right)\right]\nonumber\\
+ \mathcal{W}^{FSI}\left[\left(\testdisps,\testvelf, \testpf\right),\left(\disps,\velf, \pf\right)\right]\nonumber\\
+\mathcal{W}^{FPI}\left[\left(\testvelf, \testpf,\testvelp, \testdispp, \testpp\right),\left(\velf, \pf,\velp, \dispp, \pp\right)\right]\nonumber\\
+\mathcal{W}^{PS,c}\left[\left(\testdisps, \testdispp, \testlagmultpss^{\mathcal{A}} \right),\left(\disps, \dispp,\lagmultpss^{\mathcal{A}}\right)\right]
+\mathcal{W}^{PS,f}\left[\left(\testvelp, \testdispp,\testpp\right),\left(\velp, \dispp,\pp\right)\right]
= 0.
\end{align}
This semi-discrete form is discretized in time by a one-step-$\theta$ scheme, with equal $\theta$ for each present time derivative and solved for each timestep $n$ separately in the time interval
$[\btime,\etime]$. This leads to a system of nonlinear equations of the form 
$\tns{\mathcal{R}} = \zerovec$ for each timestep, which is solved by a Newton-Raphson based procedure:
\begin{align}
\label{eq:linearizations}
\tns C = \partiald{\tns{\mathcal{R}}}{\disctns x}, \quad
\tns{C}_{n+1}^i\cdot \Delta  \disctns x_{n+1}^{i+1} = -\tns{\mathcal{R}}_{n+1}^i, \quad
\disctns x_{n+1}^{i+1} = \disctns x_{n+1}^{i} + \Delta  \disctns x_{n+1}^{i+1}.
\end{align}
The whole system is solved in a monolithic approach, where all nodal unknowns of all fields $\disctns x$ are solved and updated within the same iteration $i$.
As soon as the a convergence criteria $||\tns{\mathcal{R}}||<\epsilon$ for a sufficiently small tolerance $\epsilon$ is reached, the iterations are stopped and the next timestep is computed.
Algorithmic details on the solution procedure for the fixed-grid fluid-structure interaction can be found in \cite{schott2017} and provide the basis for the algorithm applied here.

\section{Numerical examples}
\label{sec:num_ex}
In this section, numerical examples are presented to analyze the behavior of the proposed model. 
First, a configuration for a typical leakage flow scenario is analyzed,
followed by a squeeze-out flow of two contacting bodies.
Finally, a non-return valve, with focus on the dynamic ``closing'' and ``opening'' behavior, is investigated.
All results presented in the following have been computed with the multiphysics code environment BACI (see \cite{WalletBaciCommittee2017}).

In the following, some remarks on algorithmic details are given, which are applied to solve the numerical examples.
\begin{itemize}
\item Due to the weak enforcement of the contact constraints by the Mortar methods, the distance between both involved discrete contacting interfaces is not zero at every spatial point.
As a result, tiny disconnected fluid domains would arise, which are just a numerical artifact. To avoid this issue, ``islands'' smaller than a specific size are neglected and do not 
contribute to the fluid interfaces $\fsiinterface,\fpiinterface$, but to the contact interface $\pscpiinterfaceh$.

\item All numerical examples are discretized with 3D-hexahedral trilinear elements. For 2D examples, one element layer in the third axial direction is applied.
All degrees of freedom in the third direction are set to zero and removed from the final system of equations.
\item It should be pointed out that not all contributions of the linearization matrix $\tns C$ in \eqref{eq:linearizations} are considered to solve the system for the numerical examples presented in the next section.
Especially linearizations of the fluid weak form $\mathcal{W}^F$ and interface contributions $\mathcal{W}^{FSI}, \mathcal{W}^{FPI}$, w.r.t. to the interface position are neglected and treated in a fixed-point fashion.
\item To avoid non-essential geometric operations, 
the update of the geometric intersection of computational fluid mesh and the interfaces $\Gamma^{FS}$, $\Gamma^{FP}$ and $\Gamma^{PS,c}$  during the nonlinear iterations
is only performed as long as the displacement increments are above a specified tolerance.
\end{itemize}

\noindent To avoid multiple repetition, in the following, some common setup details for the examples shown here are depicted first.

\begin{itemize}
\item The initial state of all examples in the following is the zero-state: $\dispsB = \velsB = \velfB = \disppB = \velpsB = \velp = \zerovec$.
\item The dependence of fluid resistance and deformation in the poroelastic media is modeled by an adaption of the Kozeny-Carman formula (see e.g. \cite{Coussy:04}).
For the assumption of isotropy, the following relation holds for the material permeability:
\begin{align}
\matpermeabp = \matpermeabpscalar \unity = \initmatpermeabpscalar \frac{1-\porosityB^2}{\porosityB^3} \frac{\left(\Jp\porosity\right)^3}{1-\left(\Jp\porosity\right)^2}\unity,
\label{eq:kozcar}
\end{align}
where $\initmatpermeabpscalar$ is the initial scalar permeability.
 \item The proportionality factor of the Beavers-Joseph condition is computed based on \cite{beavers1967} and modified for the 3D case:
 \begin{align}
\kappa  = \frac{\sqrt{\text{tr}(\matpermeabp)}}{\sqrt{3} \alpha_{BJ} \viscf},
\label{eq: bjcoeff}
\end{align}
where $\alpha_{BJ}$ is a positive constant which has to be verified experimentally.
\item The dual shape functions of the contact Lagrange multiplier $\phi^j$ are based on the boundary elements of the poroelastic domain, where the index $j$ specifies a specific computational node on this discrete boundary.
\end{itemize}

In the following, if not necessary, there is no distinct separation between solid $\disps$ and poroelastic displacements $\dispp$, fluid $\velf / \pf$ and poroelastic $\velp / \pp$ velocities / pressure,
as this eases the visualization and the interpretation of the computed results.

\subsection{Leakeage test}
\begin{figure}[tp]
\centering
\definecolor{fluidc}{rgb}{0.5,0.8,1}
\definecolor{poroc}{rgb}{1.0,0.2,0.0}
\definecolor{solidc}{rgb}{1.0,0.7,0.2}
\begin{center}
\setlength{\unitlength}{0.8mm}
\begin{picture}(170,80)
\put(3.5,0){\color{solidc}\mbox{\rule[0mm]{80mm}{8mm}}}
{\color{solidc}\put(28.2,11){\rule[0mm]{40mm}{40mm}}}
{\color{poroc}\put(2,10){\rule[0mm]{80mm}{0.8mm}}}
{\color{fluidc}\put(1,11){\rule[0mm]{20mm}{23.2mm}}}
{\color{fluidc}\put(75,11){\rule[0mm]{20mm}{23.2mm}}}
\linethickness{0.3mm}
\put(0,0){\line(1,0){100}}
\put(0,10){\line(1,0){100}}
\put(0,11){\line(1,0){100}}
\put(25,61){\line(1,0){50}}
\put(25,11){\line(0,1){50}}
\put(75,11){\line(0,1){50}}
\put(0,0){\line(0,1){11}}
\put(100,0){\line(0,1){11}}
%\linethickness{0.3mm}
\put(0,0){\line(0,1){40}}
\put(100,0){\line(0,1){40}}
\put(0,40){\line(1,0){25}}
\put(75,40){\line(1,0){25}}
%domains
\linethickness{0.1mm}
\put(15,15.5){\vector(-1,-1){5}}
\put(15,14){$\Omega^P$}
\put(15,21){\vector(-1,-1){10}}
\put(15,19.5){$\Gamma^{FP}$}
\put(55,40){\vector(1,-1){20}}
\put(45,40){\vector(-1,-1){20}}
\put(46,38){$\Gamma^{FS}$}
\put(55,21){\vector(1,-1){10}}
\put(55,21){\vector(1,-2){5.5}}
\put(53,19){$\Gamma^{PS}$}
\put(48,30){$\Omega^{S_1}$}
\put(30,3){$\Omega^{S_2}$}
\put(11,25){$\Omega^{F}$}
\put(85,25){$\Omega^{F}$}
\put(102,22){$\Gamma^{out}$}
\put(-11,22){$\Gamma^{in}$}
\put(48,66){$\Gamma^{p}$}
%slip boundary
\put(24.2,11){\line(0,1){50}}
\multiput(23.2,11)(0,1){50}{\put(0,0){\line(1,1){1}}}
\put(75.8,11){\line(0,1){50}}
\multiput(75.8,11)(0,1){50}{\put(0,0){\line(1,1){1}}}
\put(0,40.8){\line(1,0){25}}
\put(75,40.8){\line(1,0){25}}
\multiput(0,41.8)(1,0){25}{\put(0,0){\line(1,-1){1}}}
\multiput(75,41.8)(1,0){25}{\put(0,0){\line(1,-1){1}}}
%fixed boundary
\multiput(-1,-1)(1,0){102}{\put(0,0){\line(1,1){1}}}
\multiput(-1,-1)(0,1){12}{\put(0,0){\line(1,1){1}}}
\multiput(100,-1)(0,1){12}{\put(0,0){\line(1,1){1}}}
%neumann struct
\multiput(26,65)(2,0){25}{\put(0,0){\vector(0,-1){4}}}
\multiput(-4,12)(0,1.94){15}{\put(0,0){\vector(1,0){4}}}
%sizes
\put(0,0){\line(-1,0){14}}
\put(0,10){\line(-1,0){7}}
\put(0,11){\line(-1,0){7}}
\put(-6,5){\vector(0,-1){5}}
\put(-9,3){$h$}
\put(-6,5){\vector(0,1){5}}
\put(-6,10){\line(0,1){1}}
\put(-6,18){\vector(0,-1){7}}
\put(-9,14){$\delta$}
\put(0,40){\line(-1,0){14}}
\put(-13,20){\vector(0,-1){20}}
\put(-13,20){\vector(0,1){20}}
\put(-16,18){$b$}
\put(0,0){\line(0,-1){7}}
\put(100,0){\line(0,-1){7}}
\put(50,-6){\vector(-1,0){50}}
\put(50,-6){\vector(1,0){50}}
\put(50,-5.5){$l$}
\put(50,56){\vector(1,0){25}}
\put(50,56){\vector(-1,0){25}}
\put(50,56.5){$a$}
\put(32,28){\vector(0,1){33}}
\put(32,26){\vector(0,-1){15}}
\put(29,34){$a$}
\put(50,7){\line(0,1){7}}
\put(44,8){\vector(1,0){6}}
\put(44,4){$A$}
\put(44,13){\vector(1,0){6}}
\put(44,13.5){$A$}
\put(0,40){\line(0,1){24}}
\put(25,40){\line(0,1){24}}
\put(12.5,63){\vector(1,0){12.5}}
\put(12.5,63){\vector(-1,0){12.5}}
\put(12,64){$c$}
%legend
\linethickness{0.3mm}
\put(120,65){\line(1,0){10}}
\linethickness{0.1mm}
\multiput(120,64)(1,0){10}{\put(0,0){\line(1,1){1}}}
\put(132,63.5){zero displacement}
\linethickness{0.3mm}
\put(120,55.5){\line(1,0){10}}
\linethickness{0.1mm}
\put(120,55){\line(1,0){10}}
\multiput(120,54)(1,0){10}{\put(0,0){\line(1,1){1}}}
\put(132,55.5){zero normal displacement}
\put(132,51.5){zero normal velocity}
\put(120,43.5){$a = 5 mm$}
\put(120,33.5){$b = 4 mm$}
\put(120,23.5){$c = 2.5 mm$}
\put(120,13.5){$l = 10mm$}
\put(120,3.5){$h = 1 mm$}
\put(120,-6.5){$\delta = 100 \mu m$}
\end{picture}
\end{center}
\caption{Geometry and boundary conditions for the leakage flow.}
\label{fig:ex1}
\end{figure}
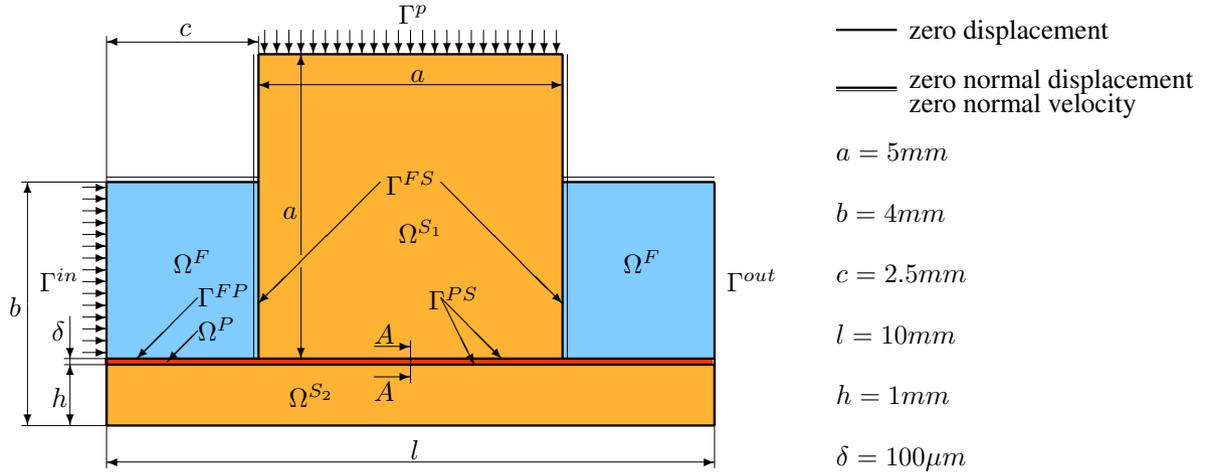

The first example analyzes the properties of the presented model for leakage flows. The setup of the problem can be seen in Figure \ref{fig:ex1}, 
where two fluid domains with different fluid pressure levels are connected solely by the fluid-saturated rough surface domain of two contacting elastic bodies.

To allow for a qualitative comparison with measurements, the dimension and material parameters of the problem are closely related to the experiment in \cite{lorenz2010}.
Therein, a water filled glass cylinder with a rubber ring on the bottom is pressed by a given force onto a rough hard solid.
A defined water column in the glass cylinder allows them to specify the fluid pressure difference in the experimental setup.
Here, we consider a plane slice of the circular leakage configuration, all dimensions and basic boundary conditions can be found in Figure \ref{fig:ex1}.

\begin{figure}[tp]
\centering
\subfigure{\includegraphics[width=0.49\textwidth]{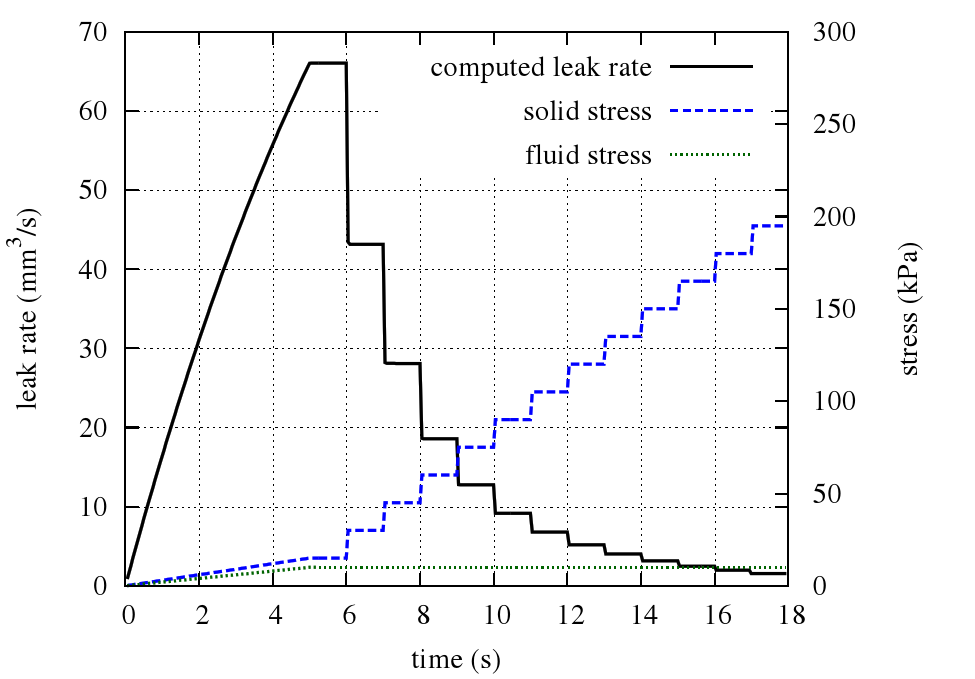}}
\subfigure{\includegraphics[width=0.49\textwidth]{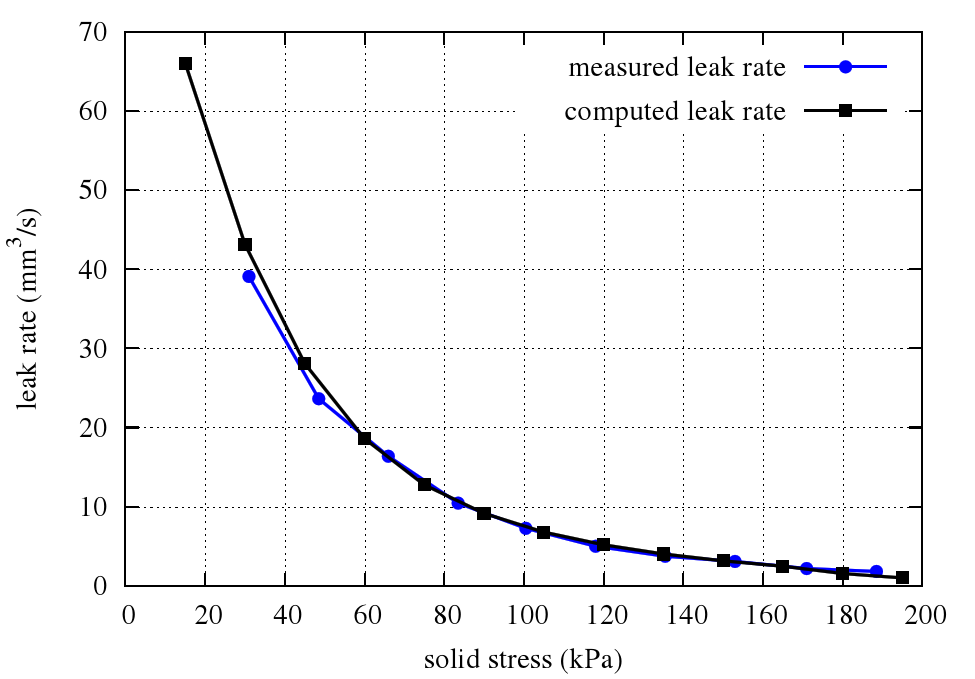}}
\caption{Applied solid stress on boundary $\Gamma^P$, fluid stress on boundary $\Gamma^{in}$ and the computed stationary leak rate in cross-section $A-A$ (left).
Comparison of the computed stationary leak rates for different solid stresses and constant fluid stress of $10 kPa$ with measured leak rates from \cite{lorenz2010} (right).}
\label{fig:ex1_leakage}
\end{figure}

The pressure difference between inflow and outflow boundary is applied by a prescribed traction-Neumann boundary condition on $\Gamma^{in}$ (see fluid stress in Figure \ref{fig:ex1_leakage} (left)) and a zero-Neumann boundary condition on $\Gamma^{out}$.
Fluid velocities on both boundaries in tangential direction are prohibited by a tangential-Dirichlet boundary condition.
To analyze the resulting leakage flow for a range of solid contact pressure, a time-dependent solid stress is prescribed on boundary $\Gamma^p$ as Neumann boundary condition (see solid stress in Figure \ref{fig:ex1_leakage} (left)).
Motion in tangential direction on the boundary $\Gamma^p$ is blocked. Interfaces $\Gamma^{FS}$ and $\Gamma^{FP}$ (with $\alpha_{BJ}=1.0$) are handled by the methods presented in Sections \ref{sec:num_fsinterface} and \ref{sec:num_fpinterface}, respectively.
As the focus of this example is on presenting the behavior of the poroelastic layer during rough surface contact, 
contact between the domains $\Omega^{S_1}$,$\Omega^{S_2}$, and $\Omega^P$ is considered directly by matching of the computational nodes and the corresponding displacements $\disps$ and $\dispp$ on the common interfaces $\Gamma^{PS}$.

In domain $\Omega^F$, water with fluid material properties at a temperature of $20^\circ C$, density $\densityf = 1000 kg/m^3$, and dynamic viscosity $\viscf = 10^{-3} Pa\cdot s$ is considered.
The material behavior of the solid domains $\Omega^{S_1}$ and $\Omega^{S_2}$ is modeled by the Neo-Hookean material model with the hyperelastic strain energy function:
\begin{align}
\label{ex1:strainenergys}
\strainenergys = c \left[\text{tr}\left(\left(\defgrads\right)^T\cdot\defgrads\right)-3\right]+\frac{c}{\beta}\left(\left(\Js\right)^{-2 \beta}-1\right), \quad
c = \frac{E}{4(1+\nu)}, \quad
\beta = \frac{\nu}{1-2 \nu}.
\end{align}
In domain $\domain^{S_1}$, the Young's modulus is $E^1 = 2.3 MPa$ and the Poisson ratio $\nu^1 = 0.49$. In domain $\domain^{S_2}$, $E^2 = 2300 MPa$ and $\nu^2 = 0.3$ are given.
The initial density is chosen to be equal to the fluid density in both solid domains $\refdensitys = \densityf = 1000 kg/m^3$.
As no direct computation of the poroelastic material parameters for these specific rough surfaces is performed, all parameters of the poroelastic layer are chosen in a physically plausible range. 
The initial porosity is set to $\porosityB = \porosity(t=0) = 0.5$ being constant in space for the poroelastic layer $\domainp$.
To describe a typical macroscopic material response for homogenized contact of rough surfaces, 
the macroscopic material behavior of the poroelastic layer is modeled by the following strain energy function (see Remark in Section \ref{rem:poro_matpar}):
\begin{align}
\label{ex1:strainenergysp}
\strainenergy{P,skel} = c \left[\text{tr}\left(\left(\defgradp\right)^T\cdot\defgradp\right)-3\right]+\frac{c}{\beta}\left(\left(\Jp\right)^{-2 \beta}-1\right)
+ \tilde{c} \left[\text{tr}\left(\left(\defgradp\right)^T\cdot\defgradp\right)-3\right]^{\alpha}, \nonumber\\
c = \frac{E^P}{4(1+\nu^P)}, \quad \beta = \frac{\nu^P}{1-2 \nu^P}.
\end{align}
The parameters are: $E^P = 0.25MPa$, $\nu^P = 0.0$, $\tilde{c}^P = 0.25MPa$, $\alpha^P = 8$.
The additional contributions to the strain energy function are
\begin{align}
\label{ex1:addstrainenergysp}
\strainenergy{P,vol} = \kappa^P \left[ \frac{(1-\porosity)\Jp}{1-\porosityB}-1-\text{ln}\left(\frac{(1-\porosity)\Jp}{1-\porosityB}\right)\right], \quad
\strainenergy{P,pen} = \eta^P\left[ \frac{\Jp \porosity}{\porosityB}-\frac{1}{\porosityB}-\text{ln}\left(\frac{\Jp \porosity}{\porosityB}\right)\right],
\end{align}
with the parameters $\kappa^P = 0.8MPa$ and $\eta^P = 1kPa$.
Considering the resulting leak rate of the rough microstructure in \cite{lorenz2010}, the initial material permeability is set to $\initmatpermeabpscalar = 4.6\cdot 10^{-4} mm^2$.

For time discretization, the backward Euler scheme $(\theta = 1.0)$ with a timestep length of $\Delta t = 0.05 s$ is applied.
As the computation is performed with one layer of 3-dimensional hexahedral elements in direction orthogonal to the 2-dimension plane, all computed leak rates are divided by the thickness in this direction.

To allow for a comparison of the computed leak rate with measured data in \cite{lorenz2010} for ``sandpaper 120'', measured leak rates are divided by the average circumference to compare leak rate per unit depth.
The computed leak rates are calculated in cross section $[A-A]$ of the poroelastic layer (see Figure \ref{fig:ex1}):
\begin{align}
\label{eq:leakrate}
\text{leak rate} = \int_{[A-A]}{\porosity\left(\velp - \velps \right)\cdot\normal^{[A-A]}}\text{d}[A-A].
\end{align}
In Figure \ref{fig:ex1_leakage} (left), the computed temporal instationary leak rates are presented.
After an instationary phase until $t=5s$, the fluid stress is kept constant.
The solid stress is increased discontinuously and kept constant for $\Delta t_{const} = 1s$.
Figure~\ref{fig:ex1_leakage} (right) shows a comparison of the computed stationary leak rates (last computed leak rate for each solid stress level), with the measured data in \cite{lorenz2010}.
The excellent agreement shows that the influence of elastic deformation due to the contact stress on the fluid flow in the rough layer can be modeled by the presented poroelastic model.

\begin{figure}[tp]
\centering
\includegraphics[width=0.7\textwidth]{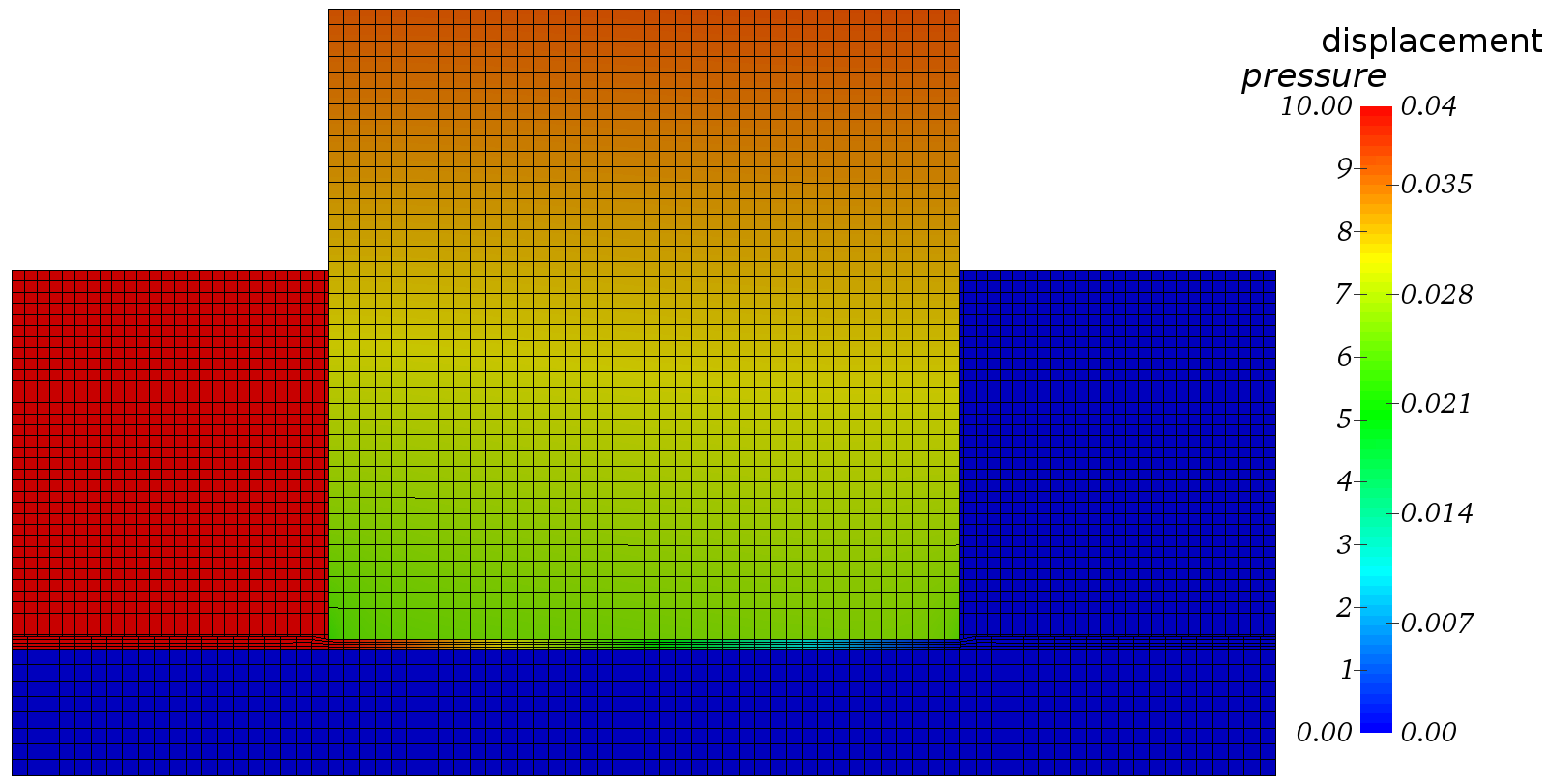}
\caption{Pressure solution for the domains $\Omega^F$ and $\Omega^P$, displacement solution for domain $\Omega^{S_1}$ and $\Omega^{S_2}$ at solid stress of $90 kPa$ and t = 11s. The black lines indicate the discretization with trilinear hexahedral elements.}
\label{fig:ex1_visualization}
\end{figure}

Figure \ref{fig:ex1_visualization} shows the overall pressure solution for the fluid domains as well as the displacement solution in the structural domains.
As intended, there is no observable pressure gradient in the fluid domain $\domainf$ and the entire pressure drop takes place in the poroelastic layer.

\begin{figure}[tp]
\centering
\includegraphics[width=1.0\textwidth]{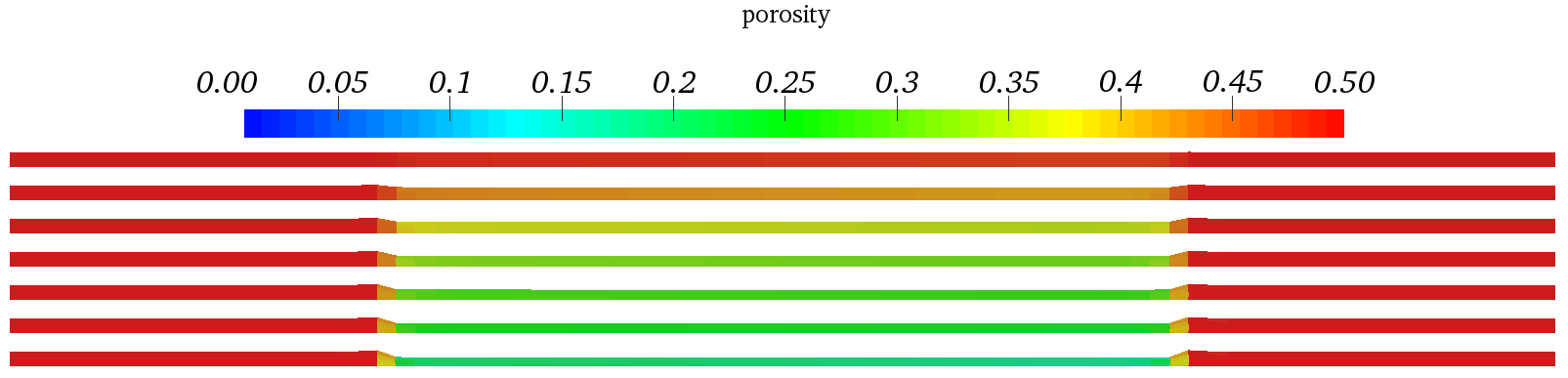}
\caption{Porosity of the deformed poroelastic layer: from top to bottom solid stress = $15,45,75,105,135,165,195 kPa$ (element-wise constant visualization).}
\label{fig:ex1_porosity}
\end{figure}

Figure \ref{fig:ex1_porosity} shows a visualization of the deformation of the poroelastic layer and the porosity for different solid stress levels.
It can be observed that for small solid stress, hardly any deformation can be observed and the porosity is close to the initial porosity.
Increasing the solid stress leads to a compression of the layer and for the relatively small fluid resistance, to a reduction of the porosity.
The smaller cross section and the increased flow resistance (change in permeability) leads to the reduction of the leak rate for higher solid stress.

\subsection{Rough surface contacting stamp}
\begin{figure}[tp]
\centering
\input{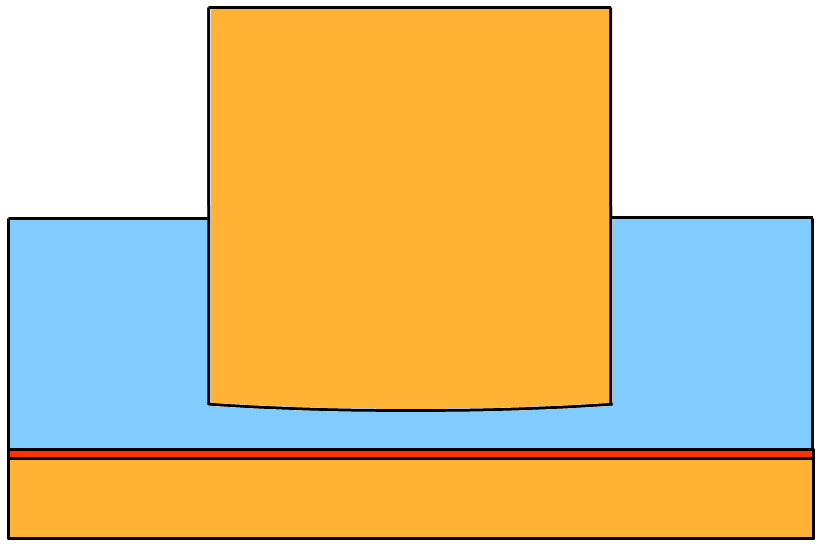}
\caption{Geometry and boundary conditions of the rough surface contacting stamp example.}
\label{fig:ex2}
\end{figure}

We analyze the rough surface contact behavior of an elastic stamp with a circular contacting surface impacting on a stiff but elastic foundation.
The setup consists of a stamp in solid domain $\domain^{S_1}$ and the foundation with solid domain $\domain^{S_2}$ coated with a poroelastic layer $\domainp$ to consider surface roughness.
Both bodies are embedded in a fluid $\domainf$. The geometry with all dimensions as well as basic boundary conditions can be found in Figure \ref{fig:ex2}.
To allow a distinct evaluation of the stamp impact as well as the lift off behavior, 
undesired vertical fluid loads acting on the stamp are prevented by a vertical linear slide.
On the interfaces $\Gamma^{FS}$, $\Gamma^{FP}$ (with $\alpha_{BJ}=1.0$), $\Gamma^{PS}$, and $\Gamma^{PS,c}$, the interface conditions are incorporated by the methods presented in \ref{sec:num_fsinterface}, \ref{sec:num_fpinterface}, and \ref{sec:num_psinterface}.

\begin{figure}[tp]
\centering
\subfigure{\includegraphics[width=0.49\textwidth]{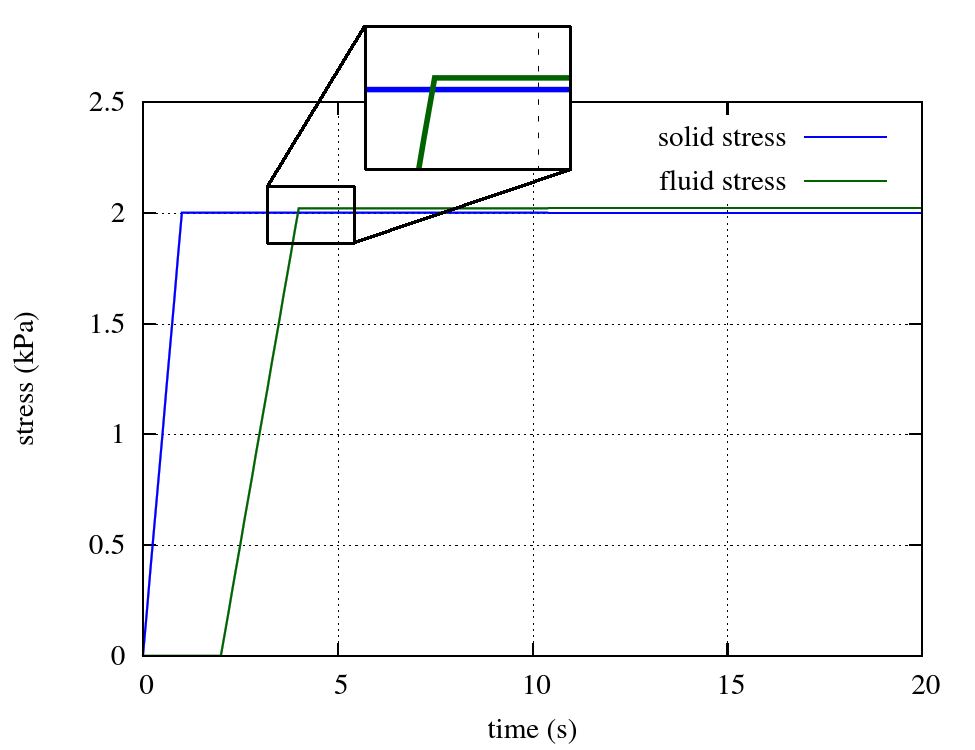}}
\subfigure{\includegraphics[width=0.49\textwidth]{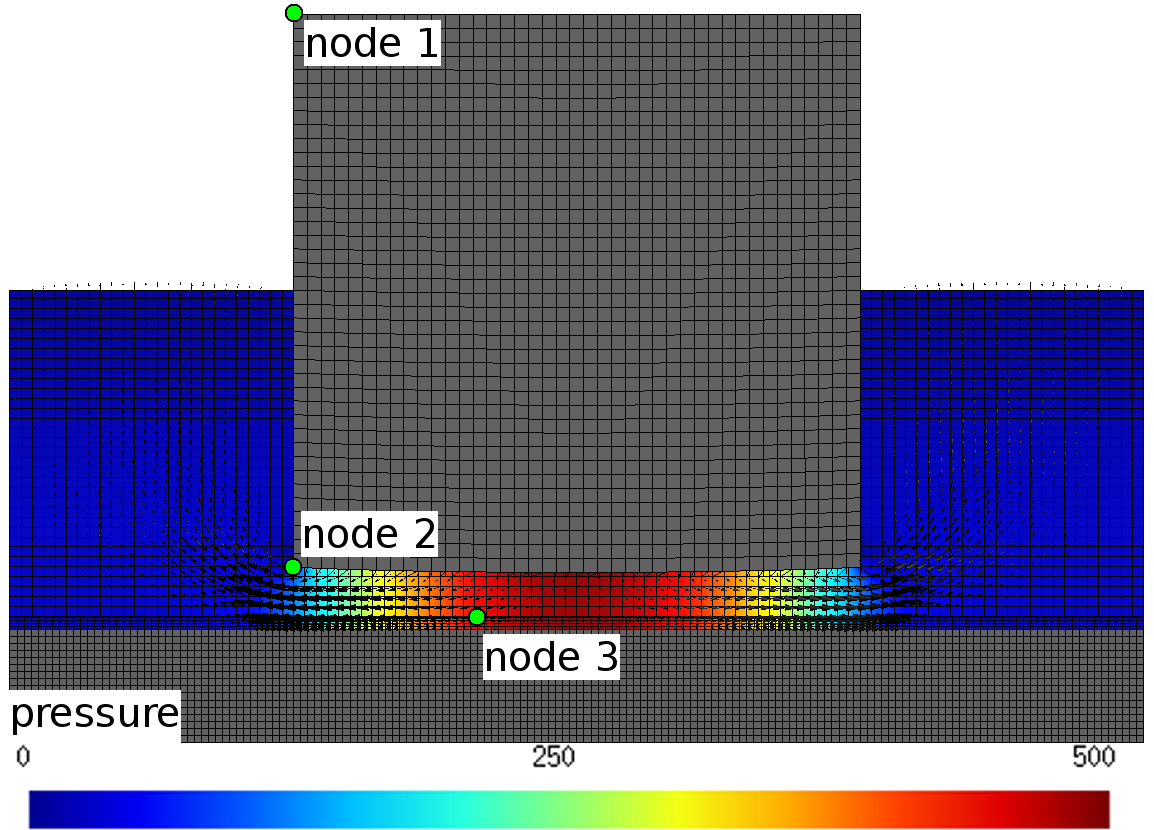}}
\caption{Time dependent solid and fluid stress on the boundaries $\Gamma^{S,N}$ and $\Gamma^{F,N}$ (left).
Fluid pressure, velocity (black arrows) and displacement (deformed domain) solution for $t=0.2s$ and visualization of the computational mesh (right).}
\label{fig:ex2_load}
\end{figure}

Starting from this initial configuration, an increasing solid stress of maximum $2kPa$ is applied on boundary $\Gamma^{S,N}$, which leads to a squeeze-out motion of the fluid, and finally contact occurs. 
Afterwards, the fluid stress is increased on boundaries $\Gamma^{F,N}$ with a maximum value of $2.02kPa$, which is slightly above the maximum solid stress.
The time-dependent solid and fluid stress are shown in Figure \ref{fig:ex2_load} (left).
As soon as the local fluid pressure in the poroelastic layer exceeds the contact stress, the bodies will lift off in this local position.
As a higher fluid stress than solid stress is applied, the whole stamp lifts off and finally moves upwards.

The fluid material parameters are $\densityf = 1.0 kg/m^3$ and $\viscf = 1.0 Pa\cdot s$.
The material behavior of the solid domains as well as the macroscopic material behavior of the poroelastic layer are modeled by the Neo-Hookean material model with the hyperelastic strain energy function:
\begin{align}
\label{ex2:strainenergysp}
\strainenergys = \strainenergy{P,skel} =c \left[\text{tr}\left(\left(\defgrads\right)^T\cdot\defgrads\right)-3\right]+\frac{c}{\beta}\left(\left(\Js\right)^{-2 \beta}-1\right), \nonumber\\
c = \frac{E}{4(1+\nu)}, \quad
\beta = \frac{\nu}{1-2 \nu}.
\end{align}
The Young's modulus $E$ and the Poisson ratio $\nu$ in solid domain $\domain^{S_1}$ is: $E^1=20kPa, \nu^1=0.3$, in domain $\domain^{S_2}$: $E^2=2000kPa, \nu^2=0.3$, and in the poroelastic domain
$\domain^P$: $E^P=10kP, \nu=0.0$. The additional volume and penalty contributions to the strain energy function of the poroelastic domain with parameters $\kappa^P = 1000kPa$, $\eta^P = 1.0Pa$ are:
\begin{align}
\label{ex2:addstrainenergysp}
\strainenergy{P,vol} = \kappa^P \left[ \frac{(1-\porosity)\Jp}{1-\porosityB}-1-\text{ln}\left(\frac{(1-\porosity)\Jp}{1-\porosityB}\right)\right], \quad
\strainenergy{P,pen} = \eta^P\left[ \frac{\Jp \porosity}{\porosityB}-\frac{1}{\porosityB}-\text{ln}\left(\frac{\Jp \porosity}{\porosityB}\right)\right].
\end{align}
The initial porosity and permeability is $\porosityB = 0.5$ and $\initmatpermeabpscalar = 10^{-2} m^2$, respectively. The backward Euler scheme is applied for time discretization,
with an appropriate timestep length of $\Delta t = 0.1s, 0.01s \,\text{or}\, 0.001s$, depending on the system dynamics.

In Figure \ref{fig:ex2_load} (right) the spatial discretization (consisting of one layer of 3D-trilinear hexahedral elements) is visualized by the black lines, 
the solid and poroelastic elements ``cut out'' the non-physical part of the non-matching fluid elements.
Furthermore, the overall computed solution for $t=0.2s$ is visualized, where fluid outflow on $\Gamma^{F,N}$ occurs due to the structural induced pressure gradient.

\begin{figure}[tp]
\centering
\input{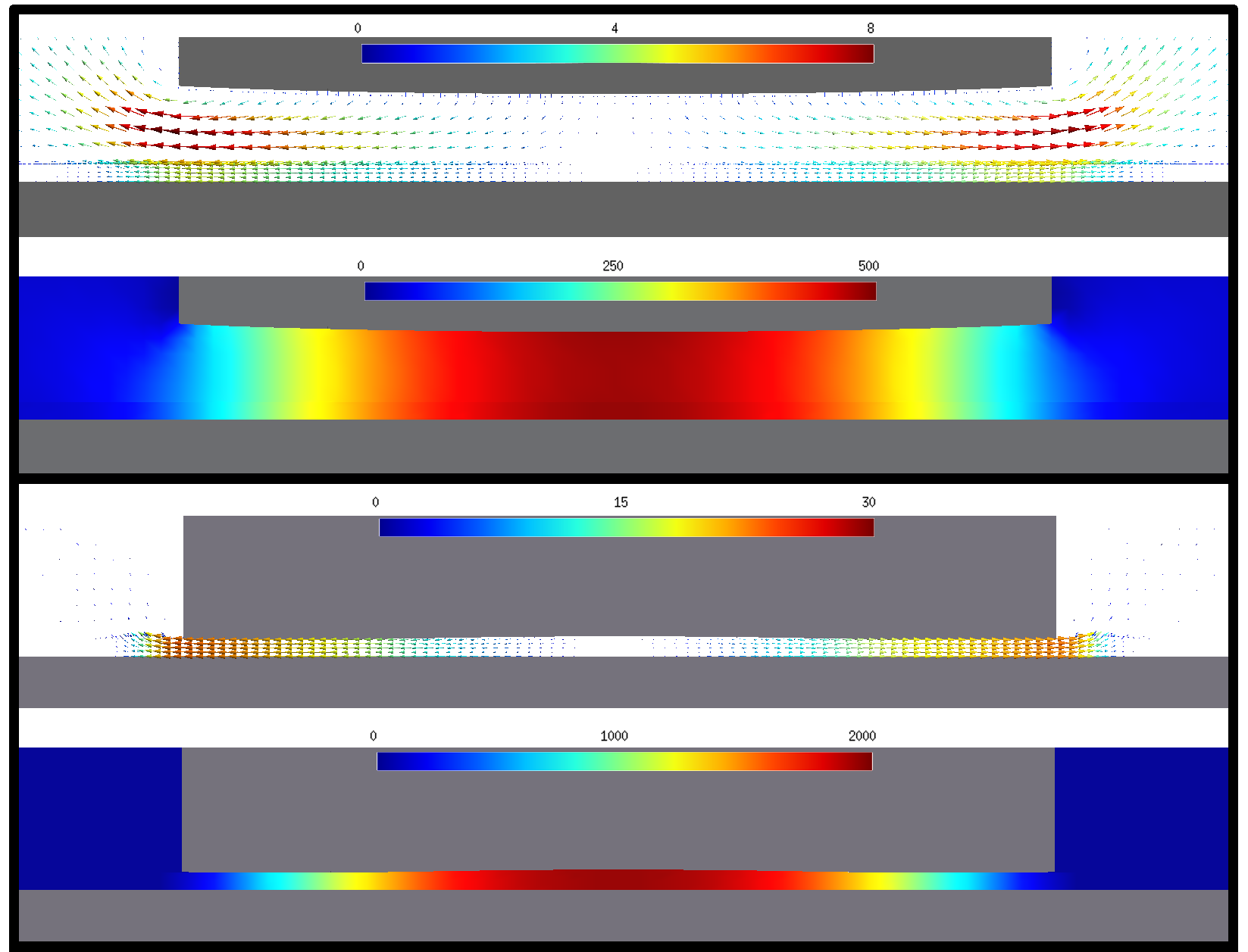}
\caption{Fluid pressure and velocity (arrows) and displacement (deformed domain) solution in contacting phase at time $t=0.2$ (top) and $t=0.71$ (bottom).}
\label{fig:ex2_full1}
\centering
\input{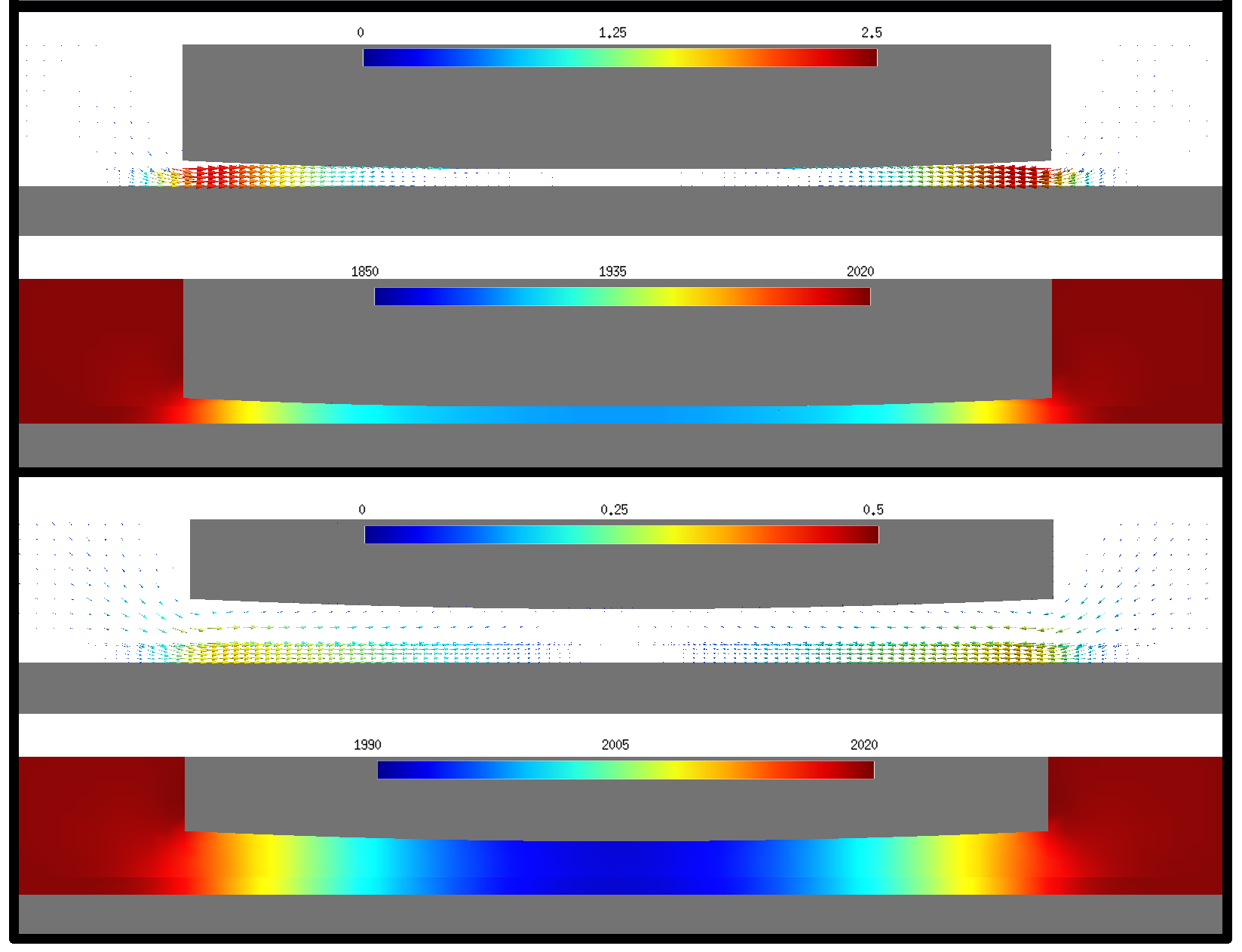}
\caption{Fluid pressure and velocity (arrows) and displacement (deformed domain) solution in lift-off phase at time $t=4.005$ (top) and $t=19.945$ (bottom).}
\label{fig:ex2_full2}
\end{figure}

An examination of the contacting phase at two different points in time is provided in Figure \ref{fig:ex2_full1}, which shows a detailed view of the contacting zone. 
It should be pointed out that in all following figures, the velocity amplitude is specified by the arrow length and the color code but not by the density of arrows.
The zone with an increased density of arrows indicates the poroelastic layer, arising from the finer spatial resolution compared to the fluid domain.
At $t=0.2s$, the motion of the stamp due to structural stress on $\Gamma^{S,N}$ displaces the fluid from the contacting zone.
As the gap between stamp and foundation is still larger than the roughness height, the significant part of the fluid mass transport happens in the free fluid domain $\domainf$.
This is a result of the similar pressure gradient in the roughness layer and the free fluid. At $t=0.71s$, there is already partial contact between the solid stamp and the poroelastic layer.
The fluid pressure increased significantly, as the fluid mass transport happens mainly in the poroelastic roughness layer with higher flow resistance than pure fluid in $\domainf$.
Compared to the velocities at $t=0.2s$, the maximum value of the fluid velocity increased due to the higher pressure gradient.
Until $t=1.0s$, a compression of the poroelastic roughness, which leads to a reduction of the porosity and an outflow of fluid mass, can be observed, due to the increasing solid stress.
The increase of fluid stress, starting at $t=2.0s$, causes the inverse behavior with an inflow into the contacting area.
Details of this lift-off phase can be seen in Figure \ref{fig:ex2_full2}.
At $t=4.005s$, the prescribed fluid stress reached its maximum.
Due to the flow resistance in the rough surface layer, the pressure in the contacting zone is still significantly lower than the prescribed fluid stress.
As the lift-off occurs from outside to inside, most of the fluid flow occurs on both outer regions.
This leads to a very small pressure gradient and negligible flow in the center of the contacting region.
Advancing in time, the pressure in the center increases until the overall fluid force on the stamp exceeds the prescribed value of the solid force, and the solid bodies detach completely.
In $t=19.945$, both bodies moved apart from each other, with a distance greater than the roughness layer height.

\begin{figure}[htpb]
\centering
\subfigure{\includegraphics[width=0.49\textwidth]{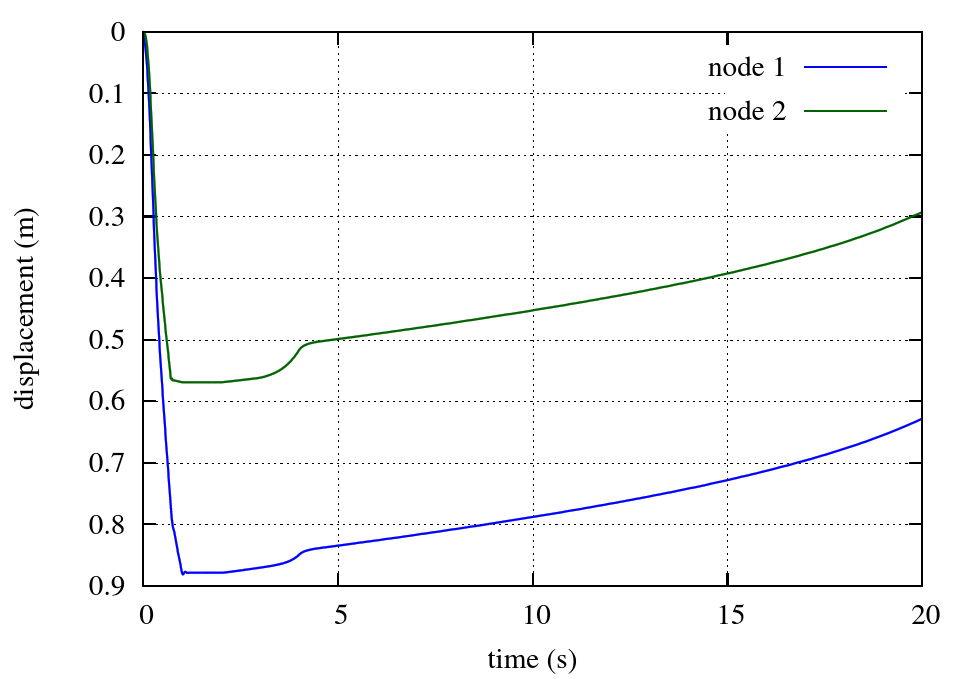}}
\subfigure{\includegraphics[width=0.49\textwidth]{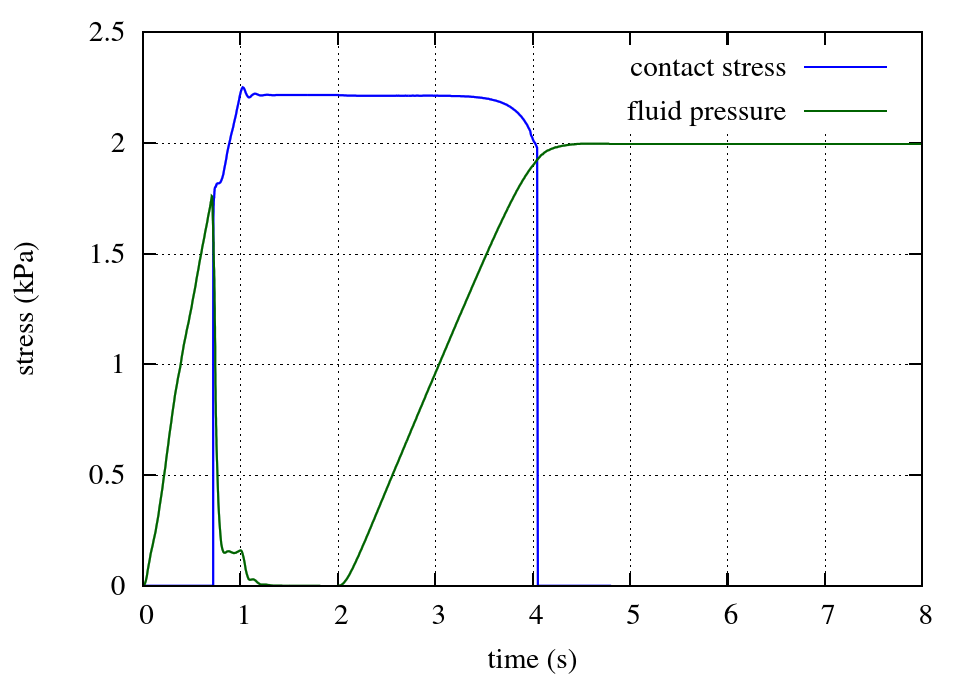}}
\caption{Left: Displacement in vertical direction of two selected computational nodes (see Figure \ref{fig:ex2_load} (right) for the selected nodes), 
Right: contact stress and fluid pressure at selected node $3$ (see Figure \ref{fig:ex2_load}).}
\label{fig:ex2_dispstress}
\end{figure}
\begin{figure}[htpb]
\centering
\includegraphics[width=0.9\textwidth]{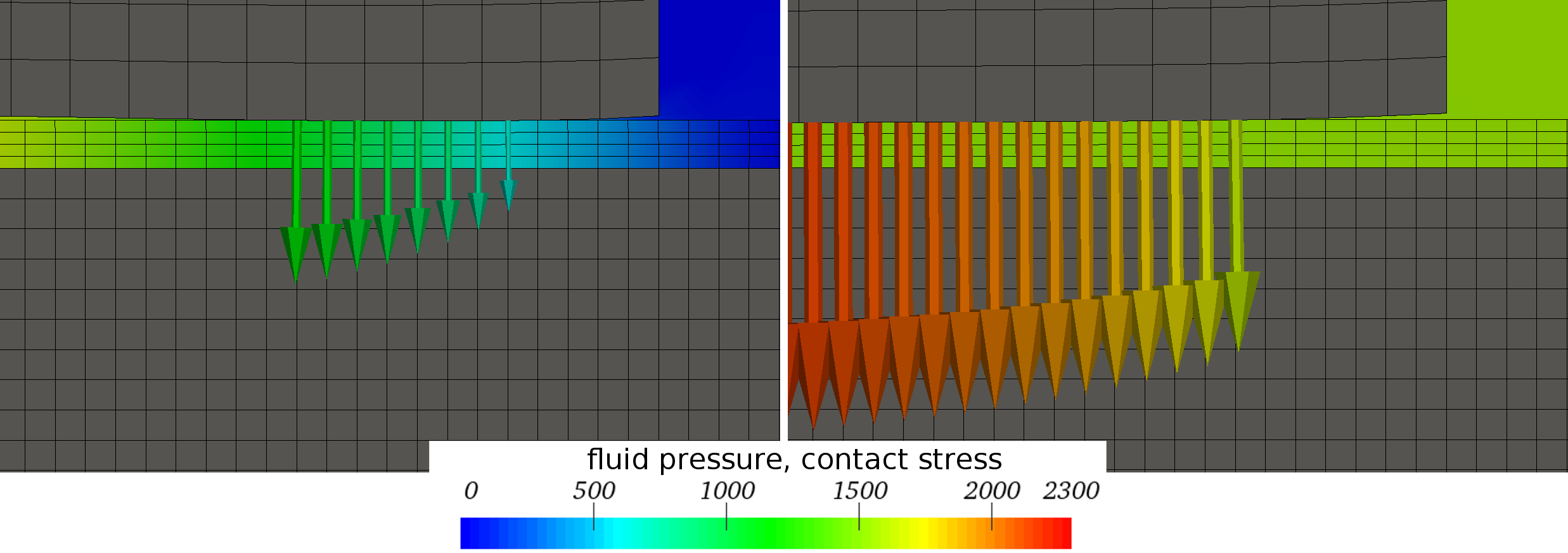}
\caption{Visualization for the right half of the fluid pressure (background color) and the contact stress (arrows) for two points in time (left) $t = 0.702s$, (right) $t=3.515s$.}
\label{fig:ex2_activeinactive}
\end{figure}

In Figure \ref{fig:ex2_dispstress} (left), the computed displacements for two selected nodes, at the left of the stamp, are plotted.
Due to the solid stress, both nodes moved down in vertical direction, until node $2$ contacts with solid body $\domain^{S_2}$ and there is just a small motion due to the compression of the poroelastic layer.
It can be observed that node $1$ moved for a longer period in time and reaches a larger displacement maximum, due to the elastic deformation of the stamp.
As there is an increase in fluid pressure for $2.0s\leq t \leq 4.0s$, both points start moving upwards again.
After the velocity in both nodes increases due to the local deformation in this area, a smaller velocity which is similar in both points can be observed.
Finally, the rising distance between both bodies leads to an increase of the fluid force on the structural bodies and therefore an acceleration of the stamp.

Figure \ref{fig:ex2_dispstress} (right) shows a comparison of the fluid pressure and the contact stress at the selected computational node $3$.
Starting from zero, the fluid pressure increases linearly due to the linearly rising prescribed solid stress on $\Gamma^{S,N}$.
As soon as contact occurs at this local point, the contact stress raises from zero to the actual level of the fluid pressure and increases linearly to the maximum stress in this computational node.
The inherent reduction of impact velocity and, therefore, the reduced fluid mass displacement results in a significant reduction of fluid pressure.
Until $t=1.0s$ a smaller pressure is still present due to the increasing solid stress. 
Starting from $t=2.0s$, the fluid stress is increased, which leads to a local lift-off of the outer parts of the stamp, which also reduces the contact stress in the point.
As soon as the fluid pressure reaches the contact stress, this point gives up the contact constraint and the entire stress between the solid bodies is exchanged via the fluid.

In Figure \ref{fig:ex2_activeinactive}, the area of the ``active'' contact constraint is visualized.
Nodes of the interfaces, where the contact stress arrows are visible, are contained in the set of active Lagrange multipliers $\lagmultpss^{\mathcal{A}}$.
It can be seen that on the borders of this area, the value of fluid pressure and contact stress are very close.
For $t=0.702$, which is during the contacting phase, one further aspect should be mentioned:
It can seen that at this point in time there is no contact in the center of the stamp. 
This results from the elastic deformation of the stamp, due to the maximum pressure in the center.
This fluid island vanishes later by fluid mass flow through the poroelastic roughness layer.

\subsection{Non-return valve}
\begin{figure}[tp]
\centering
\input{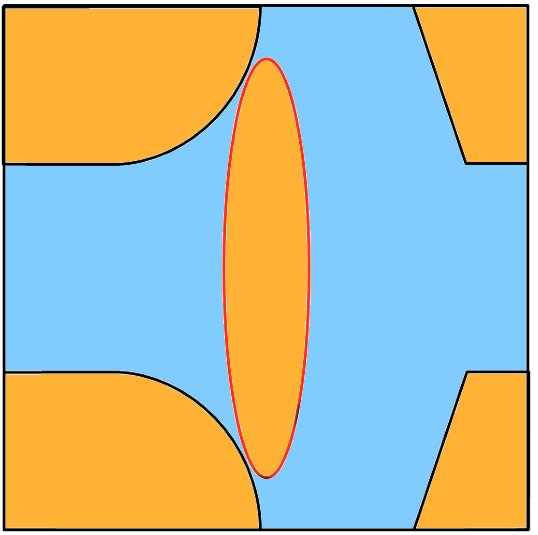}
\caption{Boundary conditions and geometry of the non-return valve.}
\label{fig:ex3}
\end{figure}

In the third example, a non-return valve is considered. 
The elementary valve that is analyzed blocks the flow in one direction, but enables the flow in the other direction.
All basic boundary conditions and the geometry can be found in Figure \ref{fig:ex3}.
The valve consists of an ellipsoidal shaped membrane $\domainp \cup \Omega^{S_1}$ and a solid support $\Omega^{S_2}$.
Not occupied by these domains and filled by fluid is the domain $\domainf$.
In the following, its subdomains to the left and right of the membrane are referred to as inflow and outflow domain, respectively.
Interfaces, boundary conditions, and dimensions are only defined once, but are still appropriate for the upper and lower part as the valve is symmetric.
On the Neumann boundaries $\Gamma^{F,N_1}$ and $\Gamma^{F,N_2}$, the fluid traction is prescribed, which leads to a flow through the valve.
To depict the functional principle, the desired behavior for two different flow configurations is described.
In the case that the fluid pressure on $\Gamma^{F,N_1}$ is higher than on $\Gamma^{F,N_2}$ (``open direction''), the elastic membrane
deforms and increases the size of the smallest constriction between $\domainp$ and $\Omega^{S_2}$.
This reduces flow resistance through the valve for an increasing pressure difference between inflow and outflow.
If the fluid pressure on $\Gamma^{F,N_2}$ is higher than on $\Gamma^{F,N_1}$ (``blocking direction''), deformation of the elastic membrane reduces the size of the constriction and therefore increases flow resistance.
At a specific pressure difference, membrane $\Omega^{S_1} \cup \Omega^{P}$ and support $\Omega^{S_2}$ come into contact, and the entire leakage flow has to pass through the rough layer.
This flow rate is intended to be small compared to the flow rate in ``open direction''.

As fluid, water with a density $\densityf = 1000 kg/m^3$ and a dynamic viscosity $\viscf = 10^{-3} Pa \cdot s$ is considered.
A Neo-Hookean material (see equation \eqref{ex1:strainenergys}) is applied to model the solid behavior with
 Young's modulus and Poisson ratio of $E^1=2.3 MPa, \nu^1=0.49$ and $E^2=2.3MPa, \nu^2=0.3$ in $\Omega^{S_1}$ and $\Omega^{S_2}$, respectively.
The initial density is equal to the fluid density $\densitys=\densityf=1000 kg/m^3$.
The poroelastic domain $\domainp$ is specified by a spatially constant initial porosity $\porosityB = \porosity(t=0) = 0.5$ and a spatially constant initial material 
permeability $\initmatpermeabpscalar=4.6\cdot 10^{-5}mm^2$.
The macroscopic material response of the poroelastic layer is given by the strain energy functions \eqref{ex1:strainenergysp} and \eqref{ex1:addstrainenergysp} with parameters
$E^P = 0.25MPa$, $\nu^P = 0.0$, $\tilde{c}^P = 0.25MPa$, $\alpha^P = 8, \kappa^P = 0.8MPa$, and $\eta^P = 1kPa$.
For discretization in time, the backward Euler scheme is applied with timestep lengths of $\Delta t=2\cdot 10^{-4}s,$ $5\cdot10^{-5}s$, $2.5\cdot 10^{-5}s$, or $1.25\cdot 10^{-5}s$, depending on 
the dynamic response of the system.

\begin{figure}[tp]
\centering
\includegraphics[width=0.49\textwidth]{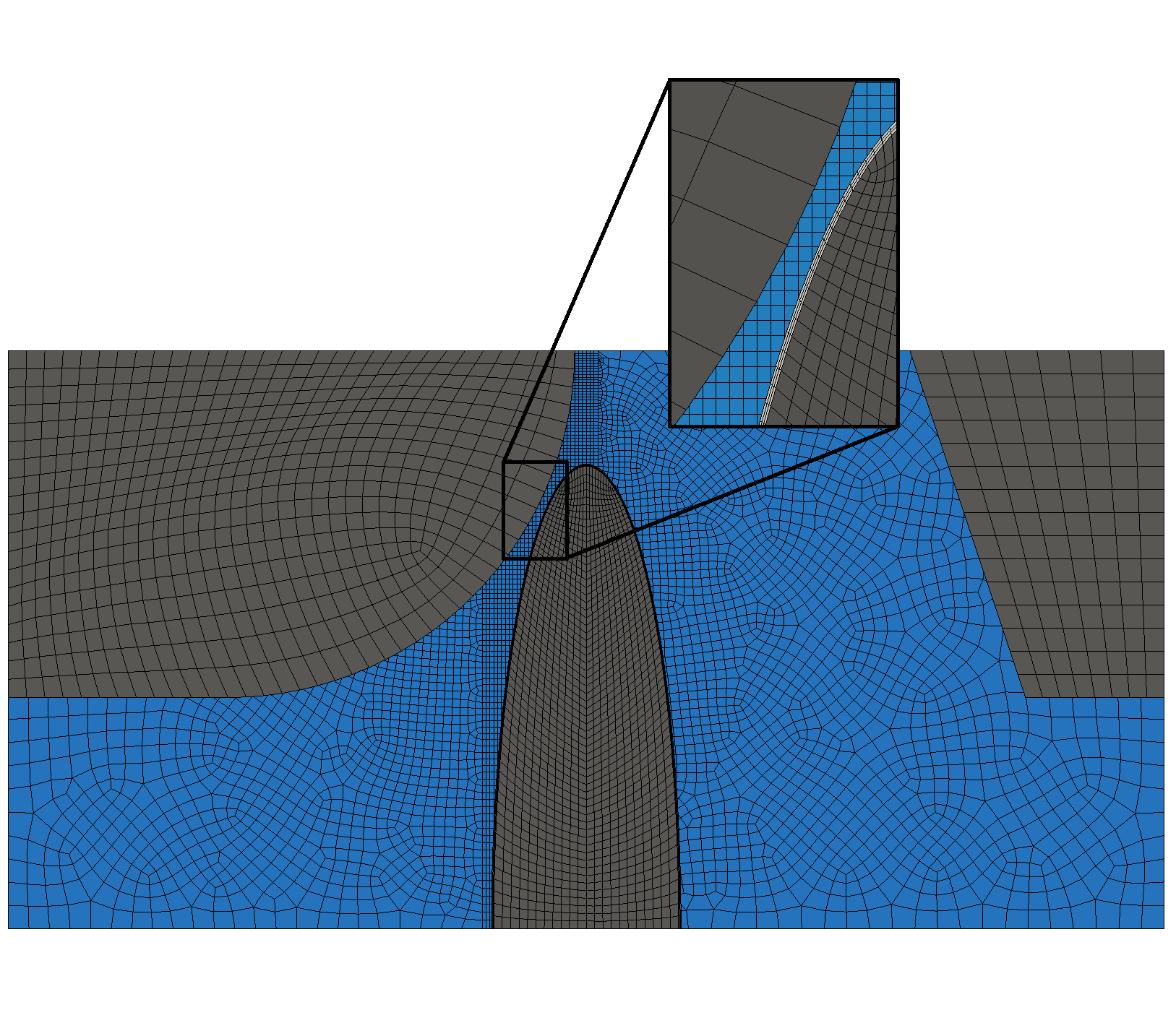}
\includegraphics[width=0.49\textwidth]{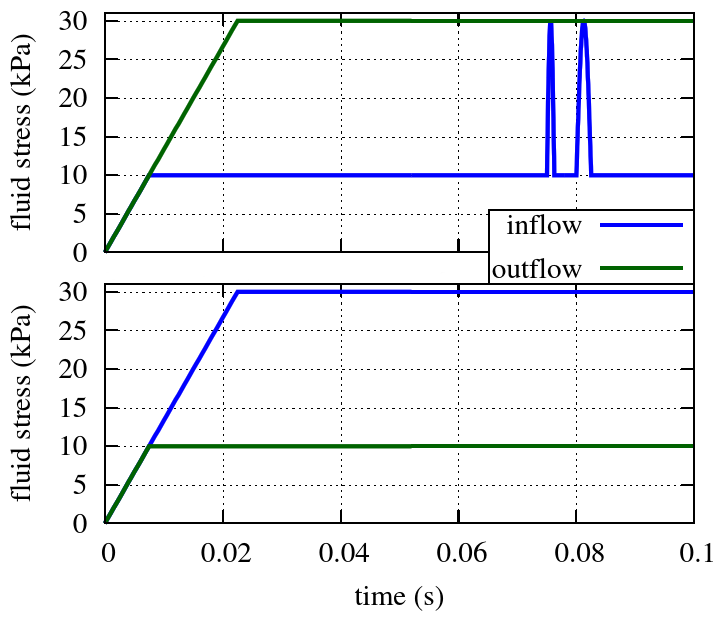}
\caption{Computational mesh for all physical fields and detailed view of the smallest constriction. Hexahedral elements are indicated by black lines (left).
Prescribed time-dependent Neumann fluid stress on inflow boundary $\Gamma^{F,N_1}$ and outflow boundary $\Gamma^{F,N_2}$. Upper diagram shows the applied load case to analyze a dynamic ``valve closing'' process.
The lower diagram shows the prescribed load to analyze the valve behavior in ``open direction'' (right).
}
\label{fig:ex3_discret_loads}
\end{figure}

\begin{figure}[htbp]
\centering
\includegraphics[width=1.0\textwidth]{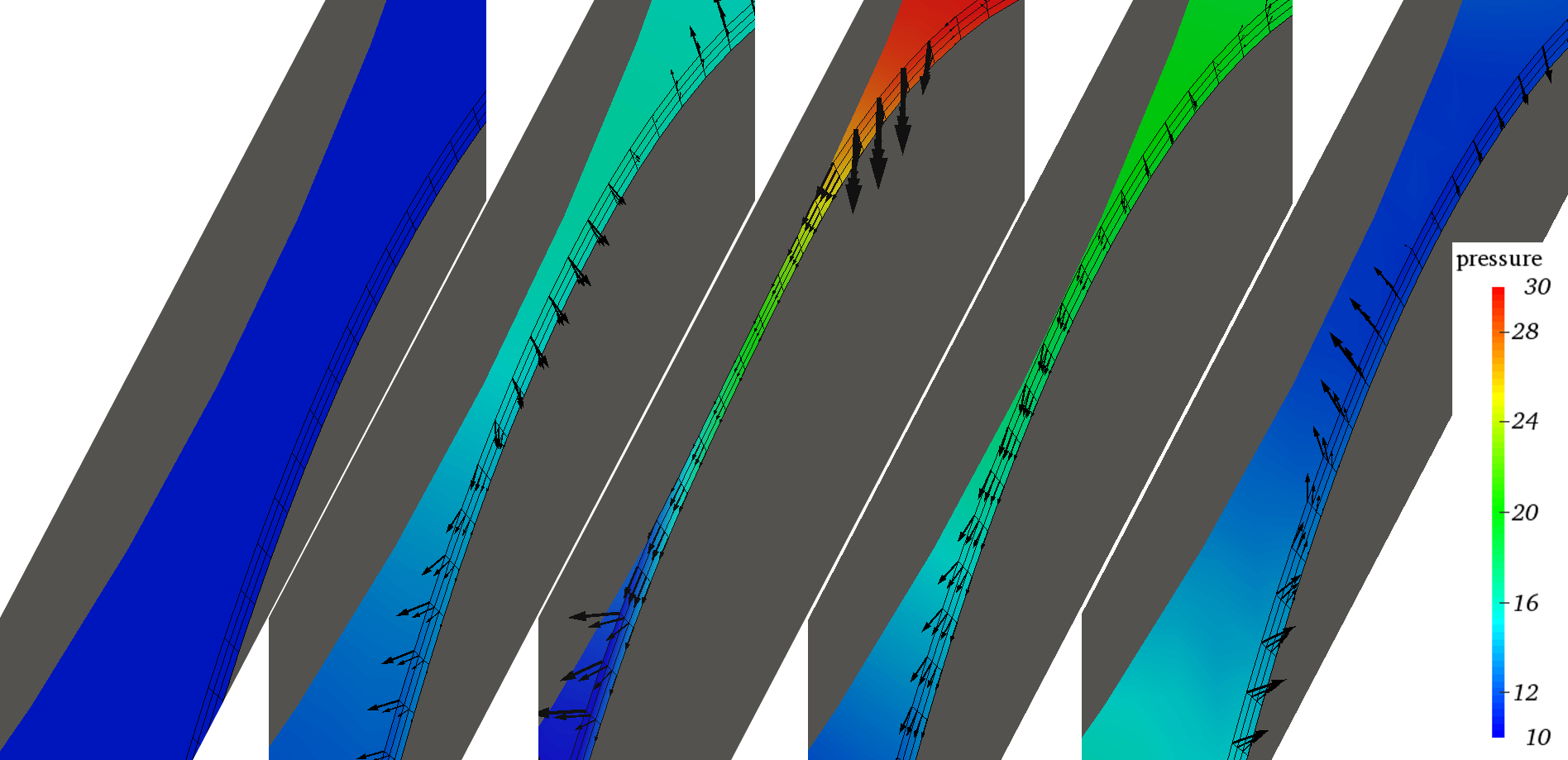}
\caption{``Valve closing'': Detailed view of the smallest constriction. 
Fluid pressure and convective velocity (black arrows) $\left(\velp-\velps\right)$ in the poroelastic domain during dynamic valve closing process for five instances in time. The black lines indicate the computational mesh of the poroelastic layer.
From left to right $t=0.008s$, $t=0.012s$, $t=0.07s$, $t=0.077s$, and $t=0.0827s$.}
\label{fig:ex3_detail_closing}
\end{figure}

The spatial discretization of all physical domains, consisting of one layer of 3D-trilinear hexahedral elements, is visualized in Figure \ref{fig:ex3_discret_loads} (left). 
The poroelastic layer $\domainp$ with height $\delta$ is discretized by three layers of elements
, which can be recognized in detail in Figure \ref{fig:ex3_detail_closing}.
Due to the symmetric configuration, only the half domains are discretized and consulted for the computations. On the arising symmetry boundary, the flow in normal direction is prohibited.
In Figure \ref{fig:ex3_discret_loads} (right), the prescribed fluid stress for two different cases is plotted.
The first case is designed to analyze the dynamic ``valve closing'' behavior.
Additionally it includes two sinusoidal-shaped peaks to determine the dynamic reaction of the valve on spurious pressure variations.
The load peaks are initiated at $t = 0.075s$ and $t=0.08s$ for a time span of $\Delta t_1 = 0.00125s$ and $\Delta t_1 = 0.0025s$, respectively.
Case two analyzes the flow in ``open direction'' for constant fluid stress difference between inlet and outlet boundary.

\begin{figure}[htbp]
\centering
\includegraphics[width=0.49\textwidth]{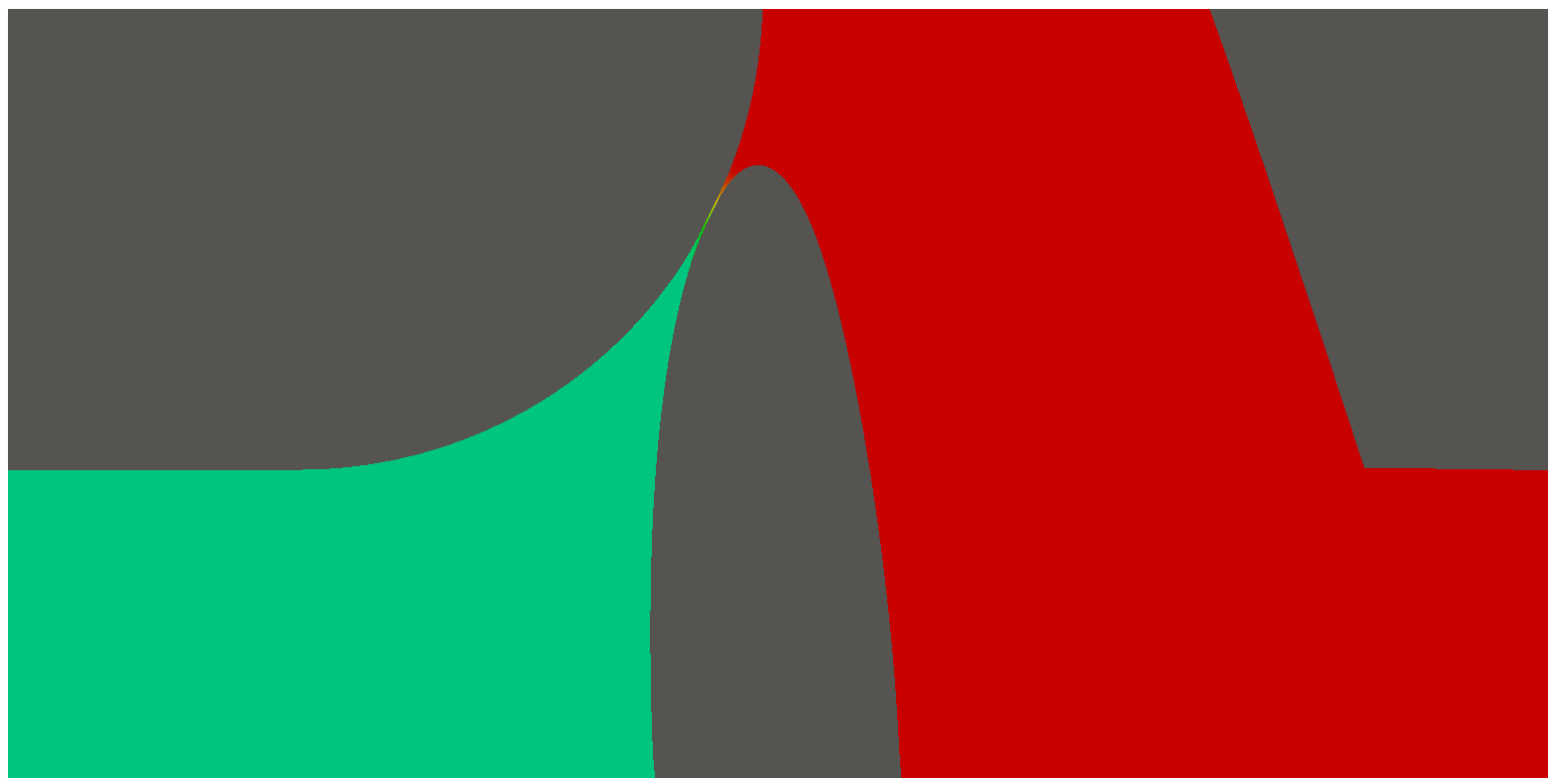}
\includegraphics[width=0.49\textwidth]{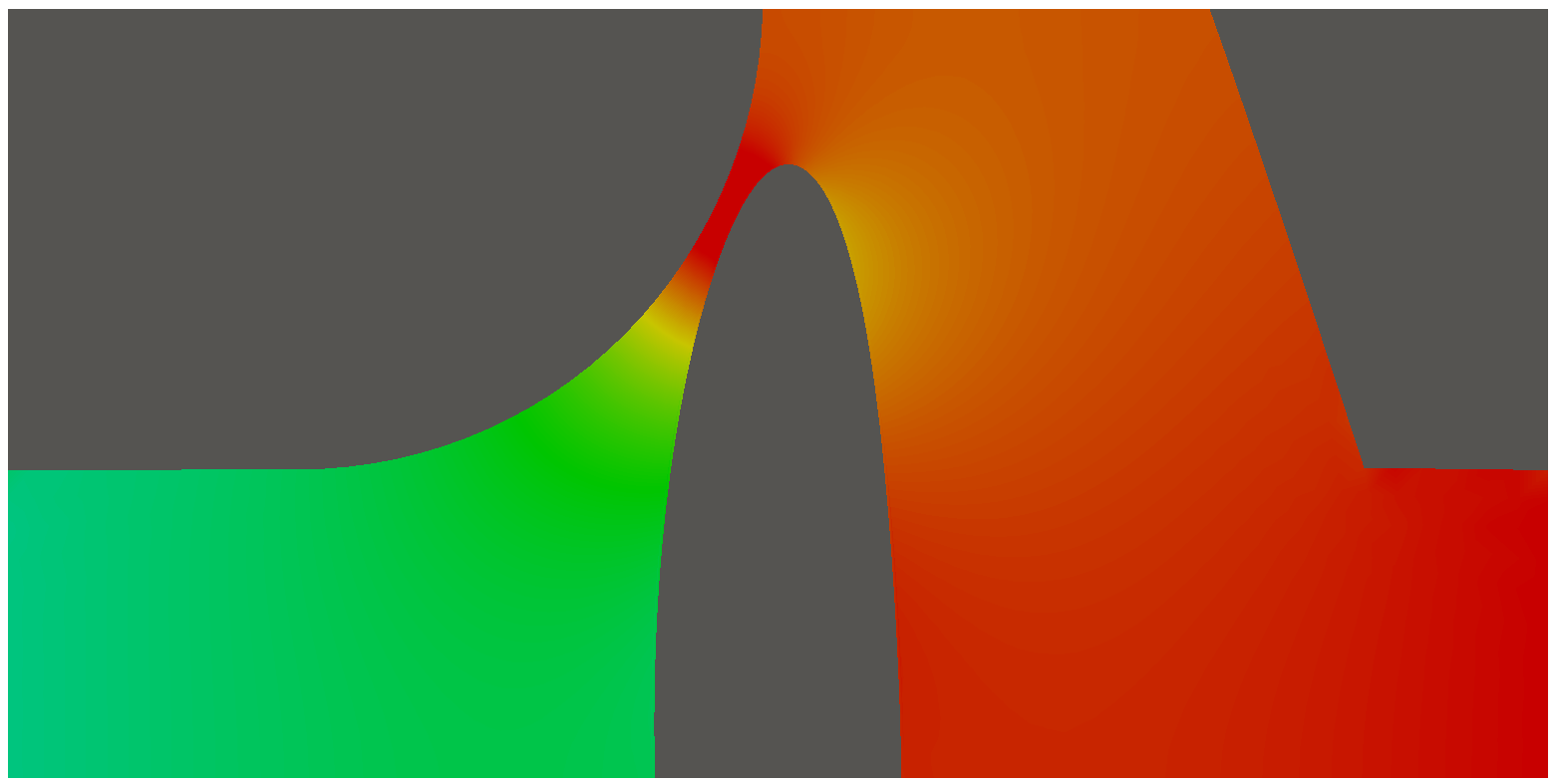}
\includegraphics[width=0.4\textwidth]{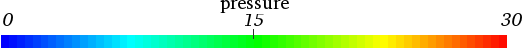}
\caption{``Valve closing'': Computed fluid pressure and solid deformation, left $t=0.07s$, right $t=0.0836s$.}
\label{fig:ex3_full_closing}
\end{figure}
\begin{figure}[htbp]
\centering
\includegraphics[width=0.49\textwidth]{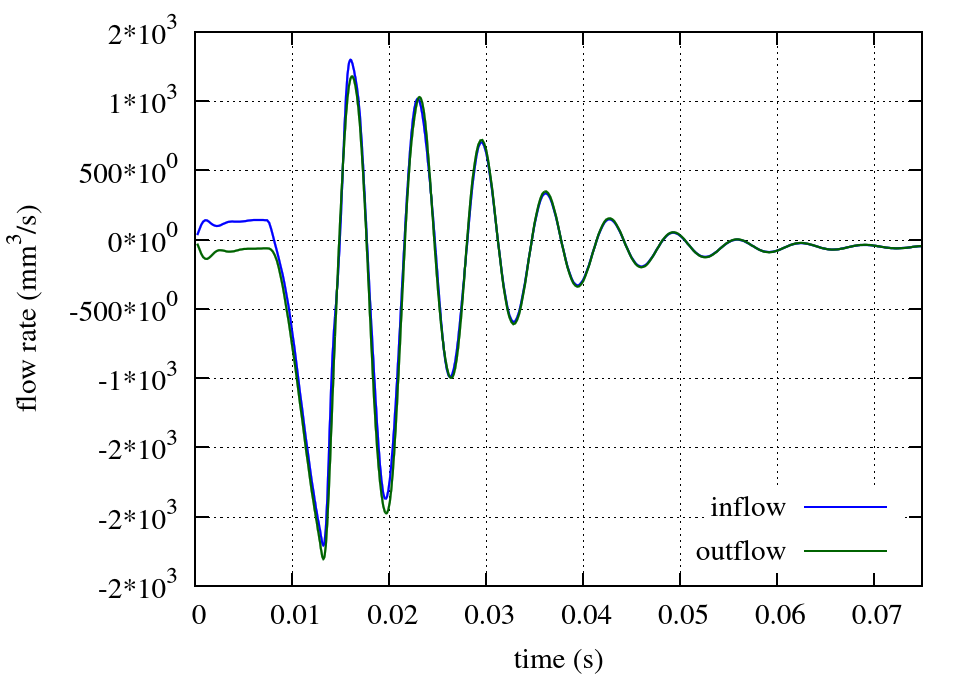}
\includegraphics[width=0.49\textwidth]{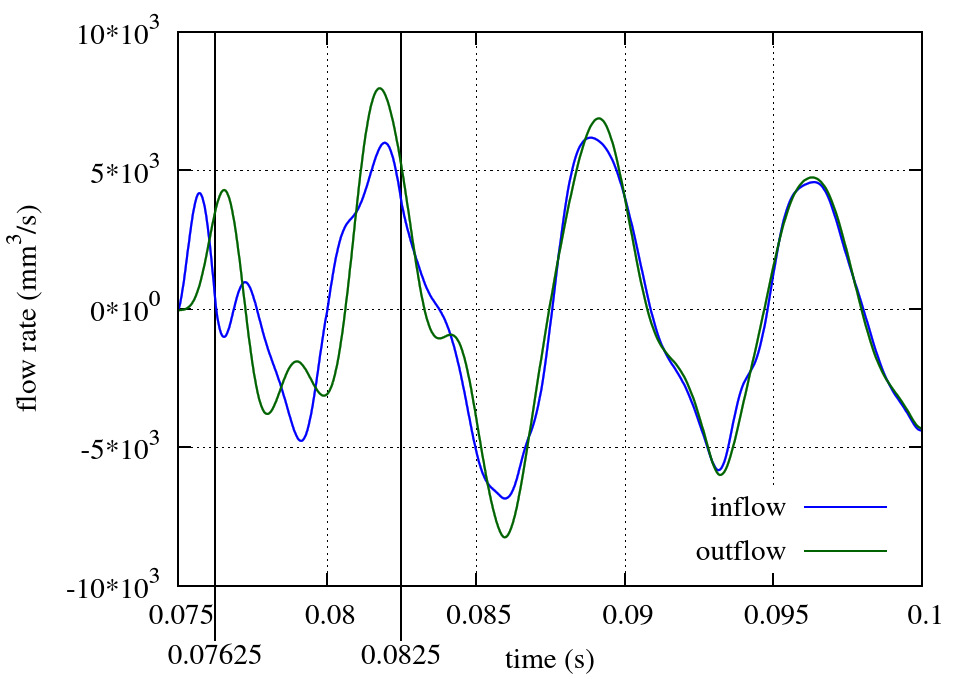}
\caption{``Valve closing'': Flow rate per unit depth on the inflow boundary $\Gamma^{F,N_1}$ and the outflow boundary $\Gamma^{F,N_2}$.}
\label{fig:ex3_flowrate_closing}
\end{figure}

First the computed results for the case ``valve closing'' are presented.
Figure \ref{fig:ex3_full_closing} shows the pressure field and deformation of the valve for two instances in time.
At $t=0.07s$, an approximately stationary solution for the flow through the closed valve can be observed. 
Contact occurs between the membrane and the support.
The overall leakage flow through the valve is small.
Thus, pressure in the inflow and outflow domain is almost constant and corresponds approximately to the applied boundary conditions.
The pressure drop occurs mainly in the poroelastic layer.
A lifting of the membrane from the support and therefore a temporal opening of the valve at $t=0.0836s$ occurs, due to the second applied sinusoidal-shaped fluid stress peak at the inflow boundary.
Here, contrary to the stationary case ($t=0.07s$), pressure gradients in the flow domain can be observed. This occurs from the fluid motion induced by deformation of membrane and support.

Figure \ref{fig:ex3_detail_closing} shows a detailed view onto the smallest constriction between membrane and support.
Due to the simultaneous increase of fluid stress (up to a value of $10 kPa$) on inflow and outflow boundary, mainly pure compression of all solid domains occurs, which leads to negligible fluid velocities (see $t=0.008s$).
At $t=0.012s$, the membrane moves towards the solid support induced by the fluid pressure.
This leads to a fluid flow through the roughness layer away from the narrowest position of the fluid domain.
For the approximately stationary solution at $t=0.07s$, a flow from the high pressure outflow domain through the poroelastic layer into the low pressure inflow domain can be observed.
This flow corresponds to the leak rate of the investigated valve.
The first fluid stress peak ($\Delta t_1 = 0.00125s$) leads to a loss of contact. 
At $t=0.077s$, the maximum distance between membrane and support solid for this first peak occurs.
Here, an increase of fluid velocity through the expanded poroelastic layer, due to the vanishing contact stress, as compared to the stationary situation can be observed.
The second load peak leads to a pronounced lift-off of the membrane. 
During the lift-off phase at $t=0.0827s$, fluid mass enters the smallest constriction.
This process is portioned between the pure fluid domain and the poroelastic domain.

To quantify the performance of the non-return valve, the flow rates through inflow $\Gamma^{N,1}$ and outflow $\Gamma^{N,2}$ boundary are computed.
In \eqref{eq:flowrate}, a unique normal vector $\normal^{\Gamma^{N,1}->\Gamma^{N,2}}$, pointing in designated flow direction of the ``open'' valve, is applied for both boundaries.
\begin{align}
\label{eq:flowrate}
\text{flow rate} = \int_{\Gamma^{N,i}}{\velf \cdot \normal^{\Gamma^{N,1}->\Gamma^{N,2}}}\text{d}\Gamma^{N,i}, \qquad i = 1,2
\end{align}
Figure \ref{fig:ex3_flowrate_closing} (left) shows the computed flow rates during the dynamic closing process. 
Until $t=0.075s$, compression of the solid bodies, due to the pressure increase, leads to inflow on both boundaries.
Due to the dynamic impact of the membrane on the solid support, recurring deformation of the membrane is initiated.
This leads to a change in volume in the inflow and outflow domain and therefore to decaying oscillations of both measured flow rates.
The average stationary flow/leak rate computed between $t=0.066s$ and $t=0.0724s$ is $52.4 mm^3/s$.
Figure \ref{fig:ex3_flowrate_closing} (right) shows the flow rates during and after the prescribed load peaks.
Although the first stress peak does not lead to a significant lift-off of the membrane, its deformation and the associated change of volume in the inflow and outflow domain
leads to high flow rates on both boundaries.
Finally, due to the longer excitation phase, the second fluid load peak leads to more pronounced flow rate peaks, which are decaying in time.

\begin{figure}[tp]
\centering
\includegraphics[width=0.4\textwidth]{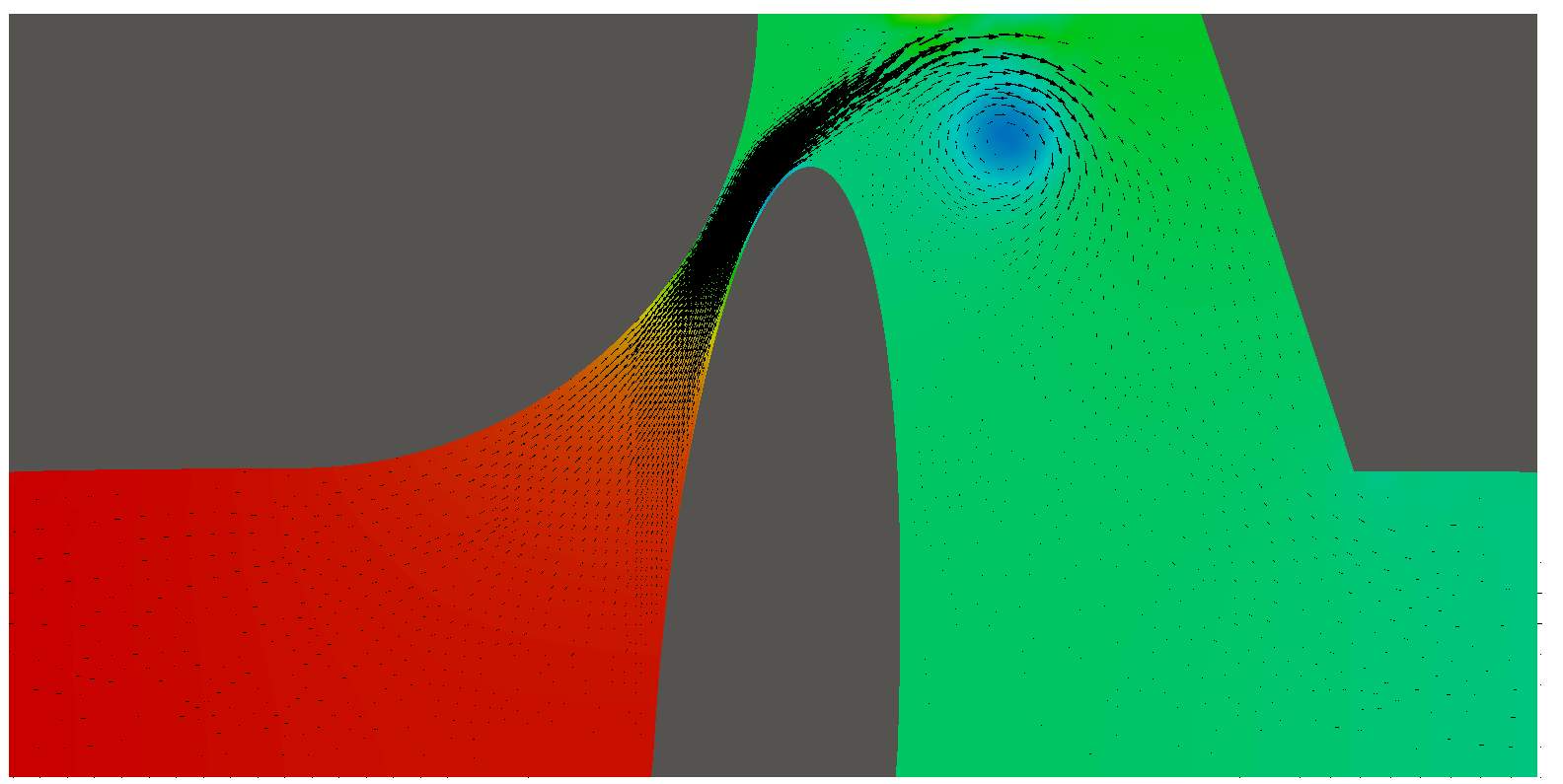}
\includegraphics[width=0.4\textwidth]{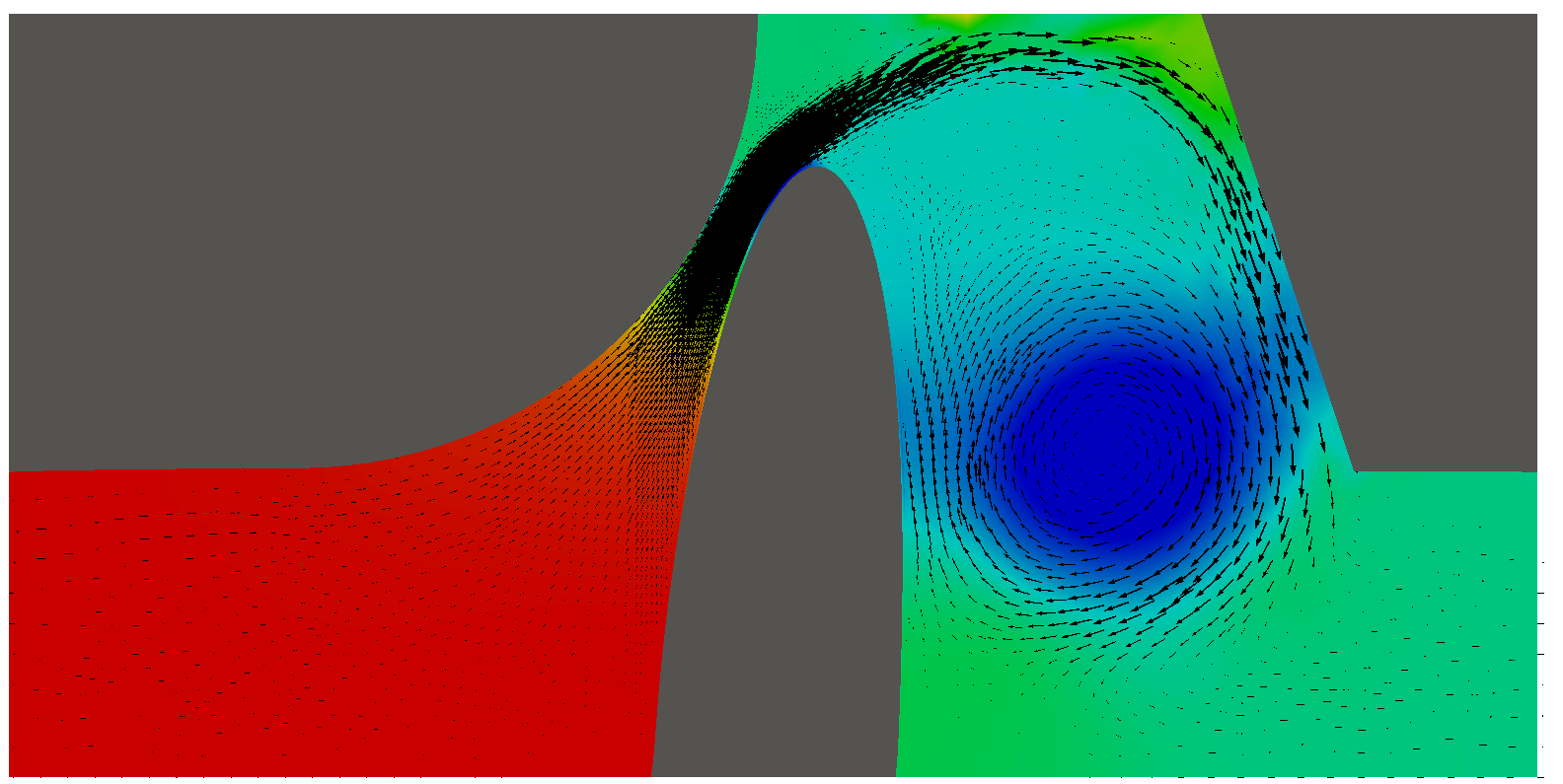}
\includegraphics[width=0.4\textwidth]{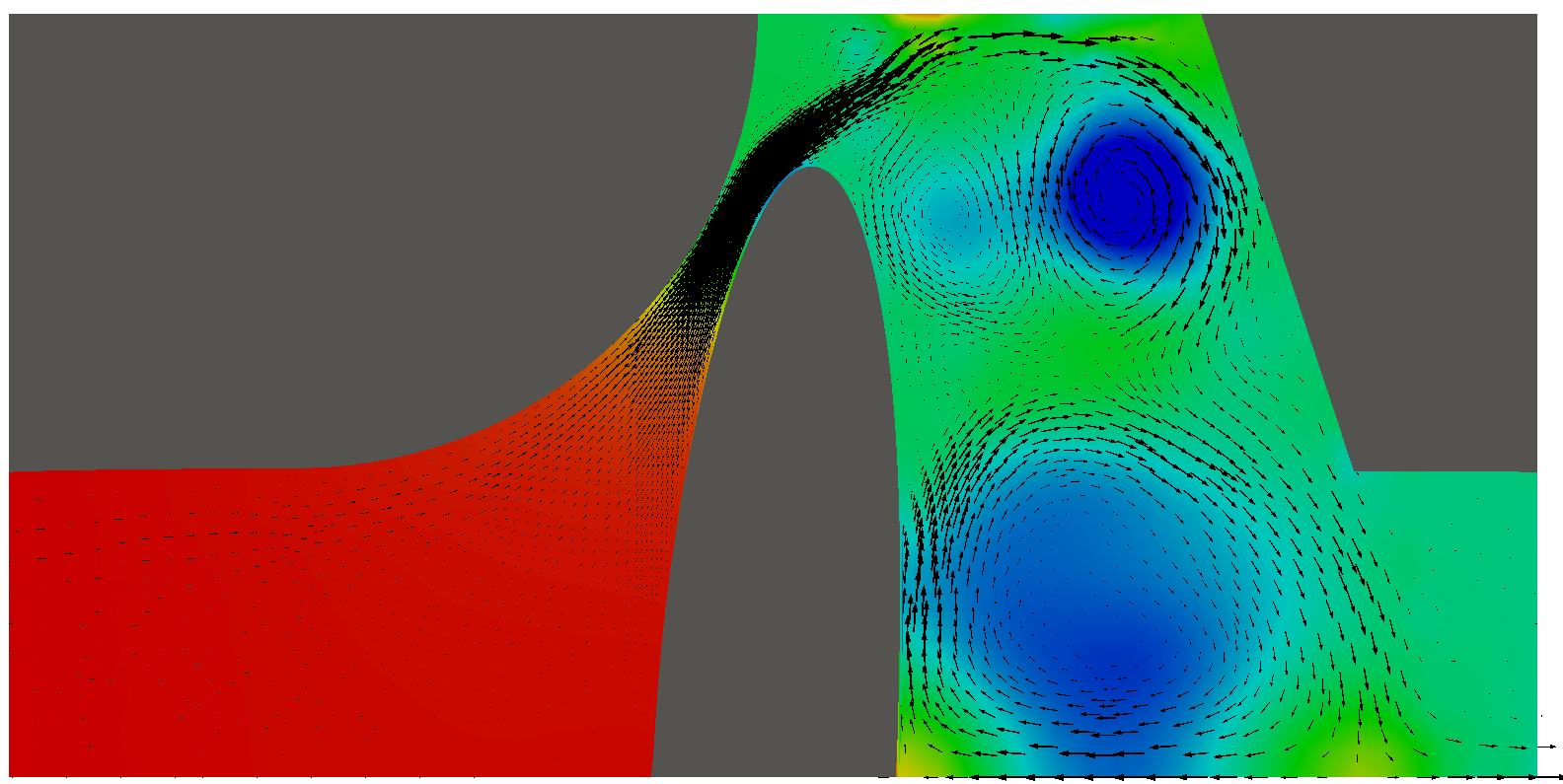}
\includegraphics[width=0.4\textwidth]{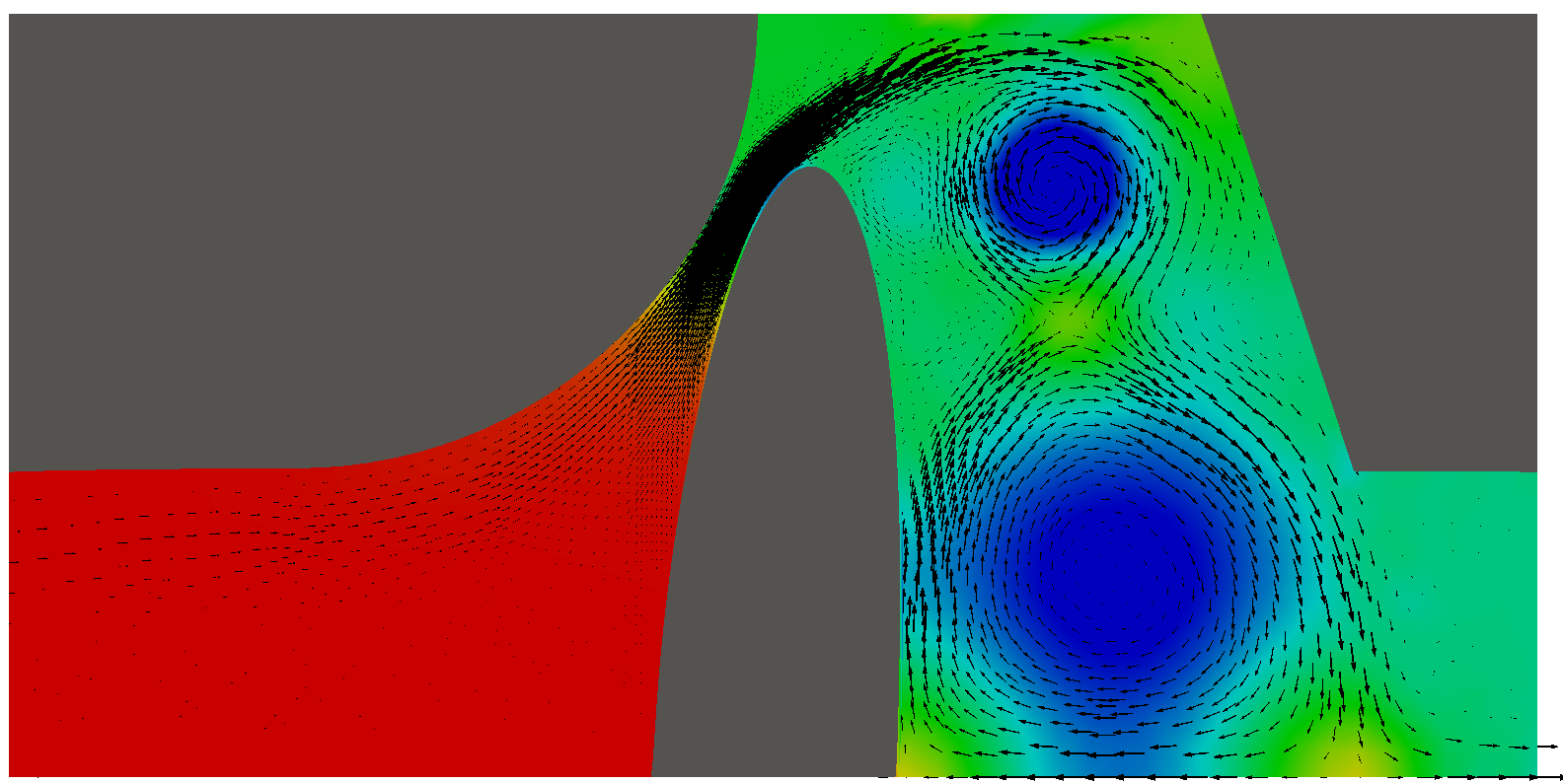}
\includegraphics[width=0.4\textwidth]{fig/ex3_open_legend.png}
\caption{``Open direction'': Fluid pressure and velocity (black arrows) at four different steps in time. From top-left to bottom-right $t_1=0.026s$, $t_2=0.035s$, $t_3=0.05$ and $t_4=0.09$.}
\label{fig:ex3_solution_open}
\end{figure}

Figure \ref{fig:ex3_solution_open} depicts the computed solution for the second case (``open direction'') at four different instances in time.
The present pressure difference deforms the membrane and increases the size of the smallest constriction.
At $t=0.026$, the evolution of a vortex behind the membrane can be observed.
The motion of this vortex towards the symmetry plane of the valve can be seen at $t=0.035s$.
For $t=0.05$ and $t=0.09s$, a complex non-stationary flow pattern can be observed.
It should be pointed out that assuming a symmetric flow field behind the membrane might influence the computed flow rates for this open valve configuration.

\begin{figure}[tp]
\centering
\includegraphics[width=0.4\textwidth]{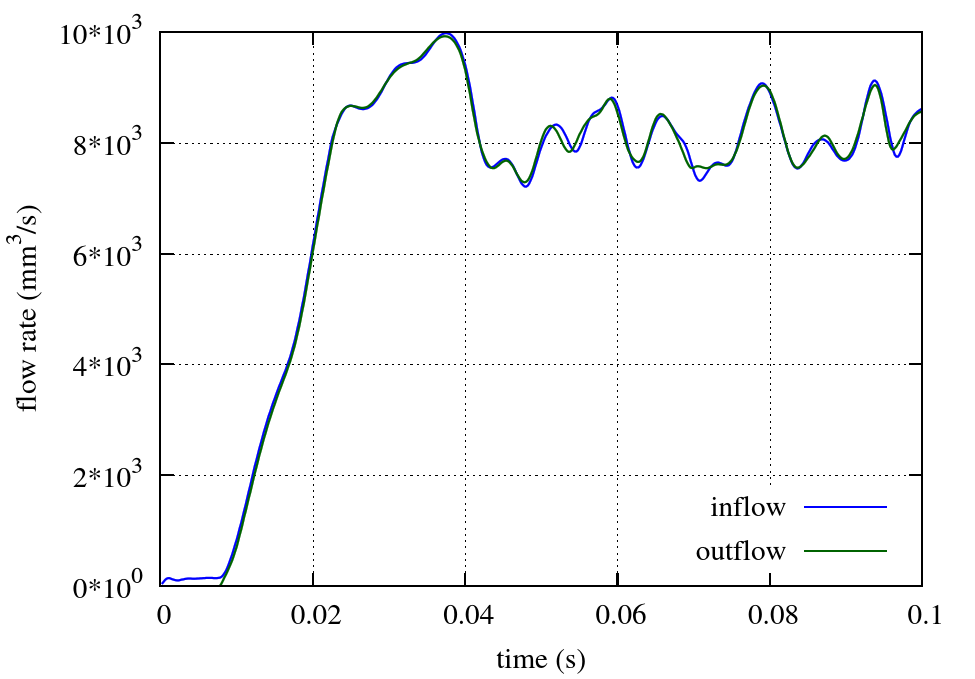}
\includegraphics[width=0.4\textwidth]{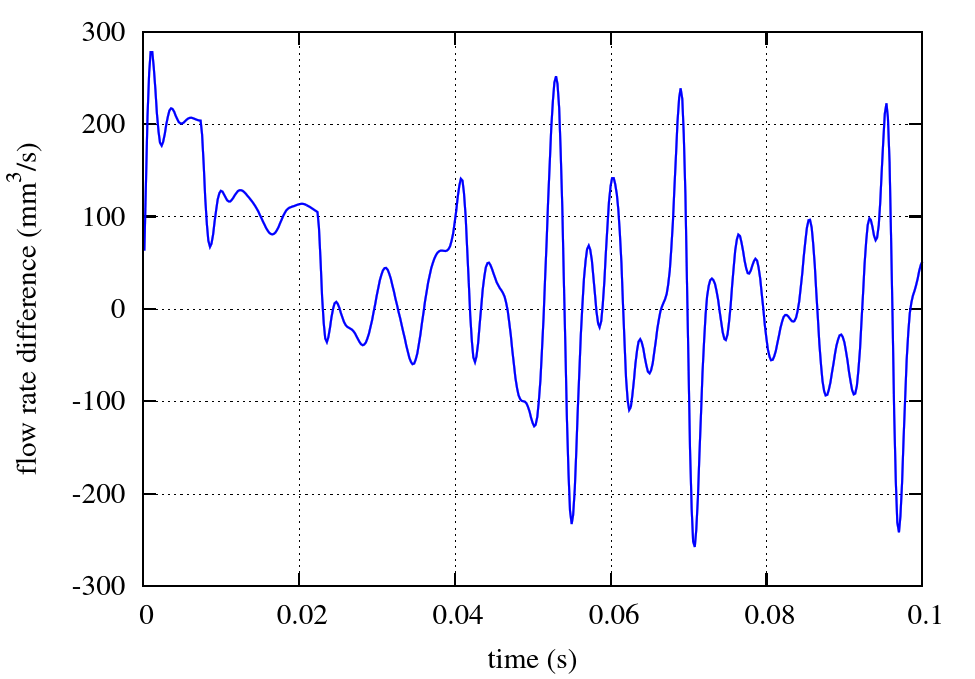}
\caption{``Open direction'': Flow rate per unit depth on the inflow boundary $\Gamma^{F,N_1}$ and the outflow boundary $\Gamma^{F,N_2}$ (left) and the flow rate difference between inflow and outflow (right).}
\label{fig:ex3_flowrat_open}
\end{figure}

In Figure \ref{fig:ex3_flowrat_open} (left), the computed flow rates on the inflow and outflow boundary are shown.
After $t=0.0075s$, the flow through the valve develops and similar flow rates for the inflow and outflow boundary can be observed. 
Due to the previously described development of fluid vortices and its influence on the membrane deformation, no stationary flow rate can be observed.
The average flow rate, evaluated between $t=0.04s$ and $t=0.1s$, is $8087 mm^3/s$.
Figure \ref{fig:ex3_flowrat_open} (right) shows the difference between the inflow and outflow flow rate. This difference occurs due to the compression of the solid support and the membrane.
At the initial phase, pure inflow into the system occurs due to the pressure increase until $t=0.025 s$.
Later in time, oscillations with a vanishing mean value occur.
This corresponds to a sequence of compression and expansion phases of the solid domains due to the time-dependent fluid loads.

\section{Conclusion}
\label{sec:conclusion}
In this contribution, we presented the first numerically and physically consistent model for fluid-structure-contact interaction.
In order to obtain a consistent physical model it incorporates effects originating from the rough microstructure of potentially contacting surfaces into the fluid-structure-contact interaction (FSCI) framework. 
In our model, the domain saturated by fluid and including all solid asperities of the rough surfaces is homogenized and finally results in a poroelastic layer.
By coating potentially contacting elastic bodies with this poroelastic layer, the effects of rough surface contact submersed in a fluid can be considered without resolving the surface microstructure in the macroscopic problem setup.

The governing equations for the involved structural, fluid, and poroelastic domains as well as appropriate coupling conditions on the interfaces were depicted.
The numerical treatment of the coupled problem based on the FEM was presented, 
where the CutFEM with ``ghost penalty'' stabilization was applied to solve the incompressible Navier-Stokes equations in the physical fluid domain on a non-matching computational mesh 
allowing to consistently going to the limiting case of a zero gap without numerical tricks.
This enables the numerical method to allow for a smooth transition throughout topological changes in the fluid domain for contacting and non-contacting solid bodies.
Due to the non-matching of interfaces and computational fluid mesh, all interface conditions were enforced weakly by Nitsche-based methods.
Furthermore, the dual mortar method was applied to incorporate contact of elastic bodies into the rough FSCI model.

Finally, three different numerical examples were presented to show the capabilities of the presented model.
The first was a typical configuration to analyze the behavior of the rough FSCI model for lubrication flow in comparison with measured data.
The second configuration focused on the contacting and ``lift-off'' behavior of two colliding bodies as well as the ``squeeze-out'' flow behavior.
In the final numerical example, the dynamic analysis of a non-return valve in ``closing'' and ``opening'' flow direction was presented.

\section*{Acknowledgements}
\label{sec:acknowledgement}
The authors C.A.,~A.P.~and W.A.W.~gratefully acknowledge the support by the protject KonRAT: ``Komponenten von Raketentriebwerken f\"ur Anwendungen in
Transportsystemen der Luft- und Raumfahrt'', work package 2400 ``Fluid-structure-interaction (FSI)'' of the Ludwig B\"olkow Campus.

\bibliography{bib.bib}

\begin{thebibliography}{10}
\providecommand{\url}[1]{\texttt{#1}}
\providecommand{\urlprefix}{URL }
\expandafter\ifx\csname urlstyle\endcsname\relax
  \providecommand{\doi}[1]{doi:\discretionary{}{}{}#1}\else
  \providecommand{\doi}{doi:\discretionary{}{}{}\begingroup
  \urlstyle{rm}\Url}\fi

\bibitem{barber2006}
Barber RW, Emerson DR. Challenges in modeling gas-phase flow in microchannels:
  from slip to transition. \emph{Heat Transfer Engineering}  2006;
  \textbf{27}(4):3--12.

\bibitem{reynolds1886}
Reynolds O. On the theory of lubrication and its application to {M}r.
  {B}eauchamp {T}ower's experiments, including an experimental determination of
  the viscosity of olive oil. \emph{Philosophical Transactions of the Royal
  Society of London}  1886; \textbf{177}:157--234.

\bibitem{christensen1971}
Christensen H, Tonder K. The hydrodynamic lubrication of rough bearing surfaces
  of finite width. \emph{Journal of Lubrication Technology}  1971;
  \textbf{93}(3):324--329.

\bibitem{patir1978}
Patir N, Cheng H. An average flow model for determining effects of
  three-dimensional roughness on partial hydrodynamic lubrication.
  \emph{Journal of Lubrication Technology}  1978; \textbf{100}(1):12--17.

\bibitem{tripp1983}
Tripp J. Surface roughness effects in hydrodynamic lubrication: the flow factor
  method. \emph{Journal of Lubrication Technology}  1983;
  \textbf{105}(3):458--465.

\bibitem{bayada1989}
Bayada G, Chambat M. Homogenization of the {S}tokes system in a thin film flow
  with rapidly varying thickness. \emph{ESAIM: Mathematical Modelling and
  Numerical Analysis}  1989; \textbf{23}(2):205--234.

\bibitem{prat2002}
Prat M, Plourabou{\'e} F, Letalleur N. Averaged {R}eynolds equation for flows
  between rough surfaces in sliding motion. \emph{Transport in Porous Media}
  2002; \textbf{48}(3):291--313.

\bibitem{jai2002}
Jai M, Bou-Said B. A comparison of homogenization and averaging techniques for
  the treatment of roughness in slip-flow-modified {R}eynolds equation.
  \emph{Journal of Tribology}  2002; \textbf{124}(2):327--335.

\bibitem{bou2004}
Bou-Said B. Comparison of homogenization and direct techniques for the
  treatment of roughness in incompressible lubrication. \emph{Journal of
  Tribology}  2004; \textbf{126}:1--5.

\bibitem{yang2009}
Yang B, Laursen TA. A mortar-finite element approach to lubricated contact
  problems. \emph{Computer Methods in Applied Mechanics and Engineering}  2009;
  \textbf{198}(47-48):3656--3669.

\bibitem{budt2012}
Budt M, Temizer I, Wriggers P. A computational homogenization framework for
  soft elastohydrodynamic lubrication. \emph{Computational Mechanics}  2012;
  :1--19.

\bibitem{almqvist2004}
Almqvist T, Almqvist A, Larsson R. A comparison between computational fluid
  dynamic and {R}eynolds approaches for simulating transient ehl line contacts.
  \emph{Tribology International}  2004; \textbf{37}(1):61--69.

\bibitem{farhat2004}
Farhat C. Encyclopedia of computational mechanics. \emph{CFD-based Nonlinear
  Computational Aeroelasticity}  2004; \textbf{3}:459--480.

\bibitem{fernandez2011}
Fern{\'a}ndez MA. Coupling schemes for incompressible fluid-structure
  interaction: implicit, semi-implicit and explicit. \emph{SeMA Journal}  2011;
  \textbf{55}(1):59--108.

\bibitem{Gee2011}
Gee MW, K\"{u}ttler U, Wall WA. Truly monolithic algebraic multigrid for
  fluid-structure interaction. \emph{International Journal for Numerical
  Methods in Engineering}  2011; \textbf{85}(8):987--1016.

\bibitem{loon2006}
van Loon R, Anderson PD, van~de Vosse FN. A fluid--structure interaction method
  with solid-rigid contact for heart valve dynamics. \emph{Journal of
  Computational Physics}  2006; \textbf{217}(2):806--823.

\bibitem{santos2008}
Dos~Santos ND, Gerbeau JF, Bourgat JF. A partitioned fluid--structure algorithm
  for elastic thin valves with contact. \emph{Computer Methods in Applied
  Mechanics and Engineering}  2008; \textbf{197}(19):1750--1761.

\bibitem{astorino2009}
Astorino M, Gerbeau JF, Pantz O, Traor{\'e} KF. Fluid--structure interaction
  and multi-body contact: application to aortic valves. \emph{Computer Methods
  in Applied Mechanics and Engineering}  2009; \textbf{198}(45):3603--3612.

\bibitem{mayer20103}
Mayer UM, Popp A, Gerstenberger A, Wall WA. 3d fluid--structure-contact
  interaction based on a combined {XFEM} {FSI} and dual mortar contact
  approach. \emph{Computational Mechanics}  2010; \textbf{46}(1):53--67.

\bibitem{wick2014}
Wick T. Flapping and contact fsi computations with the fluid--solid
  interface-tracking/interface-capturing technique and mesh adaptivity.
  \emph{Computational Mechanics}  2014; \textbf{53}(1):29--43.

\bibitem{kamensky2015}
Kamensky D, Hsu MC, Schillinger D, Evans JA, Aggarwal A, Bazilevs Y, Sacks MS,
  Hughes TJ. {An immersogeometric variational framework for fluid--structure
  interaction: Application to bioprosthetic heart valves}. \emph{Computer
  Methods in Applied Mechanics and Engineering}  2015; \textbf{284}:1005--1053.

\bibitem{hillairet2009}
Hillairet M, Takahashi T. Collisions in three-dimensional fluid structure
  interaction problems. \emph{SIAM Journal on Mathematical Analysis}  2009;
  \textbf{40}(6):2451--2477.

\bibitem{gerard2015}
G{\'e}rard-Varet D, Hillairet M, Wang C. The influence of boundary conditions
  on the contact problem in a 3d {N}avier--{S}tokes flow. \emph{Journal de
  Math{\'e}matiques Pures et Appliqu{\'e}es}  2015; \textbf{103}(1):1--38.

\bibitem{Hocking1973}
Hocking LM. The effect of slip on the motion of a sphere close to a wall and of
  two adjacent spheres. \emph{Journal of Engineering Mathematics}  Jul 1973;
  \textbf{7}(3):207--221.

\bibitem{cawthorn2010}
Cawthorn C, Balmforth N. Contact in a viscous fluid. {P}art 1. {A} falling
  wedge. \emph{Journal of Fluid Mechanics}  2010; \textbf{646}:327--338.

\bibitem{gerard2010}
G{\'e}rard-Varet D, Hillairet M. Regularity issues in the problem of fluid
  structure interaction. \emph{Archive for Rational Mechanics and Analysis}
  2010; \textbf{195}(2):375--407.

\bibitem{davis2003}
Davis RH, Zhao Y, Galvin KP, Wilson HJ. Solid--solid contacts due to surface
  roughness and their effects on suspension behaviour. \emph{Philosophical
  Transactions of the Royal Society of London A: Mathematical, Physical and
  Engineering Sciences}  2003; \textbf{361}:871--894.

\bibitem{Tichy1995}
Tichy J. A porous media model for thin film lubrication. \emph{Journal of
  Tribology}  1995; \textbf{117}:16--21.

\bibitem{li1999}
Li WL. Derivation of {M}odified {R}eynolds equation -- a porous media model.
  \emph{Journal of Tribology}  1999; \textbf{121}(4):823--829.

\bibitem{hueber2008}
H{\"u}eber S, Stadler G, Wohlmuth BI. A primal-dual active set algorithm for
  three-dimensional contact problems with coulomb friction. \emph{SIAM Journal
  on Scientific Computing}  2008; \textbf{30}(2):572--596.

\bibitem{popp2010}
Popp A, Gitterle M, Gee MW, Wall WA. A dual mortar approach for 3d finite
  deformation contact with consistent linearization. \emph{International
  Journal for Numerical Methods in Engineering}  2010;
  \textbf{83}(11):1428--1465.

\bibitem{schott2014}
Schott B, Wall WA. A new face-oriented stabilized {XFEM} approach for 2{D} and
  3{D} incompressible {N}avier--{S}tokes equations. \emph{Computer Methods in
  Applied Mechanics and Engineering}  2014; \textbf{276}:233--265.

\bibitem{burman2015cutfem}
Burman E, Claus S, Hansbo P, Larson MG, Massing A. Cut{FEM}: Discretizing
  geometry and partial differential equations. \emph{International Journal for
  Numerical Methods in Engineering}  2015; \textbf{104}(7):472--501.

\bibitem{massing2016}
Massing A, Schott B, Wall WA. A stabilized {N}itsche cut finite element method
  for the {O}seen problem. \emph{Computer Methods in Applied Mechanics and
  Engineering}  2018; \textbf{328}:262--300.

\bibitem{nitsche1971}
Nitsche J. {\"Uber ein Variationsprinzip zur L\"osung von Dirichlet-Problemen
  bei Verwendung von Teilr\"aumen, die keinen Randbedingungen unterworfen
  sind}. \emph{Abhandlungen aus dem Mathematischen Seminar der Universit\"at
  Hamburg}  1971; \textbf{36}(1):9--15.

\bibitem{burman2014}
Burman E, Fern{\'a}ndez MA. An unfitted {N}itsche method for incompressible
  fluid--structure interaction using overlapping meshes. \emph{Computer Methods
  in Applied Mechanics and Engineering}  2014; \textbf{279}:497--514.

\bibitem{schott2017}
Schott B, Ager C, Wall W. {Monolithic cut finite element approaches for
  fluid-structure interaction}. \emph{arXiv preprint$\,$}  2018; .

\bibitem{heil2004}
Heil M. An efficient solver for the fully coupled solution of
  large-displacement fluid--structure interaction problems. \emph{Computer
  Methods in Applied Mechanics and Engineering}  2004; \textbf{193}(1):1--23.

\bibitem{Kuttler2010}
K\"{u}ttler U, Gee MW, F\"{o}rster C, Comerford A, Wall WA. Coupling strategies
  for biomedical fluid-structure interaction problems. \emph{International
  Journal for Numerical Methods in Biomedical Engineering}  2010;
  \textbf{26}:305--321.

\bibitem{verdugo2016}
Verdugo F, Wall WA. {Unified computational framework for the efficient solution
  of n-field coupled problems with monolithic schemes}. \emph{Computer Methods
  in Applied Mechanics and Engineering}  2016; \textbf{310}:335--366.

\bibitem{schrefler1998}
Schrefler BA, Lewis RW. \emph{{The Finite Element Method in the Deformation and
  Consolidation of Porous Media}}. Chichester: Wiley, 1998.

\bibitem{schrefler2001}
Schrefler BA, Scotta A. A fully coupled dynamic model for two-phase fluid flow
  in deformable porous media. \emph{Computer Methods in Applied Mechanics and
  Engineering}  2001; :3223--3246.

\bibitem{Coussy:04}
Coussy O. \emph{Poromechanics}. John Wiley and Sons: West Sussex, 2004.

\bibitem{Chapelle2010b}
Chapelle D, Moireau P. General coupling of porous flows and hyperelastic
  formulations - {F}rom thermodynamics principles to energy balance and
  compatible time schemes. \emph{European Journal of Mechanics-B\textbackslash
  Fluids}  2014; \textbf{46}:82--96.

\bibitem{Vuong2015}
Vuong AT, Yoshihara L, Wall WA. A general approach for modeling interacting
  flow through porous media under finite deformations. \emph{Computer Methods
  in Applied Mechanics and Engineering}  2015; \textbf{283}:1240 -- 1259.

\bibitem{Holzapfel2000}
Holzapfel GA. \emph{{Nonlinear Solid Mechanics}}, vol.~24. Wiley Chichester,
  2000.

\bibitem{donea2003}
Donea J, Huerta A. \emph{Finite {E}lement {M}ethods for {F}low {P}roblems}.
  John Wiley and Sons, 2003.

\bibitem{vuong2016}
Vuong AT, Ager C, Wall WA. Two finite element approaches for {D}arcy and
  {D}arcy--{B}rinkman flow through deformable porous media -- {M}ixed method
  vs. {NURBS} based (isogeometric) continuity. \emph{Computer Methods in
  Applied Mechanics and Engineering}  2016; \textbf{305}:634--657.

\bibitem{vuong2017}
Vuong AT, Rauch A, Wall WA. A biochemo-mechano coupled, computational model
  combining membrane transport and pericellular proteolysis in tissue
  mechanics. \emph{Proceedings of the Royal Society A: Mathematical, Physical
  and Engineering Science}  03 2017; \textbf{473}:20160\,812.

\bibitem{discacciati2009}
Discacciati M, Quarteroni A. {N}avier-{S}tokes/{D}arcy coupling: modeling,
  analysis, and numerical approximation. \emph{Revista Matem{\'a}tica
  Complutense}  2009; \textbf{22}(2):315--426.

\bibitem{ager2018b}
Ager C, Schott B, Winter M, Wall WA. {A Nitsche-based cut finite element method
  for the coupling of incompressible fluid flow with poroelasticity}.
  \emph{arXiv preprint$\,$}  2018; .

\bibitem{beavers1967}
Beavers GS, Joseph DD. Boundary conditions at a naturally permeable wall.
  \emph{Journal of Fluid Mechanics}  1967; \textbf{30}(01):197--207.

\bibitem{saffman1971}
Saffman PG. On the boundary condition at the surface of a porous medium.
  \emph{Studies in Applied Mathematics}  1971; \textbf{50}(2):93--101.

\bibitem{gartling1996}
Gartling D, Hickox C, Givler R. Simulation of coupled viscous and porous flow
  problems. \emph{International Journal of Computational Fluid Dynamics}  1996;
  \textbf{7}(1-2):23--48.

\bibitem{cao2010}
Cao Y, Gunzburger M, Hua F, Wang X. Coupled {S}tokes-{D}arcy model with
  {B}eavers--{J}oseph interface boundary condition. \emph{Communications in
  Mathematical Sciences}  2010; \textbf{8}(1):1--25.

\bibitem{cao2010b}
Cao Y, Gunzburger M, Hu X, Hua F, Wang X, Zhao W. Finite element approximations
  for {S}tokes--{D}arcy flow with {B}eavers--{J}oseph interface conditions.
  \emph{SIAM Journal on Numerical Analysis}  2010; \textbf{47}(6):4239--4256.

\bibitem{wohlmuth2001}
Wohlmuth BI. Discretization techniques based on domain decomposition.
  \emph{Discretization Methods and Iterative Solvers Based on Domain
  Decomposition}. Springer, 2001; 1--84.

\bibitem{braack2007}
Braack M, Burman E, John V, Lube G. Stabilized finite element methods for the
  generalized {O}seen problem. \emph{Computer Methods in Applied Mechanics and
  Engineering}  2007; \textbf{196}(4):853--866.

\bibitem{sudhakar2014}
Sudhakar Y, De~Almeida JM, Wall WA. An accurate, robust, and easy-to-implement
  method for integration over arbitrary polyhedra: application to embedded
  interface methods. \emph{Journal of Computational Physics}  2014;
  \textbf{273}:393--415.

\bibitem{burman2010}
Burman E. Ghost penalty. \emph{Comptes Rendus Mathematique}  2010;
  \textbf{348}(21-22):1217--1220.

\bibitem{Badia2009}
Badia S, Quaini A, Quarteroni A. Coupling {B}iot and {N}avier-{S}tokes
  equations for modelling fluid-poroelastic media interaction. \emph{Journal of
  Computational Physics}  2009; \textbf{228}:7986--8014.

\bibitem{dangelo2011}
D'Angelo C, Zunino P. Robust numerical approximation of coupled {S}tokes' and
  {D}arcy's flows applied to vascular hemodynamics and biochemical transport.
  \emph{ESAIM: Mathematical Modelling and Numerical Analysis}  2011;
  \textbf{45}(3):447--476.

\bibitem{bukavc2015}
Buka{\v{c}} M, Yotov I, Zakerzadeh R, Zunino P. Partitioning strategies for the
  interaction of a fluid with a poroelastic material based on a {N}itsche's
  coupling approach. \emph{Computer Methods in Applied Mechanics and
  Engineering}  2015; \textbf{292}:138--170.

\bibitem{Juntunen2009}
Juntunen M, Stenberg R. {N}itsche's method for general boundary conditions.
  \emph{Math. Comput.}  2009; \textbf{78}:1353--1374.

\bibitem{winter2017}
Winter M, Schott B, Massing A, Wall W. A {N}itsche cut finite element method
  for the {O}seen problem with general {N}avier boundary conditions.
  \emph{Computer Methods in Applied Mechanics and Engineering}  2018;
  \textbf{330}:220 -- 252.

\bibitem{puso2004}
Puso MA, Laursen TA. A mortar segment-to-segment frictional contact method for
  large deformations. \emph{Computer Methods in Applied Mechanics and
  Engineering}  2004; \textbf{193}(45):4891--4913.

\bibitem{popp2013}
Popp A, Seitz A, Gee MW, Wall WA. Improved robustness and consistency of 3d
  contact algorithms based on a dual mortar approach. \emph{Computer Methods in
  Applied Mechanics and Engineering}  2013; \textbf{264}:67--80.

\bibitem{WalletBaciCommittee2017}
Wall WA, Ager C, Grill M, Kronbichler M, Popp A, Schott B, Seitz A. {BACI: A
  multiphysics simulation environment}. \emph{Technical {R}eport}, Institute
  for Computational Mechanics, Technical University of Munich 2018.

\bibitem{lorenz2010}
Lorenz B, Persson BN. Leak rate of seals: Effective-medium theory and
  comparison with experiment. \emph{The European Physical Journal E: Soft
  Matter and Biological Physics}  2010; \textbf{31}(2):159--167.

\end{thebibliography}
\bibliographystyle{wileyj.bst}

\end{document}